\documentclass[journal]{new-aiaa}
\usepackage[utf8]{inputenc}
\usepackage{amsmath}
\usepackage{indentfirst}
\usepackage{subfig}
\usepackage{courier}
\usepackage{graphicx}
\usepackage{algorithm,algorithmic}
\usepackage{geometry}
\usepackage{dsfont}
\usepackage{color}
\usepackage{multirow}
\usepackage{verbatim}
\usepackage[justification=centering]{caption}
\usepackage{siunitx}
\usepackage{longtable,tabularx}
\usepackage{multirow}
\usepackage{hhline}
\pdfoutput=1

\newcommand{\beq}{\begin{eqnarray}}
\newcommand{\eeq}{\end{eqnarray}}
\newcommand{\bal}{\begin{eqnarray}\begin{aligned}}
\newcommand{\eal}{\end{aligned}\end{eqnarray}}

\newcommand{\mb}{\mathbf}

\newcommand{\re}{\mathds{R}}

\newcommand{\RNum}[1]{\uppercase\expandafter{\romannumeral #1\relax}}

\title{Reduced Order Modeling Framework for Combustor Instabilities Using Truncated Domain Training}
\author{Jiayang Xu\footnote{Graduate Student Research Assistant, Aerospace Engineering Department, AIAA Student Member, davidxu@umich.edu.}, 
Cheng Huang\footnote{Assistant Research Scientist, Aerospace Engineering Department, AIAA Fellow, huangche@umich.edu.}, 
Karthik Duraisamy\footnote{Associate Professor, Aerospace Engineering Department, AIAA Fellow, kdur@umich.edu.} \\
{\normalsize\itshape
University of Michigan, Ann Arbor, MI, 48109}} 

\begin{document}

\maketitle

\begin{abstract}

A multi-fidelity framework is established and demonstrated for prediction of combustion instabilities in rocket engines. The major idea is to adapt appropriate fidelity modeling approaches for different components in a rocket engine to ensure  accurate and efficient  predictions. 
Specifically, the proposed framework integrates projection-based Reduced Order Models (ROMs) that are developed using bases generated on truncated domain simulations. The ROM training is performed on  truncated domains, and thus does not require full order model solutions on the full rocket geometry, thus demonstrating the potential to greatly reduce training cost.  Geometry-specific training is replaced by the response generated by  perturbing the characteristics at the boundary of the truncated domain. This training method is shown to enhance predictive capabilities and robustness of the resulting ROMs, including at conditions outside the training range. Numerical tests are conducted on a quasi-1D model of a single-element rocket combustor and the present framework is compared to  traditional ROM development approaches. 

\end{abstract}
\section*{Nomenclature}
{\renewcommand\arraystretch{1.0}
\noindent\begin{longtable*}{@{}l @{\quad=\quad} l@{}}
{$\alpha$}&{amplification factor}\\
{$A$}&{cross sectional area}\\
{${C}_{f/o}$}&{stoichiometric fuel-to-oxidizer ratio}\\
{$\eta$}&{cumulative energy in singular values}\\
{$gr$}&{pressure oscillation growth rate}\\
{$J$}&{characteristic invariant for incoming acoustic wave}\\
{$k$}&{number of modes in proper orthogonal decomposition (POD) basis}\\
{$L_c$}&{chamber length}\\
{$l_s$}&{starting location of fuel injection}\\
{$l_f$}&{ending location of fuel injection}\\

{${\dot m }_f$}&{fuel mass flow rate}\\
{${\dot \omega }_{f}$}&{fuel consumption rate}\\
{${\dot \omega }_{ox}$}&{oxidizer consumption rate}\\
{$q'$}&{unsteady heat release rate}\\
{$\sigma$}&{singular value}\\
{$\tau$}&{time lag}\\
{$\mb{V}$}&{POD basis}\\
{$\mb{V}\bot$}&{complementary basis of $\mb{V}$}\\
{$\xi$}&{variable for fuel injection shape description}\\
{$Y_{ox}$}&{oxidizer mass fraction}\\


\end{longtable*}}
\section{Introduction}
Combustion instabilities have long been a hurdle in the development of modern rocket engines. These instabilities are  characterized by the coupling between acoustics, hydrodynamics and heat release. In propulsion systems, the triggering of combustion instabilities can lead to catastrophic engine failures, and the underlying physical mechanisms are sensitive to many parameters including the geometry and fuel-to-oxidizer ratio. Therefore, a significant number of geometric configurations and operating conditions have to be evaluated and analyzed in rocket engine design. Though the advancement of modern computational technology has enabled routine modeling of laboratory-scale rockets ~\cite{harvazinski2015coupling} and other experimental configurations such as atmospheric combustors~\cite{matsuyama2016large}, direct numerical simulations (DNS)~\cite{domingo2005dns} and large eddy simulations (LES)~\cite{ihme2012large,huang2014computational,harvazinski2015coupling,lacaze2009large,hernandez2011laboratory,srinivasan2015flame} continue to be prohibitively expensive for the simulation of full-scale engines~\cite{urbano2016exploration,staffelbach2009large,wolf2012using}. This motivates research on alternative approaches such as reduced-fidelity modeling \cite{baukal2000computational,frezzotti2014low, sirignano2013two, popov2016transverse,you2005generalized} and reduced order modeling~\cite{lieu2006reduced,lucia2004reduced}. 

Reduced-fidelity modeling approaches usually adapt certain levels of simplifications in physical and numerical models to achieve higher computational efficiency. 
Sirignano and Popov \cite{sirignano2013two} developed a two-dimensional model for transverse-mode combustion instability in a cylindrical rocket motor and further extended it to rectangular configurations~\cite{popov2016transverse}. Though reduced-fidelity modeling approaches can be efficiently used for engine design with satisfying accuracy for critical quantities of interest,  many challenges are encountered in accurately predicting crucial combustion dynamics that can trigger instability.

In addition to approaches that model  combustion dynamics by solving transport equations,  empirical approaches have been pursued. These approaches  formulate the combustion response as a function of well-characterized flow variables such as pressure and velocity. This class of techniques include the so-called flame transfer  (FTF)~\cite{durox2009experimental} and flame describing functions (FDF)~\cite{noiray2008unified}. In Ref. \cite{you2005generalized}, more factors such as variations in geometry are taken into account and the response function is successfully applied to a swirl-stabilized combustor. Though FTF/FDF approaches impose a simple model form for the flame dynamics that can be easily derived from both experimental and computational data, their applications are usually restricted to combustion system with low-amplitude perturbations with regards to the mean flow due to the fact that the derivation of the flame model is based on linear assumptions in frequency domain. Therefore, to model highly nonlinear flame dynamics which is typical in rocket engines, the validity of the FTF/FDF approach remains to be evaluated. 

In contrast, reduced order modeling (ROM) approaches pursue mathematically formal reductions, offering the promise of generalization. Projection-based ROMs~\cite{chatterjee2000introduction,lucia2003projection} have recently been applied to combustion instability problems~\cite{xu2017reduced,xu2018multi,huang2018exploration,huang2016multi,huang2017multi}.
In projection-based ROMs, the  full order model is projected onto a set of informative basis functions using proper orthogonal decomposition (POD)-Galerkin method~\cite{rowley2004model}, resulting in a low dimensional set of equations that can retain most of the modeling fidelity. Thus, the need for commonly used empirical modeling strategies such as combustion response functions can be eliminated. Preliminary explorations of projection-based ROMs on representative combustion model problems can be found in Refs. \cite{xu2017reduced,huang2018exploration,huang2016multi,huang2017multi}. It should be pointed out that POD-based ROMs  have been more widely used in non-reacting flow problems such as aerodynamics~\cite{ravindran2000reduced, rowley2004model}, aeroelasticity~\cite{amsallem2008interpolation,lieu2007adaptation} and flow control~\cite{barbagallo2012closed,barbagallo2011input}.

However,  applications of ROMs to rocket engines involve the following challenges:  
\begin{itemize}
\item A single high fidelity simulation or Full Order Model (FOM) of a full engine may be prohibitively expensive.

\item Even if full engine FOM simulations are affordable, conventional ROM construction is based on FOM simulations for different injector configurations. If there is any change to the geometry, the ROM has to be re-generated or re-parameterized, which restricts the scope of ROM applications. 


\item POD-ROMs  may be incapable of providing predictions beyond the training interval and parameters.

\item 
Since  POD basis construction involves global minimization,  the resulting basis will be  dominated by high-energy regions. The dynamics in low-magnitude regions can be under-resolved and consequently affect stability and robustness.

\end{itemize}

The present work extends the multi-fidelity framework methodology developed by Huang et al.~\cite{huang2016multi,huang2017multi,huang2019multifidelity} and assesses ROM performance on the major aspects listed above based on a quasi-1D model combustor problem, which was originally developed by Smith et al.~\cite{smith2008computational} to model the experimental work by Yu et al. on the single-element combustor~\cite{yu2012spontaneous,yu2009experimental}. It should be mentioned that additional challenges exist in ROMs of more realistic problems, where complex physics such as three dimensional turbulent combustion, vortex shedding, and flame-vortex interactions are present. While progress is being made in addressing these issues~\cite{huang2018exploration,huang2019investigations,huang2018challenges}, the present work focuses on the issue of ROM development in problems where the full order solution is not available for the entire domain.  The approach itself is generic for problems of different levels of complexity. The framework pursues a multi-domain approach where the computational domain is divided into two components: the upstream one covering the critical physics in the heat addition/acoustic-flame interaction region and the downstream part dominated by flow and acoustics dynamics. Reduced order models are used to model the critical dynamics in the upstream component. The ROM is developed and trained based on FOM solutions of the truncated domain corresponding to the upstream component with a perturbed characteristic downstream boundary condition. This training strategy is designed to incorporate rich information in the resulting ROM and proves to be effective in enhancing predictive capabilities. The truncated domain ROM is then integrated within a multi-fidelity solver~\footnote{For the current demonstration, we replace the reduced-fidelity part with the full order model for consistency of evaluations}. The multi-fidelity model is then compared to ROMs based on the FOM solutions of the entire computational domain, which will be referred as  conventional ROMs in the current paper.

The rest of this paper is organized as follows: The quasi-1D combustor setup and formulations for the FOM and ROM are described in Sec. \ref{sec formulation}. In Sec. \ref{sec framework}, the present framework and ROM training with characteristic boundary are introduced.  Numerical experiments are presented in Sec. \ref{sec result_domain}. Concluding remarks are given in Sec. \ref{sec conclusion}
\section{Formulation}\label{sec formulation}
\subsection{Full order model}
\begin{figure}
\begin{center}
	\includegraphics[width=0.8\textwidth]{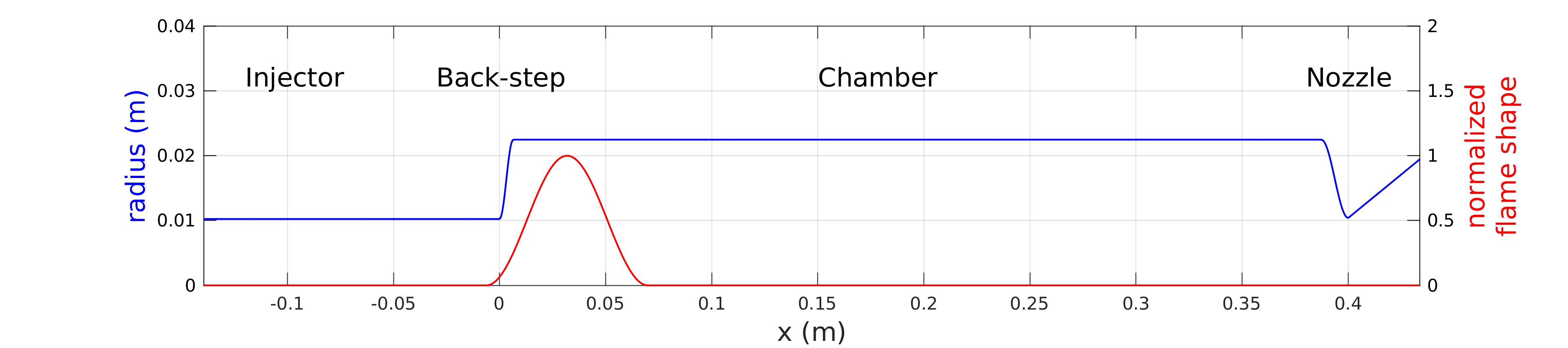}
	\caption {Computational domain and flame shape}\label{fig geometry} 
	\end{center}
\end{figure}

In this work, a quasi-1D version of the single element Continuously Variable Resonance Combustor (CVRC)~\cite{yu2012spontaneous,yu2009experimental} is used as the full order model. The geometry of the CVRC is sketched in Fig. \ref{fig geometry}. Numerical studies are performed with multiple chamber lengths, which exhibit various instability behaviors. The geometric parameters for other sections are given in Table \ref{table geometry}. To avoid invalidating the quasi-1D assumption, the back-step and the converging part of the nozzle are sinusoidally contoured in this study, in contrast to a discontinuous change in the cross-sectional area in the experimental setup. The same gas properties and operating conditions as in Ref.~\cite{xu2017reduced} are adopted and listed in Table \ref{CVRC}. 

\begin{table}[h]
	\begin{center}
	\caption {Geometry parameters}\label{table geometry}
		\begin{tabular}{ |c| c| c| c| c|}
		\hline
			Section 	&Injector	&Back-step	    &Nozzle converging part	&Nozzle diverging part\\
			\hline
			Length (m)	&0.1397		&0.0064			&0.0127				&0.034\\
			Radius (m)	&0.0102		&0.0102 to 0.0225	&0.0225 to 0.0104     &0.0104 to 0.0195\\
			\hline
		\end{tabular}
	\end{center} 
\end{table}

\begin{table}
	\begin{center}
	\caption {CVRC operating conditions}\label{CVRC} 
		\begin{tabular}{ |c| c| }
		\hline
			Parameter & Value\\ 
			\hline
			Fuel mass flow rate, kg/s 		& 0.027\\
			Fuel temperature, K 			& 300\\
			Oxidizer mass composition       & $57.6\%\: H_2O+42.4\%\:O_2$\\
			Oxidizer mass flow rate, kg/s 	& 0.32\\
			Oxidizer temperature, K			& 1030\\
			Fuel composition                & $100\%\:CH_4$\\
			Equivalence ratio				& 0.8\\
			\hline
		\end{tabular}
	\end{center}
\end{table}

The governing equations in this problem are the quasi-1D unsteady Euler equations with species transport, represented as
\begin{equation}\label{eq fom}
	\frac{{\partial \mathbf{q}}}{{\partial t}} + \frac{{\partial \mathbf{f}}}{{\partial x}} = \mathbf{s}_A+\mathbf{s}_f+\mathbf{s}_q,
\end{equation}
where 
\small  
\begin{equation}
\mathbf{q} = \left( {\begin{array}{*{20}{c}}
	\rho A  \\ 
	{\rho uA} \\ 
	{\rho{E}A} \\ 
	{\rho {Y_{ox}A}} 
	\end{array}} \right),
\mathbf{f} = \left( {\begin{array}{*{20}{c}}
	{\rho uA} \\ 
	{\left(\rho {u^2} + p\right)A} \\ 
	{\left(\rho{E} + p\right)uA} \\ 
	{\rho u{Y_{ox}}A} 
	\end{array}} \right),
\mathbf{s}_A = \left( {\begin{array}{*{20}{c}}
	0 \\ 
	{p\frac{{dA}}{{dx}}} \\ 
	0 \\ 
	0 
	\end{array}} \right),
\mathbf{s}_f = \left( {\begin{array}{*{20}{c}}
	{{{\dot \omega }_f}} \\ 
	{{{\dot \omega }_f}u} \\ 
	{{{\dot \omega }_f}\Delta {h_0}} \\ 
	{ - {{\dot \omega }_{ox}}} 
	\end{array}} \right),
\mathbf{s}_q = \left( {\begin{array}{*{20}{c}}
	0 \\ 
	0 \\ 
	{q'} \\ 
	0 
\end{array}} \right).
\end{equation}
\normalsize

In the conservative variable vector $\mathbf{q}$, $\rho$ is the density, $u$ is the velocity, $E$ is the total internal energy, $Y_{ox}$ is the oxidizer mass fraction, and $A$ is the cross sectional area. The corresponding convective fluxes are represented by the vector $\mathbf{f}$, where $p$ is the static pressure. 

In the three source terms, $\mathbf{s}_A$ is due to area variations and the latter two terms represent combustion. In the experiment~\cite{yu2009experimental}, the fuel is injected through an annular ring located at the back-step and reacts at a finite rate with the oxidizer injected. In this work, we follow the choice of Frezzotti et al.~\cite{frezzotti2015response, frezzotti2015parametric} and assume an infinitely-fast one-step combustion model. An important assumption behind the model is that when fuel is injected, it will react with the oxidizer instantaneously to form products and no intermediate species are produced, thus only one species transport equation is needed. This process is accounted for in $\mathbf{s}_f$. As suggested by Frezzotti et al.~\cite{frezzotti2015response, frezzotti2015parametric}, to reproduce the combustion region of a finite width and avoid discontinuities, the fuel injection process is modeled in a sinusoidal shape at the rate of ${\dot \omega }_f$, yielding
\bal\label{eq flame}
{{\dot \omega }_f} = \frac{{{{\dot m}_f}}}{{\int {\left( {1 + \sin \xi } \right)} dx}}\left( {1 + \sin \xi } \right),\\
\xi  =  - \frac{\pi }{2} + 2\pi \frac{{x - {l_s}}}{{{l_f} - {l_s}}},{\text{with }}{l_s} < x < {l_f},
\eal
where ${\dot m}_f$ is the total mass flow rate of fuel injection, and $l_s$ and $l_f$ are the starting and ending location of the flame, respectively. 
Due to the infinitely-fast model, the consumption rate of the oxidizer is 
\beq
{\dot \omega }_{ox} = \frac{{\dot \omega }_f}{C_{f/o}},
\eeq
where $C_{f/o}$ is the stoichiometric fuel-to-oxidizer ratio. The shape of the fuel injection and resulting flame given by Eq. \eqref{eq flame} is also shown in Fig. \ref{fig geometry} along with the CVRC geometry.  

The final source term $\mathbf{s}_q$ models the unsteady heat release and represents the response function introduced by Crocco et al.~\cite{crocco1958importance},  given in Eq. \eqref{eq alpha-tau}. The function takes into account the coupling between acoustics and combustion by expressing the unsteady part of the heat release as a function of pressure with an amplification parameter $\alpha$ and a time lag $\tau $.

\begin{equation}\label{eq alpha-tau}
q' = {\dot \omega _f}\Delta {h_0}\alpha \frac{{p\left( {x,t - \tau } \right) - \bar p\left( x \right)}}{{\bar p\left( x \right)}}.
\end{equation}


In the simulation, a steady state is first achieved with the unsteady source term $\mathbf{s}_q$ turned off. From the steady state, a low-amplitude perturbation is applied to the inlet boundary to trigger the instability at the beginning of predictive unsteady simulations (either FOM or ROM).  After the perturbation, the unstable response will become self-excited and start to grow under certain parameters. The instability behavior is described in Sec. \ref{sec result_domain}.

Numerical solutions of  Eq. \eqref{eq fom} will be used as a surrogate for the high-fidelity solution in the current work, and the corresponding simulation is referred as FOM.

\subsection{Projection-based ROM}
To derive the projection-based ROM equations, we first rewrite the full order model given by Eq. \eqref{eq fom} for each variable as
\begin{equation}\label{eq FOM1}
\frac{{\partial {\mathbf{q}^i}}}{{\partial t}} = {\mathbf{r}^i}\left( {{\mathbf{q}}\left( {x,t} \right)} \right),
\end{equation}
where the variable index $i$  corresponds to the four conserved variables respectively. The residual vector for variable $i$ consists of the corresponding flux and source terms
\begin{equation}
\mathbf{r}^i =  - \frac{{\partial \mathbf{f}^i}\left( {{\mathbf{q}}\left( {x,t} \right)} \right)}{\partial x} + \mathbf{s}^i.
\end{equation}

For each variable, defining an individual POD projection basis $\mb{V}^i \in \re^{n_{grid}\times k}$ spanning a subspace $\mathcal{V} \subset \re^{n_{grid}}$, and a complementary basis $\mb{V}^i_\bot$ spanning $\mathcal{V}_\bot$, such that $\mathcal{V} \oplus \mathcal{V}_\bot = \re^{n_{grid}}$, the following decomposition can be derived:
\beq\label{split_q}
	\mb{q}^i = \mb{V}^i \mb{q}^i_r + \mb{V}^i_\bot \mb{q}^i_\bot = \tilde{\mb{q}}^i + \hat{\mb{q}}^i.
\eeq

The two bases $\mb{V}^i$ and $\mb{V}^i_\bot$ are obtained by performing singular value decomposition (SVD) on the snapshots of the $i$-th conserved variable. Defining $k$ to be the designed size of the reduced order system for variable $i$,  $\mb{V}^i$ is constructed column-wise from the first $k$ left-singular vectors from the SVD~\footnote{The remaining left-singular vectors are $\mb{V}^i_\bot$}. The conserved variables are then approximated as
\begin{equation}\label{eq ROM0}
    {\mathbf{q}^i}(x,t) \approx {\bf{\tilde q}}^i = \mathbf{V}^i{\mathbf{q}^i_r}=\sum\limits_{n=1}^{k} {q^{i(n)}_r(t)\mb{v}^{i(n)}(x)},
\end{equation}
where $q^{i(n)}_r$ is the $n$-th element in the reduced order variable vector $\mb{q}^i_r \in \re^k$, and $\mb{v}^{i(n)}$ is the $n$-th column in the POD basis $\mb{V}^i$. Since each left-singular vector represents a spatial mode of the FOM solution, the approximation can be viewed as a superimposition of the first $k$ leading modes, with $q^{i}_r$ being the temporal coefficient in the evolution of the $i$-th mode. 

The Galerkin-projected ROM for $k$ coefficients is then
\begin{equation}\label{eq ROM1}
	\frac{{d {\mathbf{q}^i_r}}}{{d t}} = {(\mathbf{V}^i)^T}{\mathbf{r}^i}\left({\mathbf{V}\mathbf{q}_r} \right).
\end{equation}

For clarity of presentation, $\mathbf{V}\mathbf{q}_r$ is used to denote the approximated conserved variables, including ${\mathbf{V}^1\mathbf{q}^1_r},\dots,{\mathbf{V}^4\mathbf{q}^4_r}$.

In practice, to achieve further reduction in computational cost and to improve robustness, additional constructs such as sparse sampling techniques and stabilization will be required. While we are developing  such techniques~\cite{xu2017reduced,xu2018multi,huang2018exploration,ericpaper}, they will not be considered in the present work, as the focus is on multi-domain modeling.

\section{Framework Overview}\label{sec framework}

\begin{figure}
\begin{center}
	\includegraphics[width=0.6\textwidth]{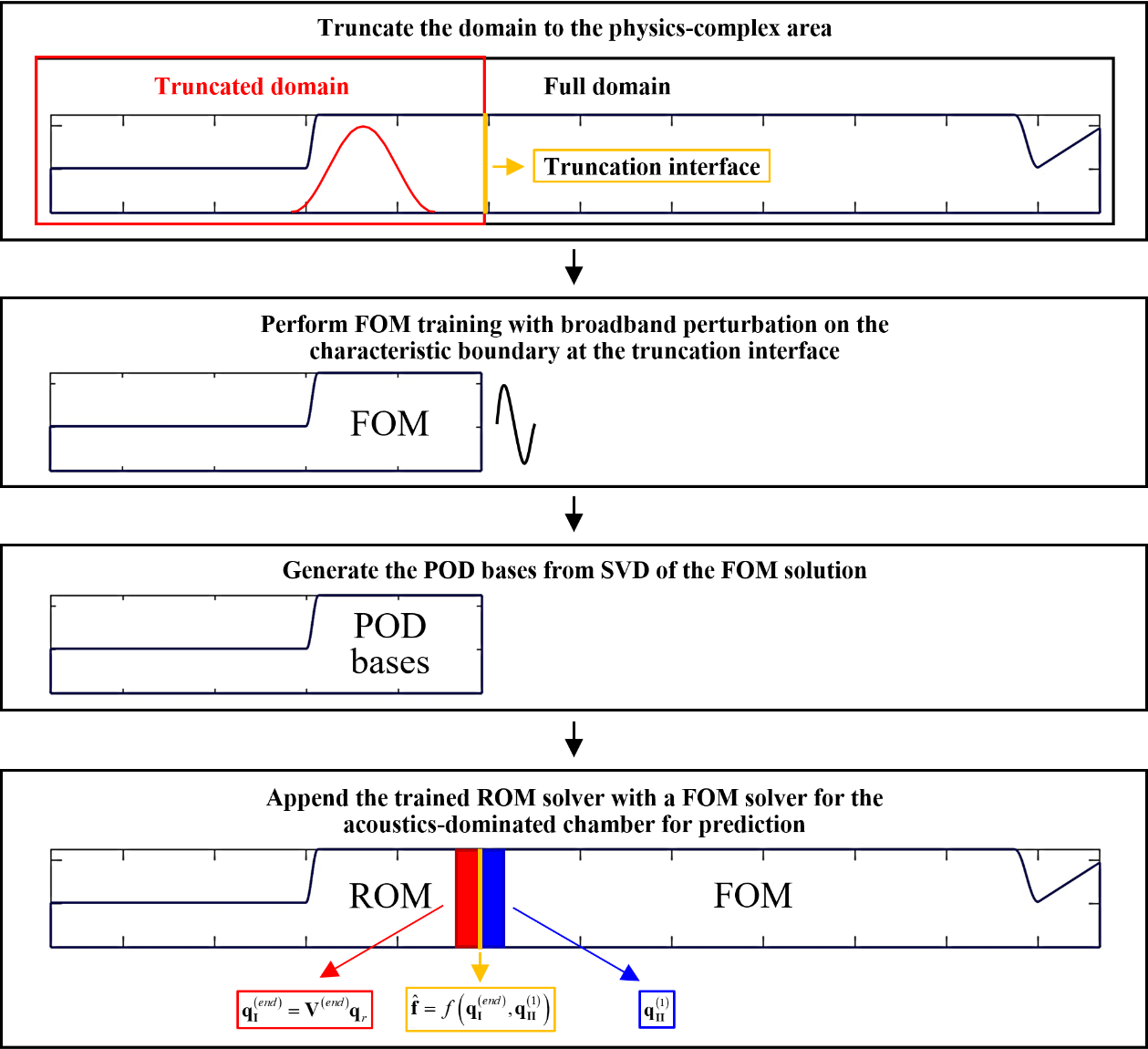}
	\caption {Schematic of framework}\label{fig frameworkschematic} 
	\end{center}
\end{figure}

In this section, we explore characteristic-based ROM training on a truncated domain and contrast it with  conventional ROM development. 
A  schematic is given in Fig. \ref{fig frameworkschematic}. The major steps include
\begin{enumerate}
  \item Truncate the domain into two sub-domains, the first containing the physics-complex area that covers the fuel injection and the flame, and the second containing the variable-length chamber dominated by acoustics. The ROM will be trained for the first sub-domain.
  \item Perform a FOM training simulation on the first sub-domain with a broadband perturbation added on the truncation interface, which is treated as a characteristic boundary. 
  \item Generate the POD bases for different variables using SVD of the solution from the training simulation. 
  \item Use the POD bases in a ROM solver for the first sub-domain and couple it with a FOM solver for the variable-length, acoustics-dominated chamber for predictions. 
\end{enumerate}

It can be noted that in the framework ROM is only developed for the first sub-domain, whereas the second sub-domain is solved in FOM. There are two reasons for this choice:
\begin{enumerate}
    \item The complex, computation-intensive area is fully covered in the first sub-domain. The accurate modeling of this domain requires high-resolution simulations. The FOM computation in the second sub-domain is expected to be much less challenging than the first domain and affordable with coarser resolution modeling approaches (e.g. coarse-mesh LES and URANS), which makes a ROM replacement unnecessary.
    \item In design evaluations, the chamber length in the second sub-domain is variable, thus a new ROM training is required for each chamber length, which violates the goal of reducing the computing cost using ROM.
\end{enumerate}

\subsection{Characteristic ROM training on reduced-domain}
 In traditional ROM formulations, POD bases are  generated using FOM solutions of the target full domain. In the proposed approach, the domain is split into two sub-domains with the interface at $x=0.096$ m.  The first sub-domain includes the injector, back-step and the leading part of the chamber where the flame is located, and the second includes the rest of the geometry, which is variable in design evaluations. 

To obtain a basis that is representative of the physics in the first sub-domain, a FOM simulation is performed in a reduced domain, treating the truncation interface as a characteristic outlet boundary, which helps to eliminate the undesirable resonant acoustic modes corresponding to the reduced-domain geometry. As demonstrated by Huang et al.~\cite{huang2017multi}, the presence of such resonant acoustic modes can significantly affect the predictive capabilities of the ROMs. A similar training procedure as in Refs. ~\cite{huang2016multi,huang2017multi} is used. The training FOM simulation is also started from the steady state as in the self-excited simulations on the full domain (Sec. \ref{sec formulation}). The difference is that instead of a single frequency inlet perturbation, a broadband perturbation is imposed on the incoming characteristic wave at the truncation boundary to cover the range of responses created by the resonant frequencies corresponding to different chamber lengths. This perturbation is imposed over  the entire duration of the FOM simulation.

To maintain consistency, the rest of the specified properties at the boundary are as follows:
\beq
\Omega_{bc} = \{J, u, T, Y_{ox}\}^T,
\eeq
where $c$ is the speed of sound, $J = -\frac{p}{\overline{\rho c}} + u$ is one-dimensional approximation of the characteristic invariant for the in-coming acoustic wave, $\overline{\rho c}$ is obtained from the steady state solution, and the other three primitive variables are extrapolated from the values of the interior cells.


\subsection{Multi-domain coupling}
Following the aforementioned simulations, the bases $\mb{V}$ for the conserved variables are obtained from the reduced domain solution snapshots, and used in POD-Galerkin projection (Eq. \eqref{eq ROM1}) to derive the ROM for the first sub-domain. The second sub-domain is computed using the full order equations \eqref{eq fom}. Due to the hybrid integration, chamber lengths can be changed without any further modifications to the ROM. The two sub-domains communicate at every time-step by exchanging interface conditions to be the value at the neighboring cell of the adjacent domain. Similar to the interior of  the computing domain, Roe's upwind flux \cite{roe1986characteristic} is used at the interface. Using subscript \RNum{1} and \RNum{2} for the first and second sub-domain, and superscript $(1)$ and $(end)$ for the first and last cell in the corresponding sub-domain respectively, the flux at the interface is given by
\beq
{\bf{\hat f}} = \frac{1}{2}\left( {\bf{f}}^{(end)}_{\text{\RNum{1}}} + {{\bf{f}}^{(1)}_{\text{\RNum{2}}}} \right) - \frac{1}{2}\left| {\frac{{\partial {\bf{f}}}}{{\partial {\bf{q}}}}\left( {{{\bf{q}}^*}} \right)} \right|\left( {{{\bf{q}}^{(1)}_{\text{\RNum{2}}}} - {{\bf{q}}^{(end)}_{\text{\RNum{1}}}}} \right),
\eeq
where $\mb{q}^*$ is the Roe-averaged state calculated from $\mb{q}^{(end)}_{\text{\RNum{1}}}$ and $\mb{q}^{(1)}_{\text{\RNum{2}}}$. $\mb{q}^{(end)}_{\text{\RNum{1}}}$ is computed from the last row of the basis and the reduced variable, $\mb{q}^{(end)}_{\text{\RNum{1}}} = \mb{V}^{(end)}\mb{q}_r$.

\subsection{Control groups for comparison}
The methodology detailed above is compared to the following  control groups using the conventional ROM approach:
\begin{description}
\item[Control group A] also uses a hybrid multi-domain solver, i.e. the first sub-domain is solved using the ROM and the second solved using the FOM. The difference from the proposed framework is in the training data generation and collection stage. In control group A, for each combination of chamber length and $\alpha$, a full-domain FOM simulation is performed instead of the proposed characteristic training on the truncated-domain. Then the solution is restricted to the first sub-domain and collected to generate the POD bases for the ROM of the sub-domain. 
\item[Control group B] uses a traditional full-domain ROM. The same FOM training simulations as in control group A are used and the POD bases are directly generated on the full-domain solution.
\end{description}

To summarize, let $n_{L_c}$ and $n_{\alpha}$ be the number of chamber lengths and amplification factors studied, respectively. In the proposed framework, $n_{\alpha}$ FOMs simulated on the reduced domain are used in total~\footnote{although as shown in the next section, this can be reduced} using characteristic perturbations. Then the first sub-domain is simulated using the ROM, second sub-domain using the FOM. In control groups A\&B, $n_{L_c}\times n_{\alpha}$ self-excited FOM simulations are conducted. In group A, the first sub-domain is solved using a ROM, and the second is solved using the FOM. In group B, the whole domain is solved using a ROM. 

\begin{figure}
\begin{center}
	\includegraphics[width=1\textwidth]{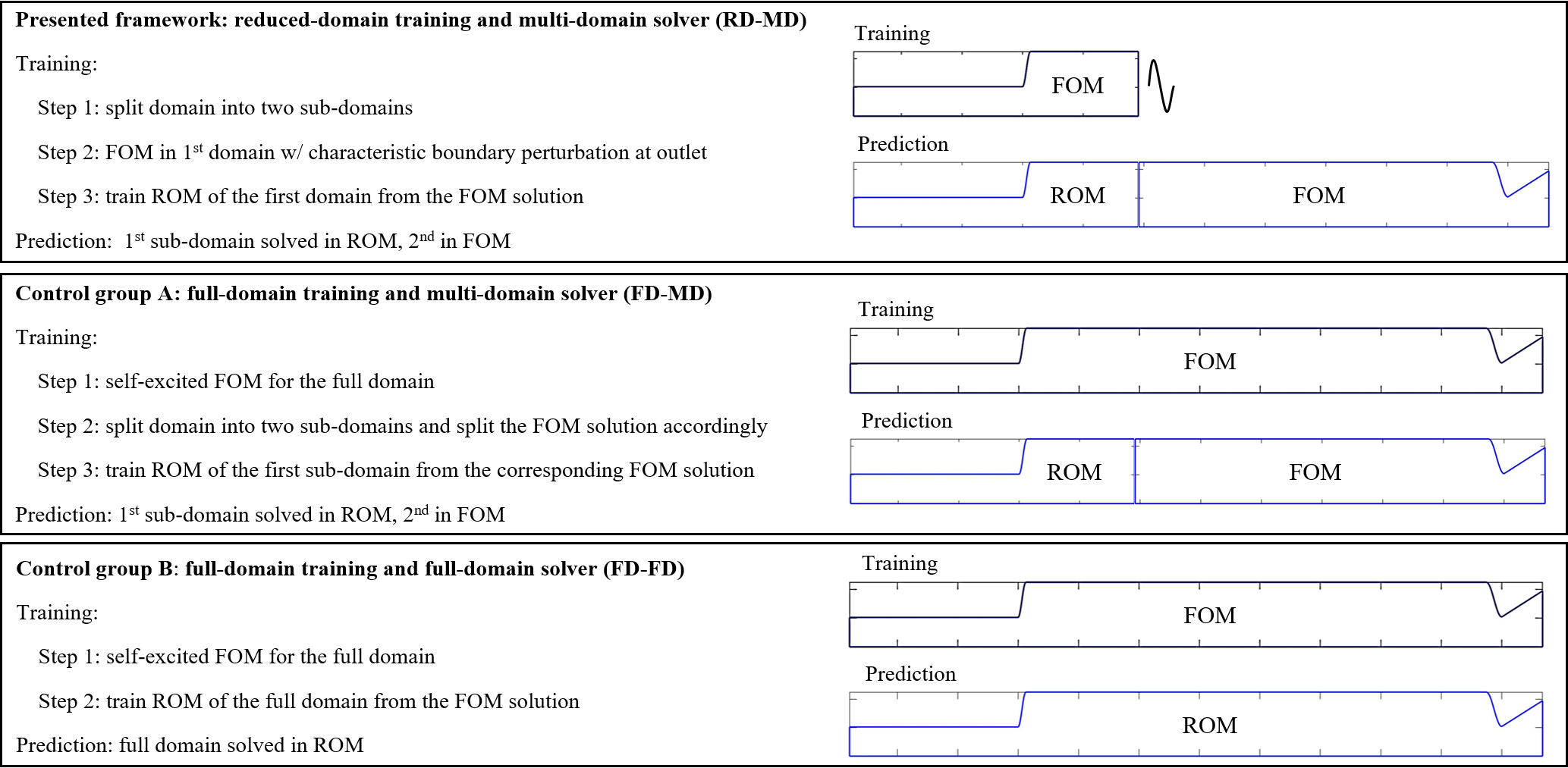}
	\caption {Schematic of training (outlined in black) and prediction (outlined in blue) stages, curves indicating perturbation added at the boundary}\label{fig flowchart} 
	\end{center}
\end{figure}

For conciseness, the proposed framework will be referred to as ``reduced-domain training and multi-domain solver (RD-MD)", control group A as ``full-domain training and multi-domain solver (FD-MD)", and control group B as ``full-domain training and full-domain solver (FD-FD)". A schematic  of the different methods is given in Fig. \ref{fig flowchart}.

\section{Results}\label{sec result_domain}
Numerical tests are conducted for chamber lengths $L_c$ ranging from 0.254 m to 0.762 m with an interval of $\Delta L_c=0.0635$ m and at $\alpha = 3.1,3.25$ and $3.4$. In the proposed training method, a FOM simulation is performed in reduced domain for each $\alpha$. 
From \textit{a priori}~\cite{grenda1995application} analysis, the longitudinal frequency of the chamber spans approximately between 700 and 2000 Hz, and thus the broadband perturbation signal is set to
\bal\label{eq train signal}
J'\left( t \right) = 0.01{J_0}\sum\limits_{i_f = 1}^{n_f} {\sin \left( {2\pi (f_0+(k-1)\Delta{f})t} \right)},
\eal
which is a superimposition of $n_f=14$ frequencies, starting from $f_0 =  700\text{ Hz}$ with an incremental interval $\Delta f = 100\text{ Hz}$. 

To train the control groups, full-domain self-excited FOM simulations are conducted for each combination of $L_c$ and $\alpha$. 
In all three methods, snapshots of the training solution are collected every 100 time-steps over a period  $t=0 - 0.05$ s, with the initial steady state set at $t = 0$. Once the basis $\mb{V}$ is obtained using the SVD of the corresponding snapshot, ROM simulations are conducted over $t=0 -0.1$ s. 

\subsection{FOM results} 
Results from the full-domain FOM simulations are presented to characterize the instability behavior. The predicted response is visualized using pressure signals obtained 0.0127 meters upstream of the converging part of the nozzle, which is a typical location selected to probe combustion instabilities in the work conducted in Ref. \cite{frezzotti2015response}. 

\begin{figure}
	\centering
	\subfloat{
		\includegraphics[width=0.5\textwidth]{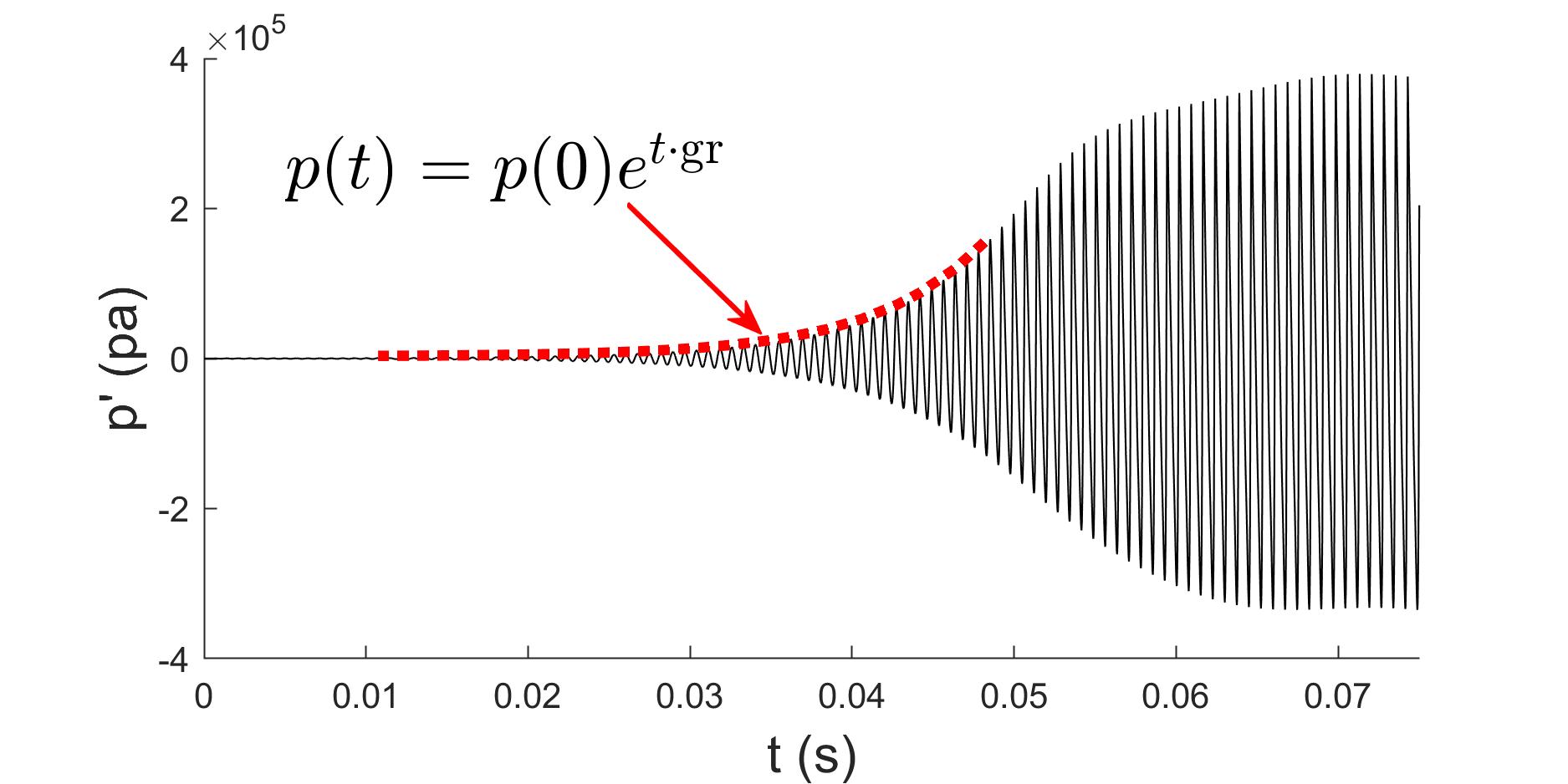}}\\
	\caption{Definition of growth rate}\label{fig growth rate} 
\end{figure}

With different combinations of parameters, two general categories of responses are identifiable: one with positive growth rate, in which the pressure oscillation grows and settles into a limit cycle oscillation (LCO),  and one with negative growth rate in which the instability  starts to decay after the perturbation ends and the flow converges to a steady state. The definition of the growth rate, $gr$, is based on the peak-to-peak amplitude of the unsteady part of the pressure signal, whose growth is fitted to a exponential function plotted in Fig. \ref{fig growth rate}, given by
\begin{equation}\label{eq gr}
    p(t)=p(0)e^{t\cdot gr}.
\end{equation}

\begin{figure}
	\centering
	\subfloat[Pressure signal at the monitored location]{
		\includegraphics[width=0.5\textwidth]{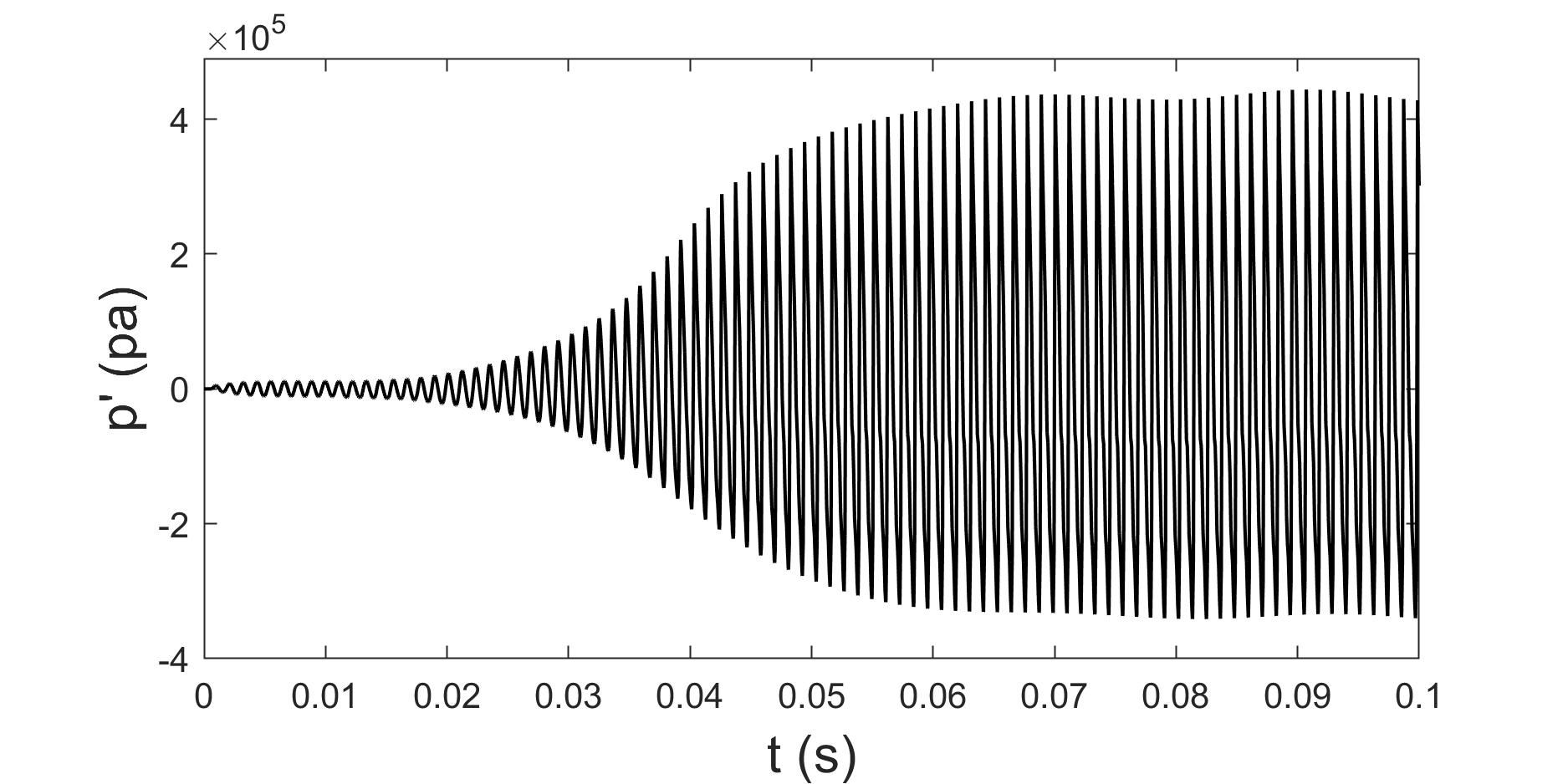}}
	\subfloat[Spatio-temporal diagram of the pressure evolution]{
		\includegraphics[width=0.5\textwidth]{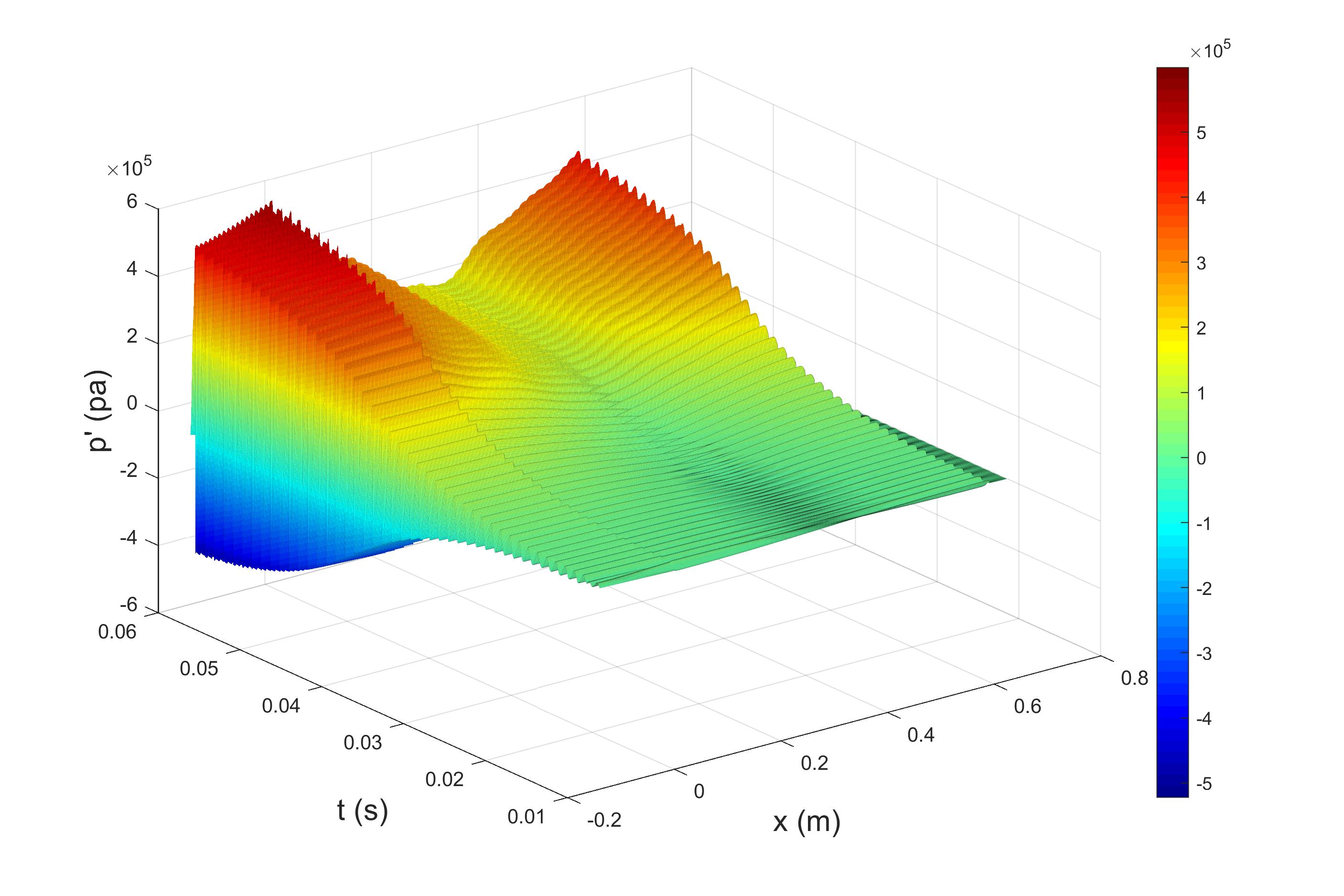}}\\
	\subfloat{
		\includegraphics[width=0.45\textwidth]{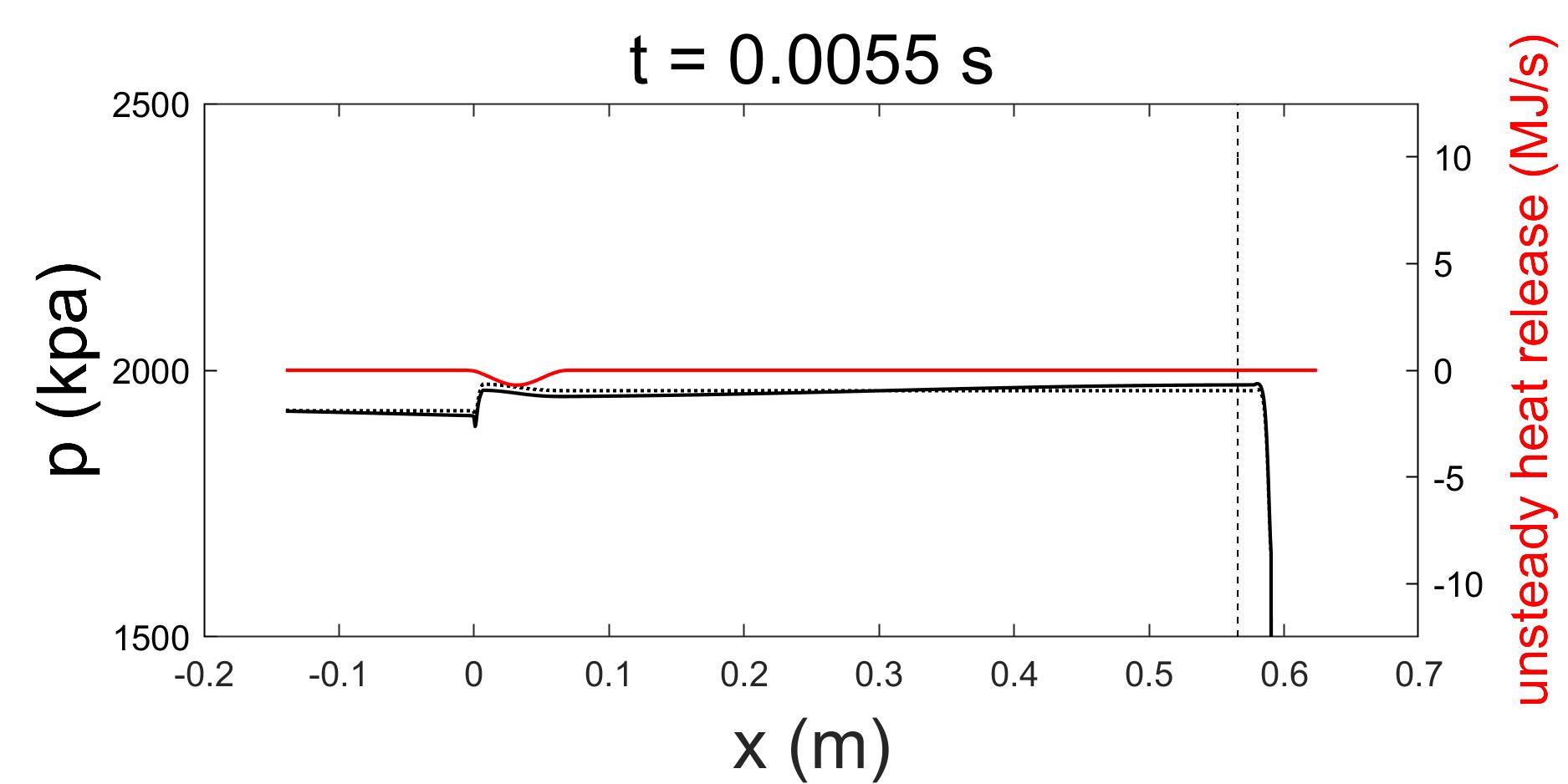}
		\includegraphics[width=0.45\textwidth]{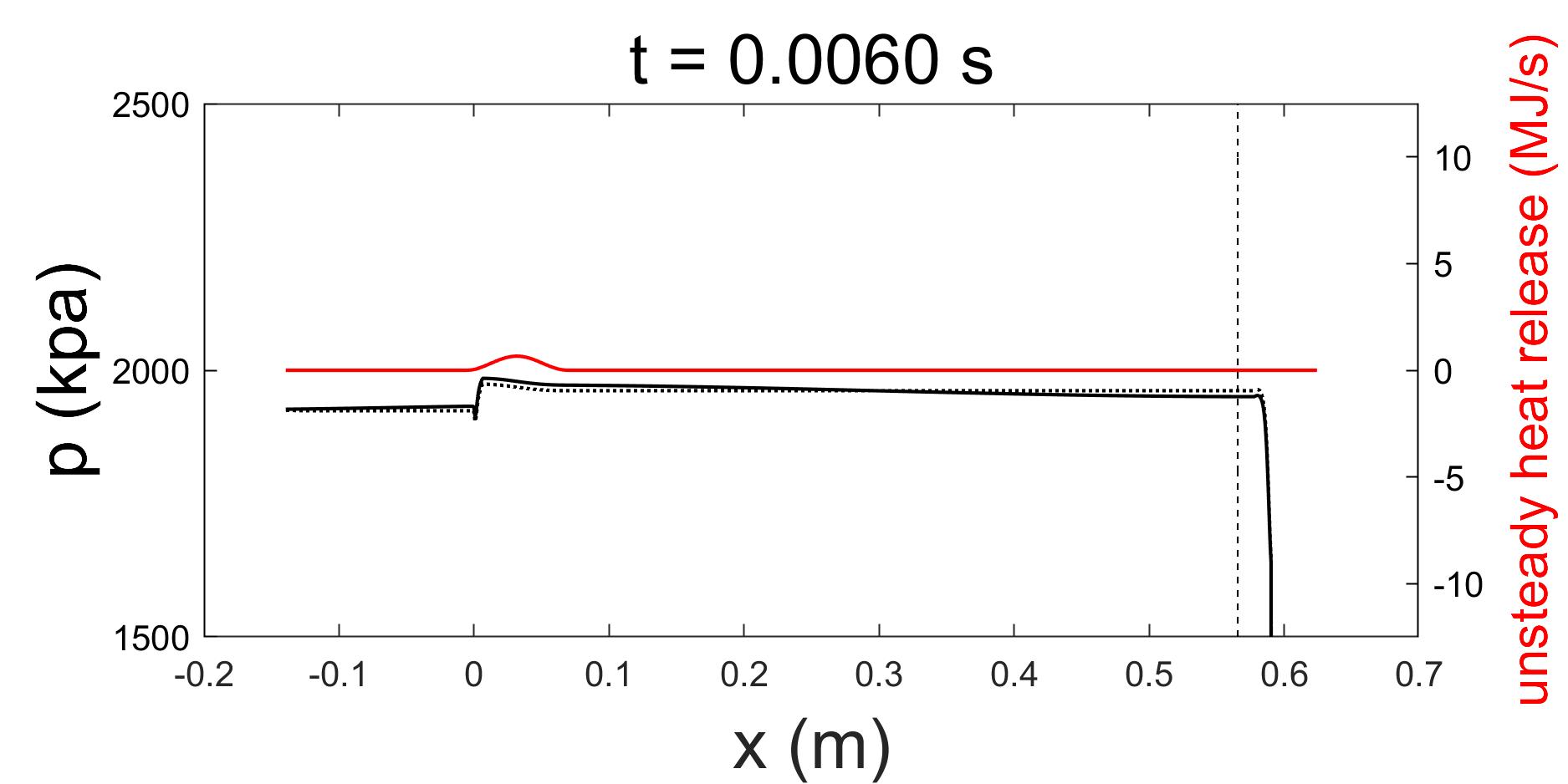}}\\
	\subfloat[Pressure and unsteady heat release profiles. Dotted line: steady state pressure, dashed line: pressure signal monitor location][Pressure and unsteady heat release profiles. \\Dotted line: steady state pressure, dashed line: pressure signal monitor location]{
		\includegraphics[width=0.45\textwidth]{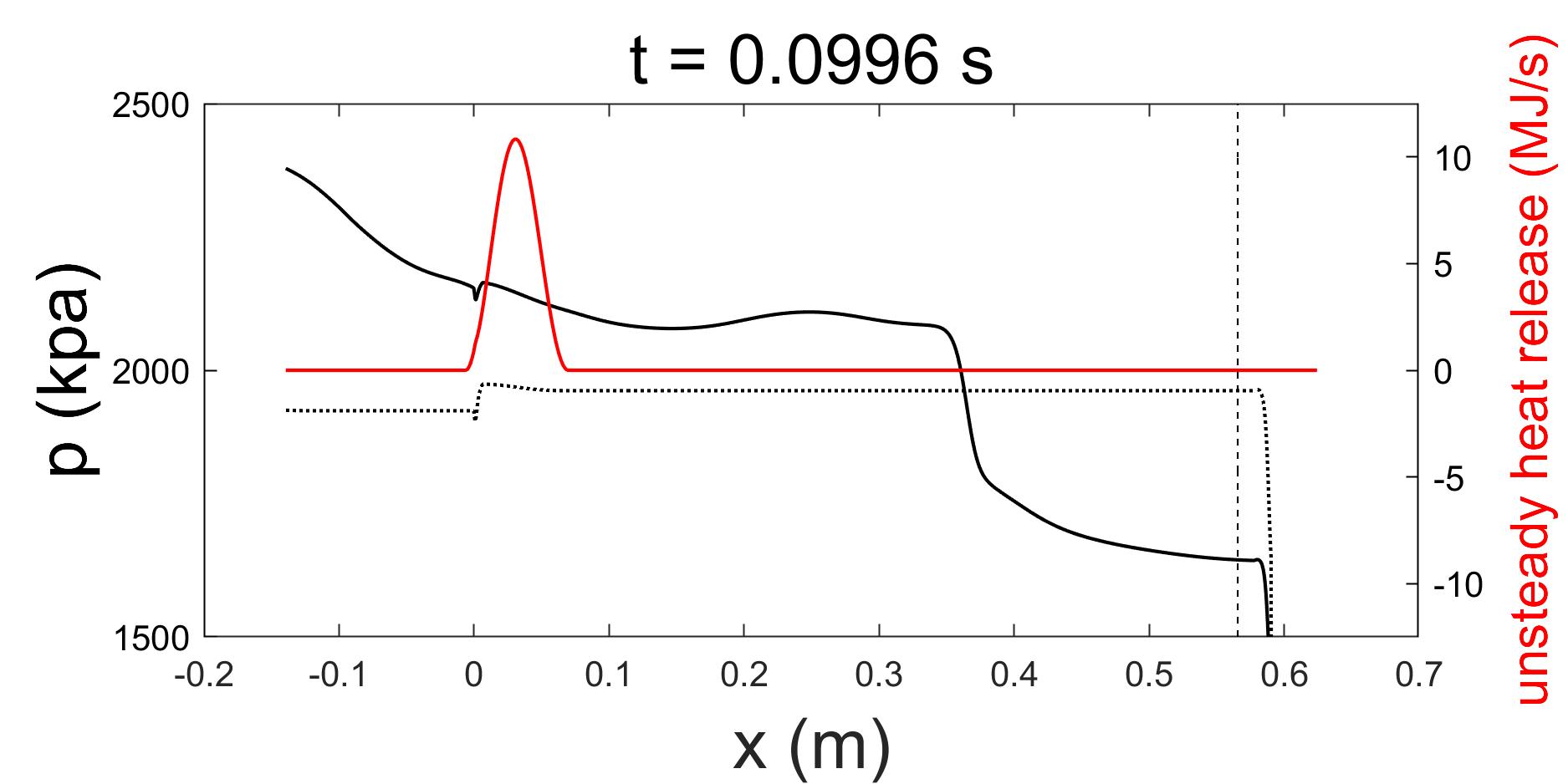}
		\includegraphics[width=0.45\textwidth]{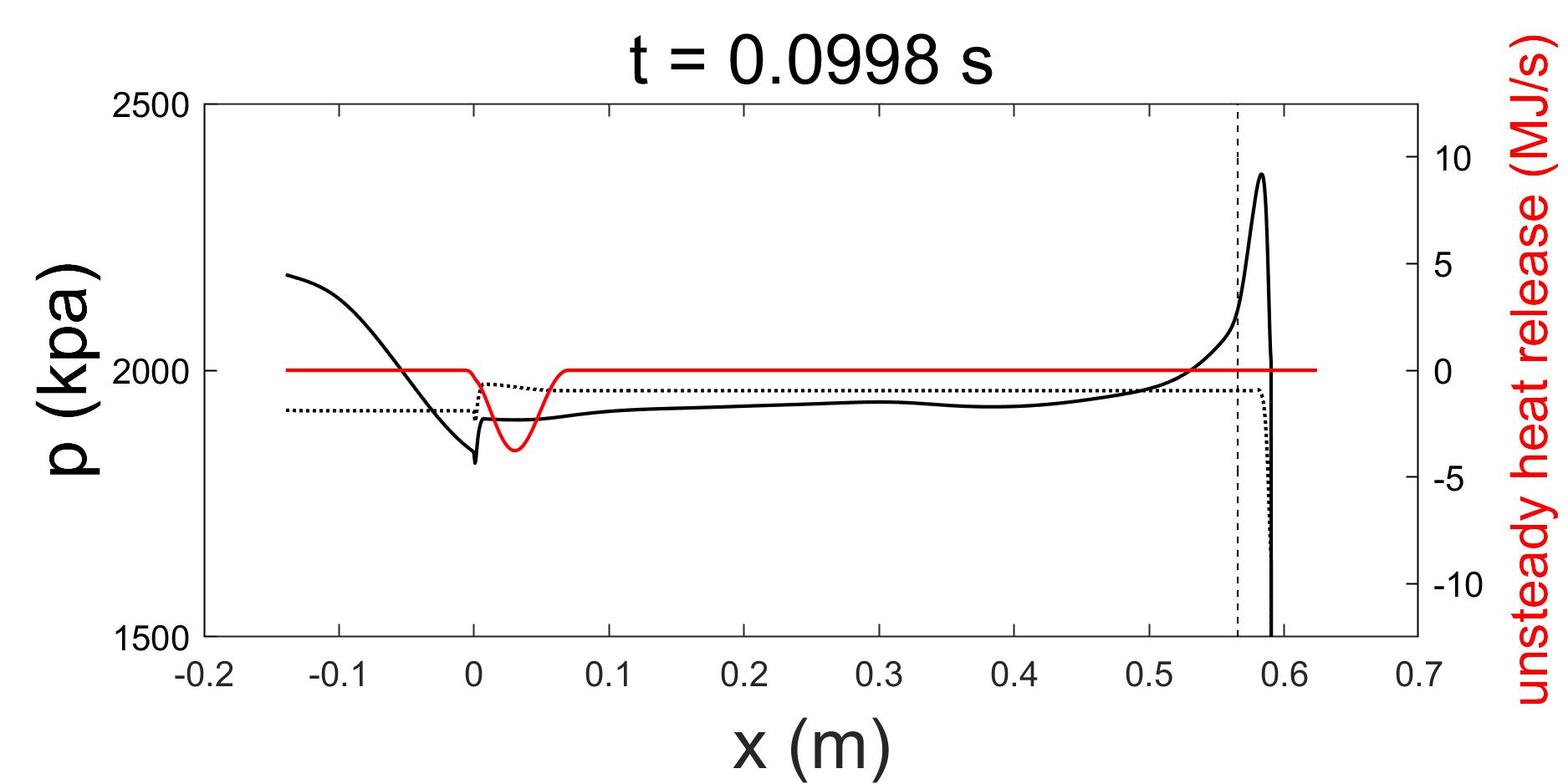}}
	\caption{Example response with positive growth rate, $\alpha=3.4, L_c=0.5715$ m.}\label{fig p_grow} 
\end{figure}

\begin{figure}
	\centering
	\subfloat[Pressure signal at the monitored location]{
		\includegraphics[width=0.5\textwidth]{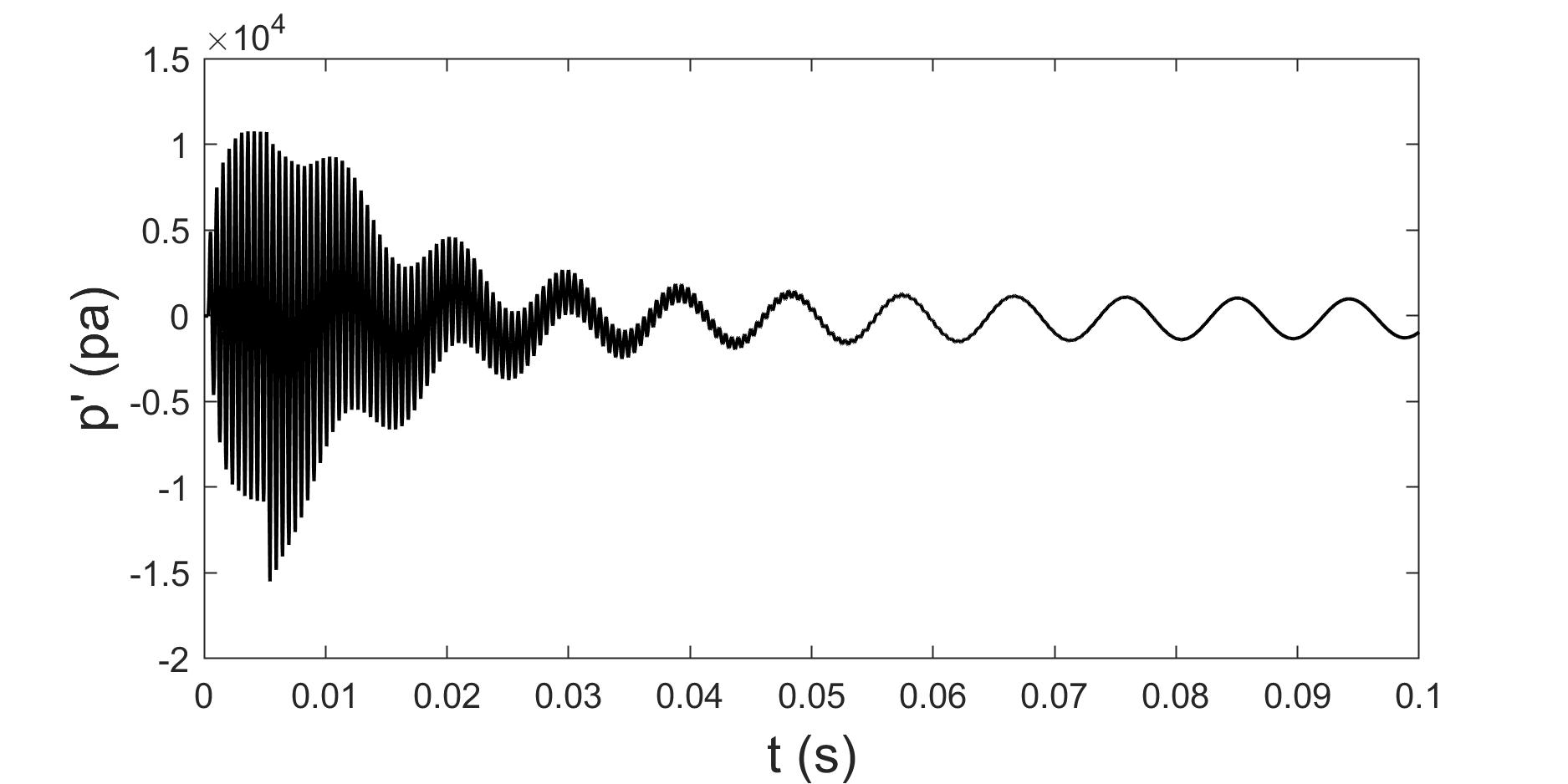}}
	\subfloat[Spatio-temporal diagram of the pressure evolution]{
		\includegraphics[width=0.5\textwidth]{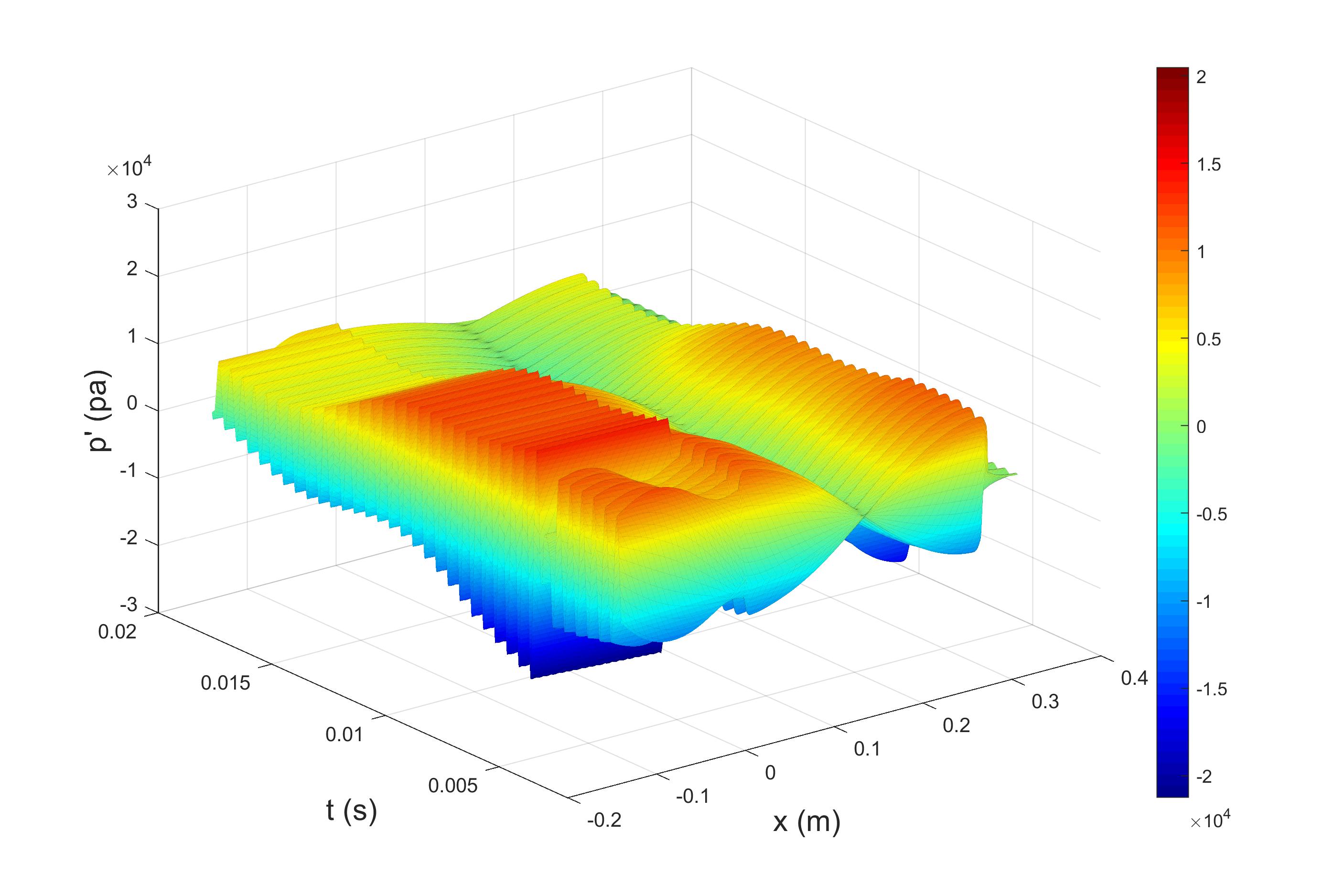}}\\
	\subfloat{
		\includegraphics[width=0.45\textwidth]{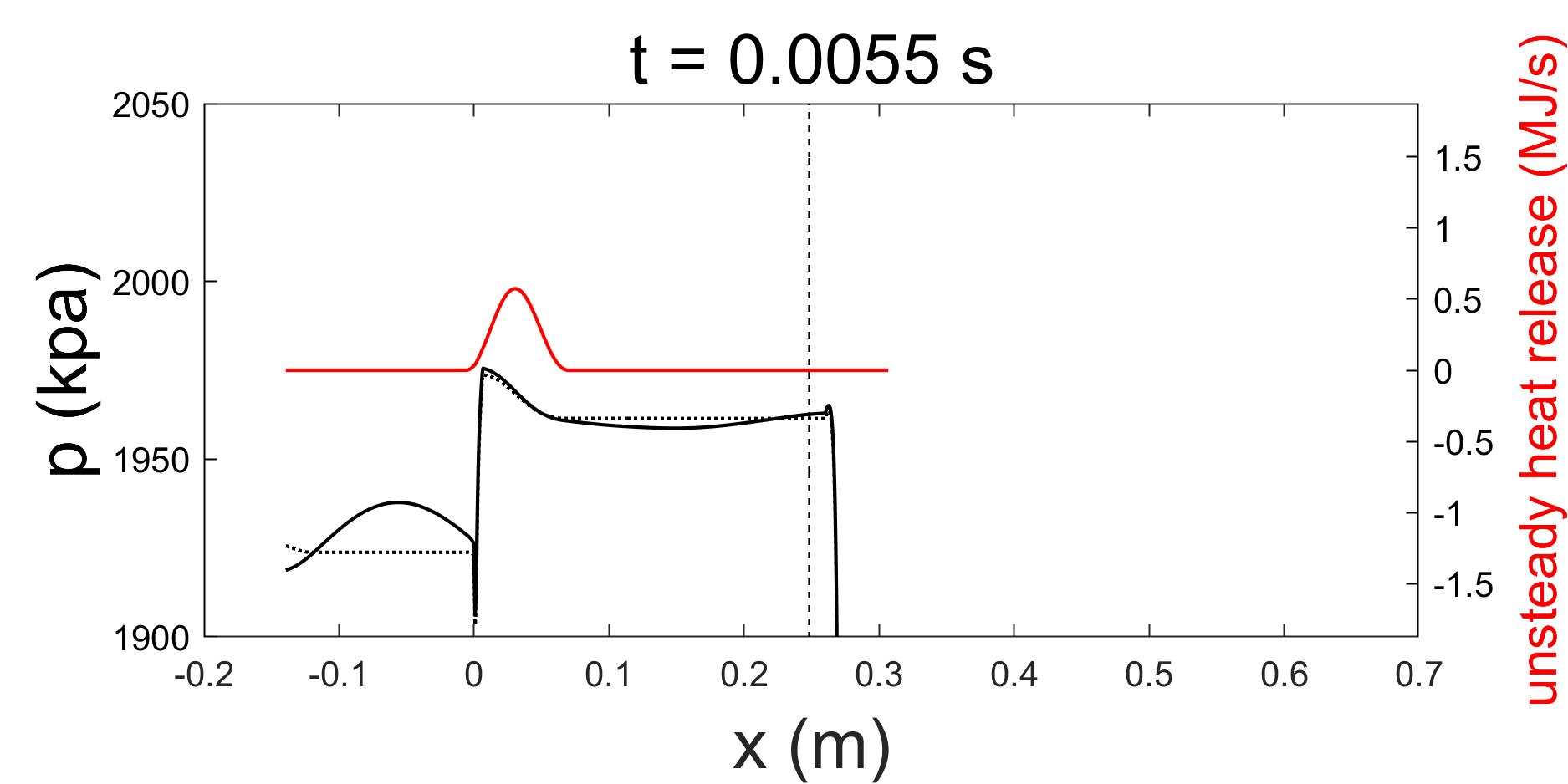}
		\includegraphics[width=0.45\textwidth]{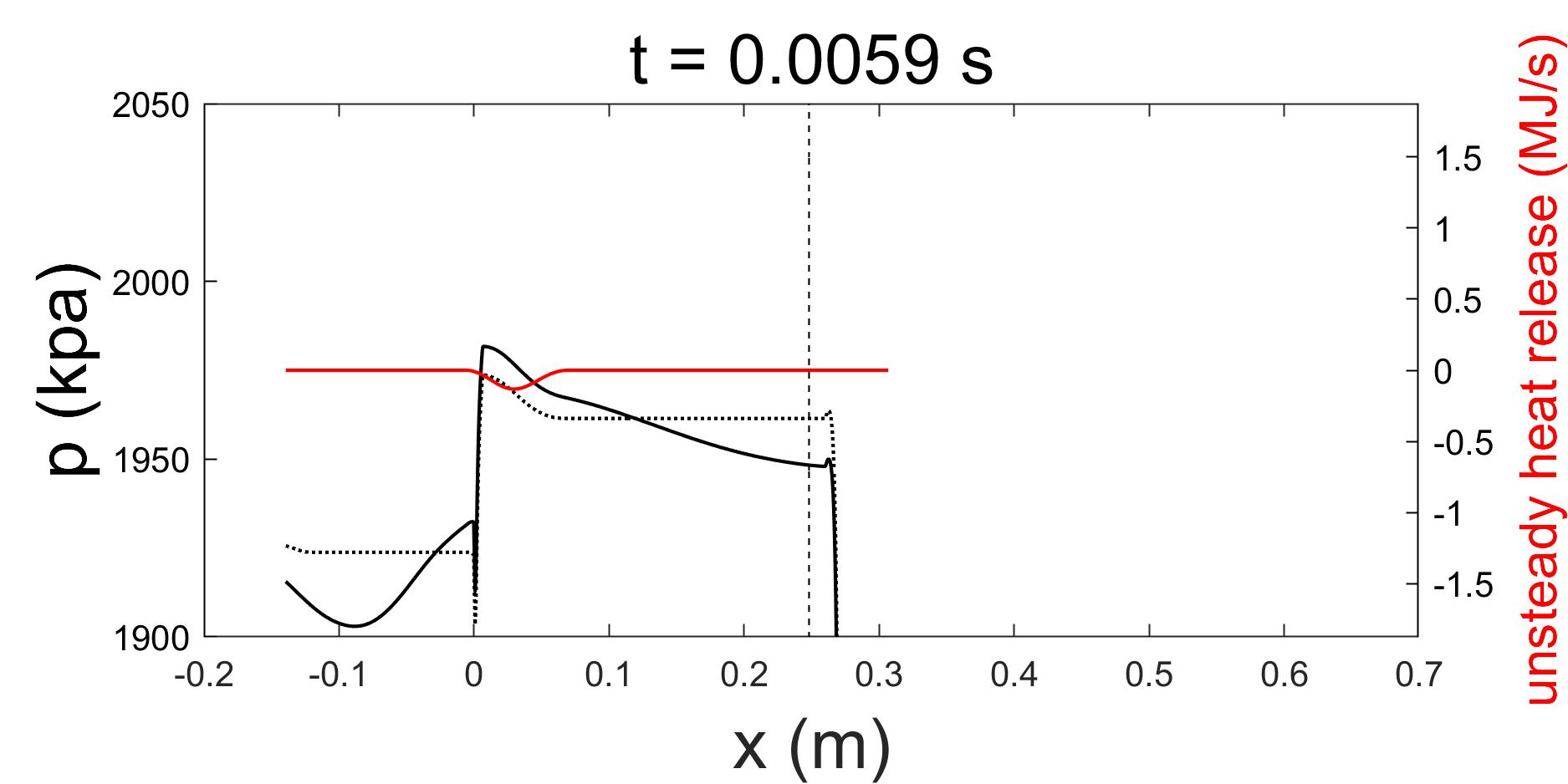}}\\
	\subfloat[Pressure and unsteady heat release profiles. Dotted line: steady state pressure, dashed line: pressure signal monitor location][Pressure and unsteady heat release profiles. \\Dotted line: steady state pressure, dashed line: pressure signal monitor location]{
		\includegraphics[width=0.45\textwidth]{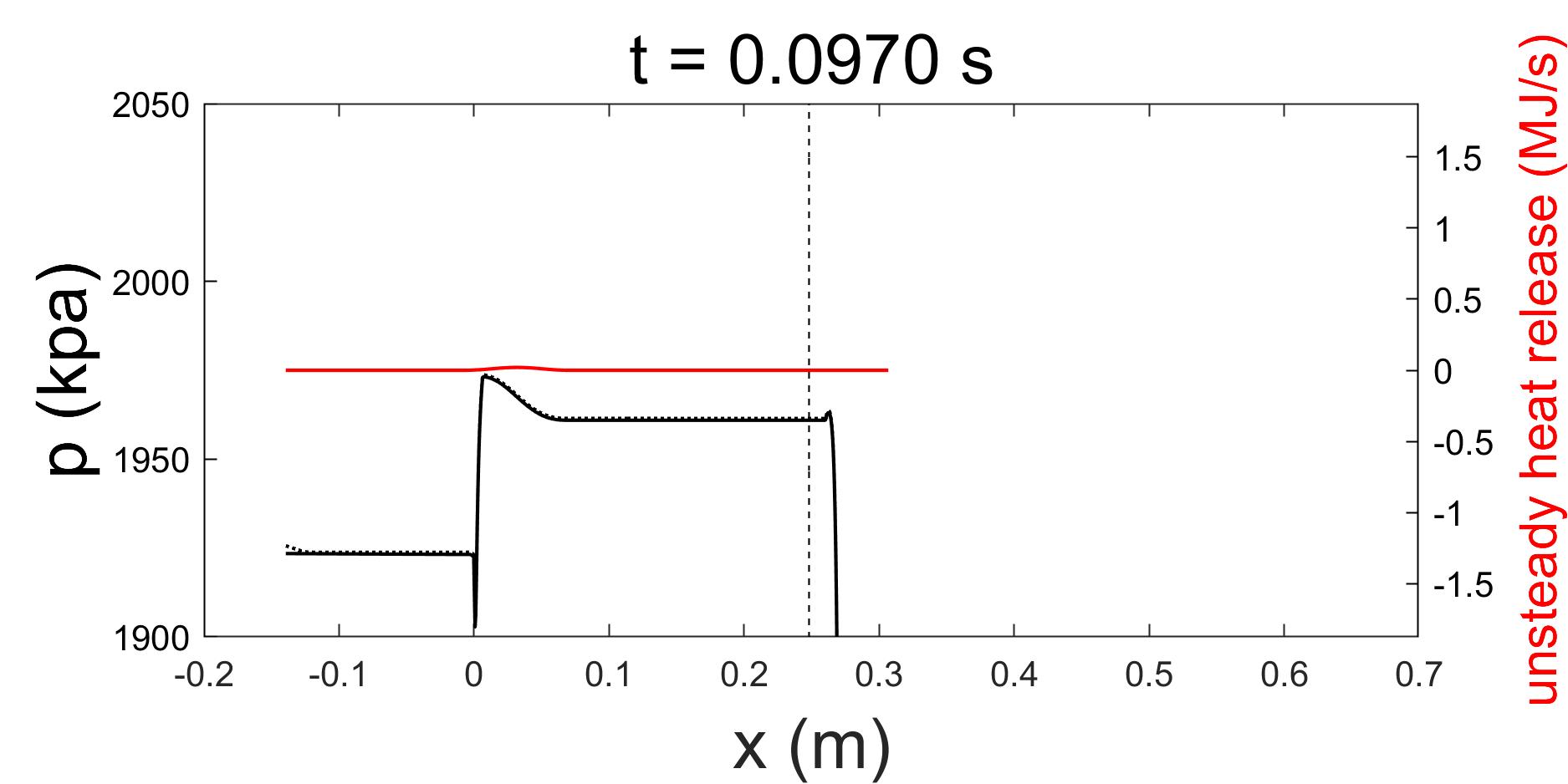}
		\includegraphics[width=0.45\textwidth]{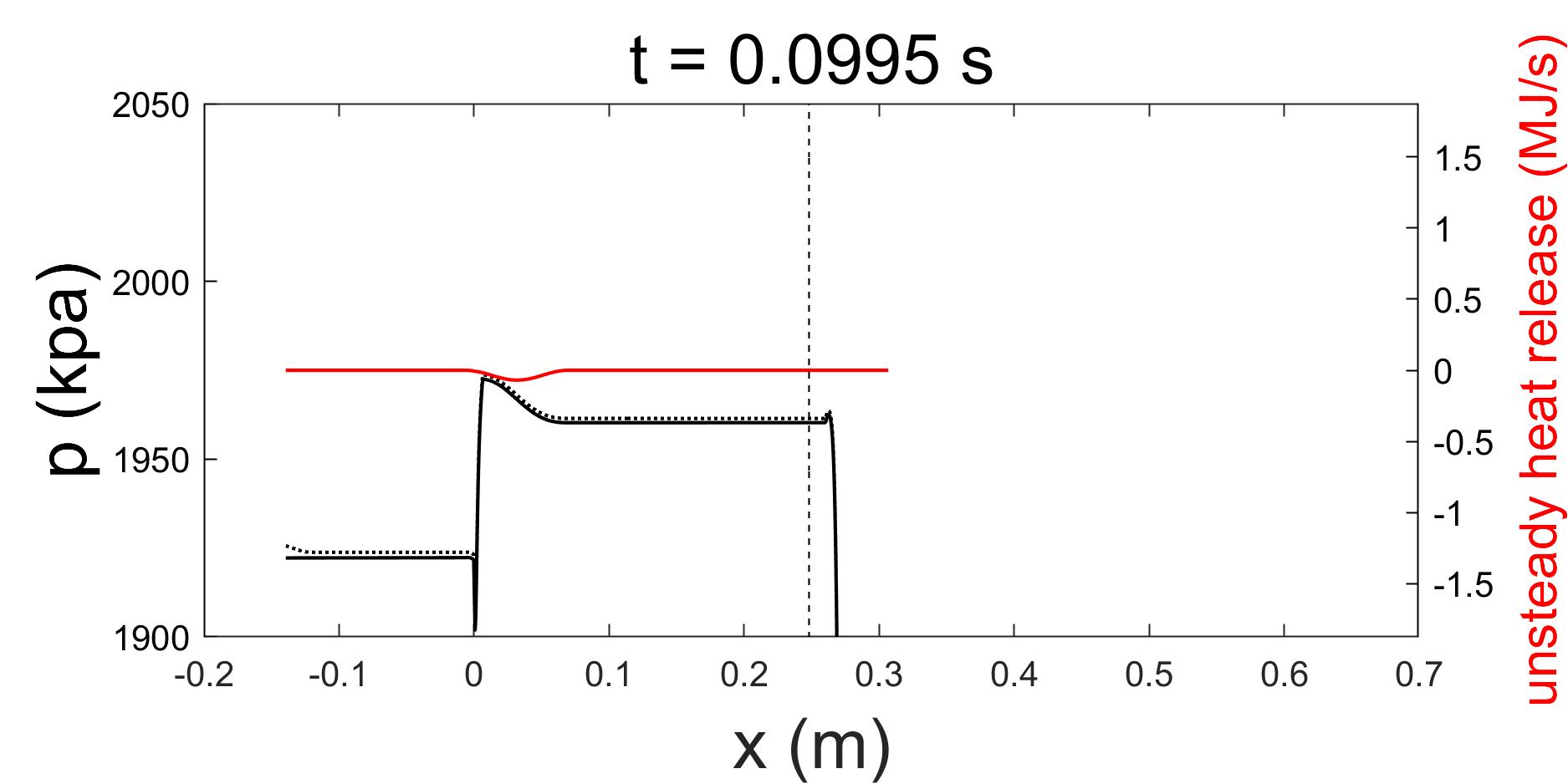}}
	\caption{Example response with negative growth rate, $\alpha=3.1, L_c=0.254$ m.}\label{fig p_decay} 
\end{figure}

To better distinguish the categories, two representative examples (at $\alpha=3.4,L_c=0.5715$ m for the growing category and $\alpha=3.1,L_c=0.254$ m for the decaying category, respectively) are presented in Fig.~\ref{fig p_grow} and \ref{fig p_decay}, where the difference in growth rate can be clearly observed via the monitored pressure signals and the spatio-temporal diagrams of the pressure evolution. Similar responses can be found in previous studies~\cite{xu2017reduced, wang2018non}. To provide more insight into the dynamics and the coupling of pressure and heat release described by Eq. \eqref{eq alpha-tau}, the pressure distribution and the unsteady heat release term are also plotted in the same figures. Each figure includes two snapshots at the beginning of the simulation, and two snapshots towards the end. It can be seen that in the case showing pressure amplitude growth (typically referred to as unstable case), the unsteady heat release has grown significantly with the pressure oscillation at the end of the simulation compared with the beginning due to the fact that the oscillations of pressure and heat release rate are in-phase, which is considered favorable to drive combustion instability. While for the other case with pressure amplitude decay (usually referred to as stable case), for some time instances, the pressure and heat release rate oscillations are out-of-phase, which is recognized as a combustion instability damping mechanism. 

\subsection{Modal decomposition}
\subsubsection{Singular values}
The singular values from a POD of snapshots of density, $\rho$ from different training methods are compared at $\alpha = 3.25, L_c = 0.254,0.508,0.762$ m. There are 472 cells in the first sub-domain, 897 in the full domain at $L_c = 0.254$ m, 1405 at 0.508 m and 1913 at 0.762 m. Since the number of snapshots is larger than the mesh size, the maximum number of modes and singular values from SVD is limited to the mesh size. 

\begin{figure}
	\centering
	\includegraphics[width=0.4\textwidth]{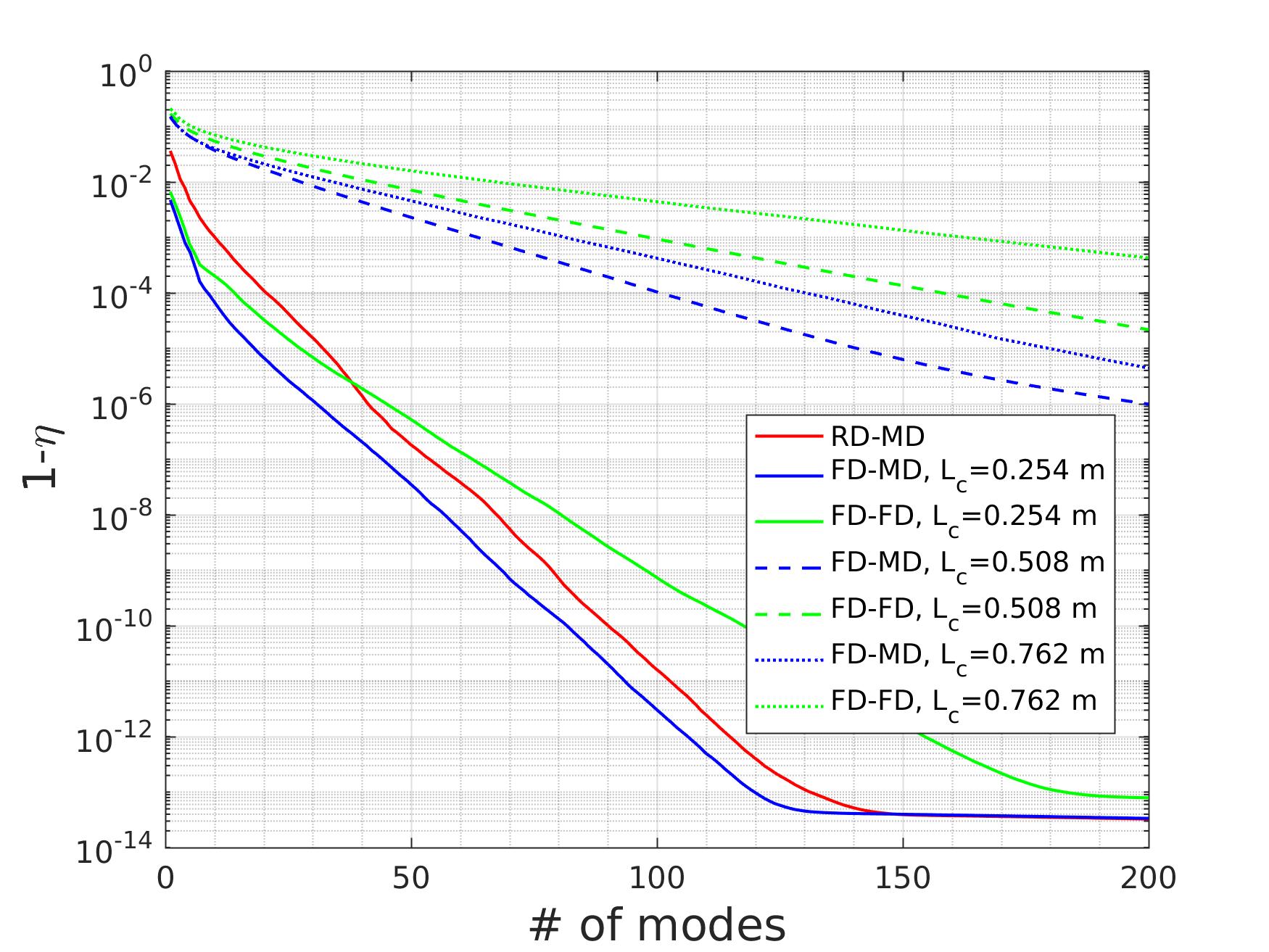}
	\caption{Complementary part of cumulative energy in the singular values for $\rho$ at $\alpha = 3.25$}\label{fig spectrum}
\end{figure}

The result is visualized by the complementary part of the cumulative energy in Fig. \ref{fig spectrum}. For $k$ modes, the cumulative energy is defined as
\beq
\eta_{k}  = \frac{{\sum\limits_{i = 1}^{k} {{\sigma _i}} }}{{\sum\limits_{i = 1}^n {{\sigma _i}} }}.
\eeq

It should be noted that there is only one curve for RD-MD as it uses the same training simulation and POD bases for all chamber lengths. It can be observed that for both FD-MD and FD-FD, the decay in the complementary part increases as the chamber length decreases. This is due to the fact that the dynamics have a higher frequency when the chamber length is shorter, which is easier to be captured by fewer modes in SVD. In all cases, the decay in FD-FD is slower than FD-MD, because it covers a larger domain with more spatial variations in geometry and physics, and more modes are therefore needed to contain the same portion of information stored in the training data. For medium-to-high $L_c$, the proposed framework gives the best decay since its training data contains more high-frequency contents. At small $L_c$, the decay in the leading modes for the two full-domain training methods become slightly better than RD-MD as the response frequency in their training data is comparable to the highest one included in the reduced-domain training, whereas there are also lower frequencies contained in the latter. Results for other variables and $\alpha$ follow a similar pattern.

\subsubsection{Mode shapes}
\begin{figure}
	\centering
	\subfloat[RD-MD]{
		\includegraphics[width=0.33\textwidth]{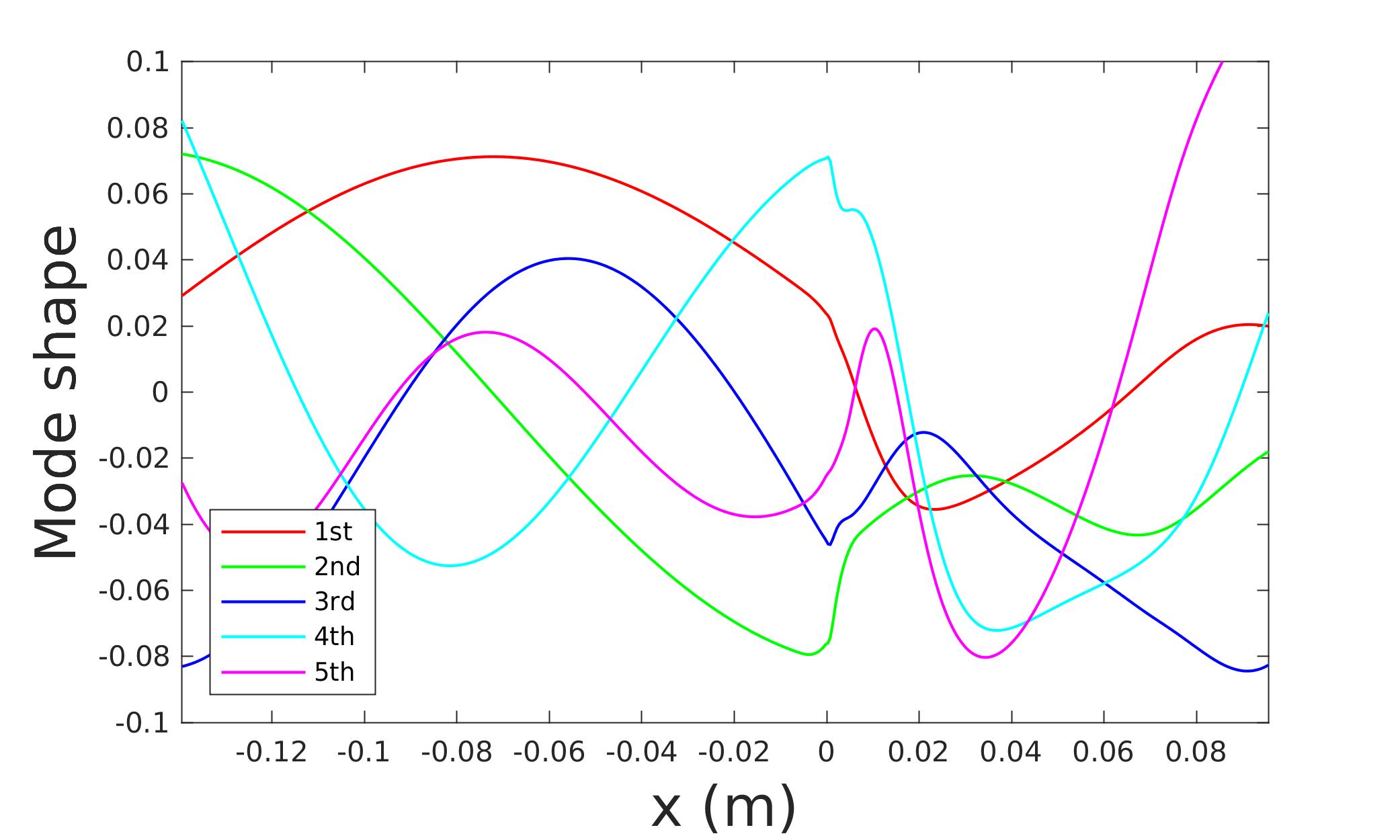}}
	\subfloat[FD-MD]{
		\includegraphics[width=0.33\textwidth]{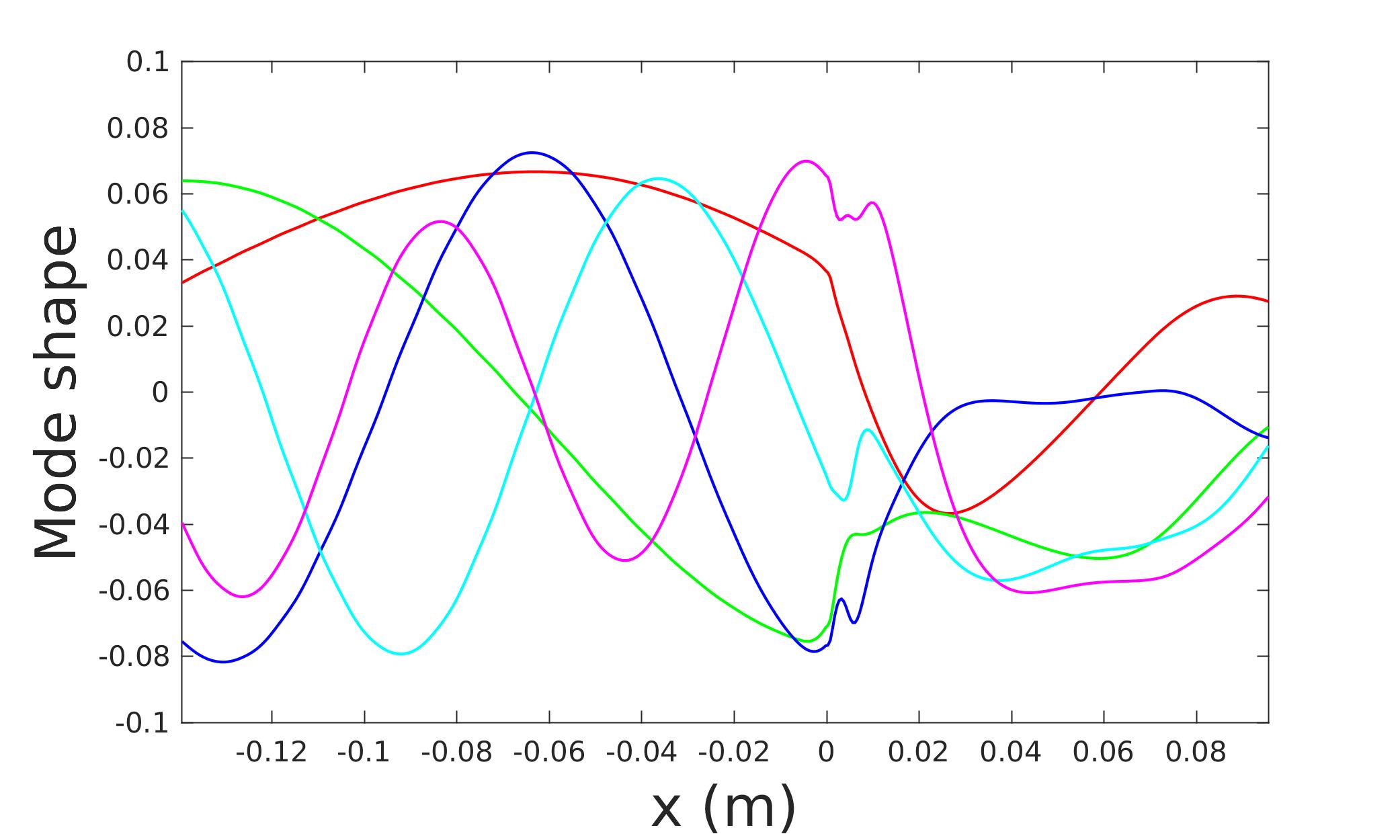}}
	\subfloat[FD-FD]{
		\includegraphics[width=0.33\textwidth]{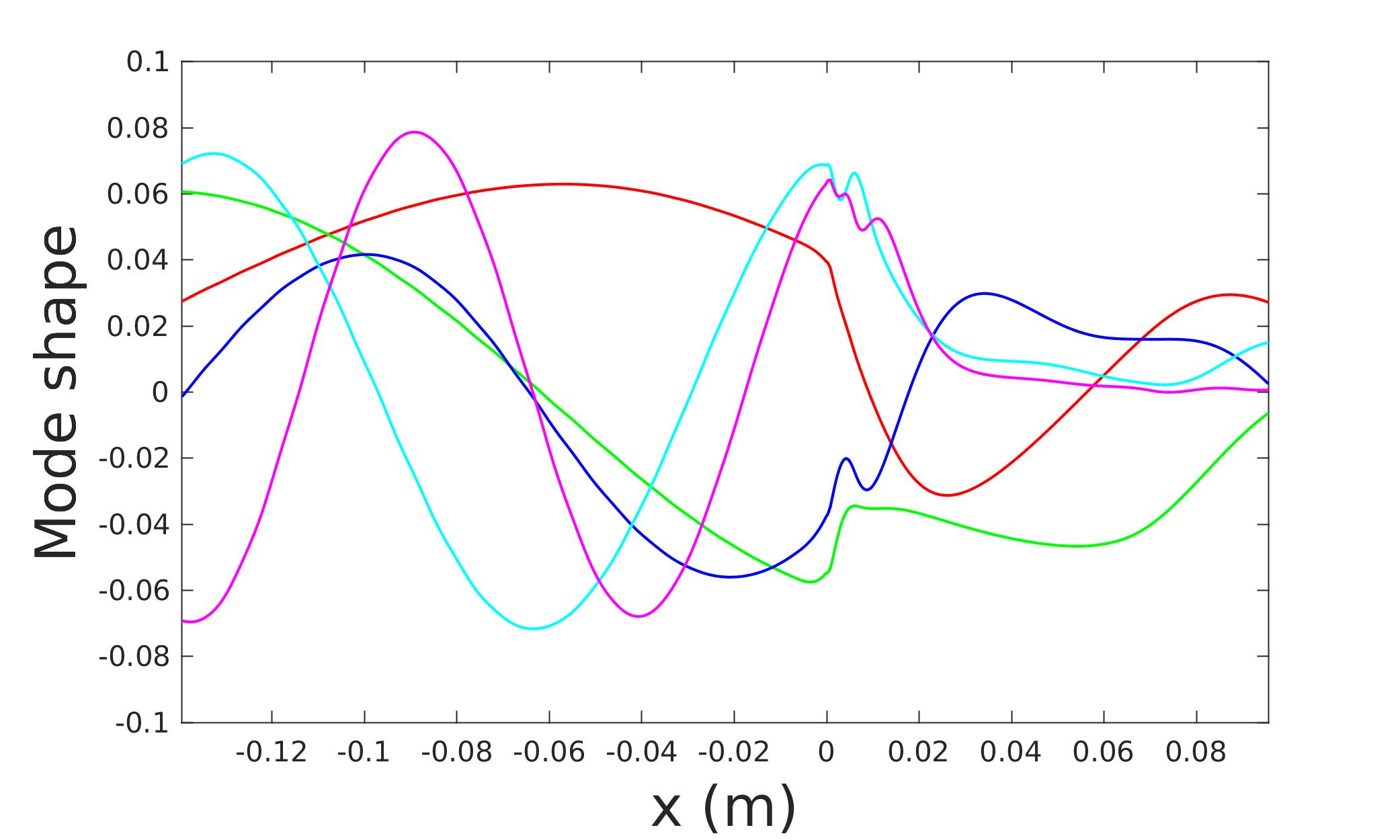}}
	\caption{Leading spatial modes for $\rho$ at $\alpha=3.25, L_c=0.508$ m}\label{fig mode} 
\end{figure}

Comparison of just the singular values does not provide enough information on the compatibility of the basis and physical features, since each training method is different. To provide further insight, the first five most energetic  modes for $\rho$ at $L_c = 0.508$ m, $\alpha = 3.25$ are shown in  Fig. \ref{fig mode}. It should be noted that the basis for FD-FD is on the full domain and only truncated in the plot to the first sub-domain for  comparison purposes. The first mode for all methods appears to be identical. For lower energy modes, the basis for the RD-MD is quite different from the other two methods. The basis for the FD-MD and FD-FD are also different for $k>4$, which suggests that performing POD on different domains can result in different mode shapes, even when the training data is collected from the same simulation. This is because that the SVD process for POD bases generation solves a global minimization problem that will yield a basis with minimal least square projection error for the input snapshots. Thus,  the scope of the minimization problem is changed with the size of the domain. There is hence no guarantee that the same local mode shapes will be obtained, though the input snapshots are the same locally.   Consistent observation has been made in Ref.~\cite{huang2016analysis}, where the mode shape in full domain and different sub-domains of a 3D simulation of the CVRC is compared (equivalent to the FD-MD and FD-FD comparison).

To better demonstrate the evolution of the modes, the FOM solution at $\alpha=3.25, L_c=0.508$ m is projected onto the POD bases. Due to orthogonality, the projection can be performed independently and the resulting reduced variables are given by
\beq
q_r^{(i)}(t)=({\mb{v}}^{(i)})^T\mb{q},
\eeq
where ${\mb{v}}^{(i)}$ is the $i$-th POD mode. 

\begin{figure}
	\centering
	\subfloat[RD-MD]{
		\includegraphics[width=0.33\textwidth]{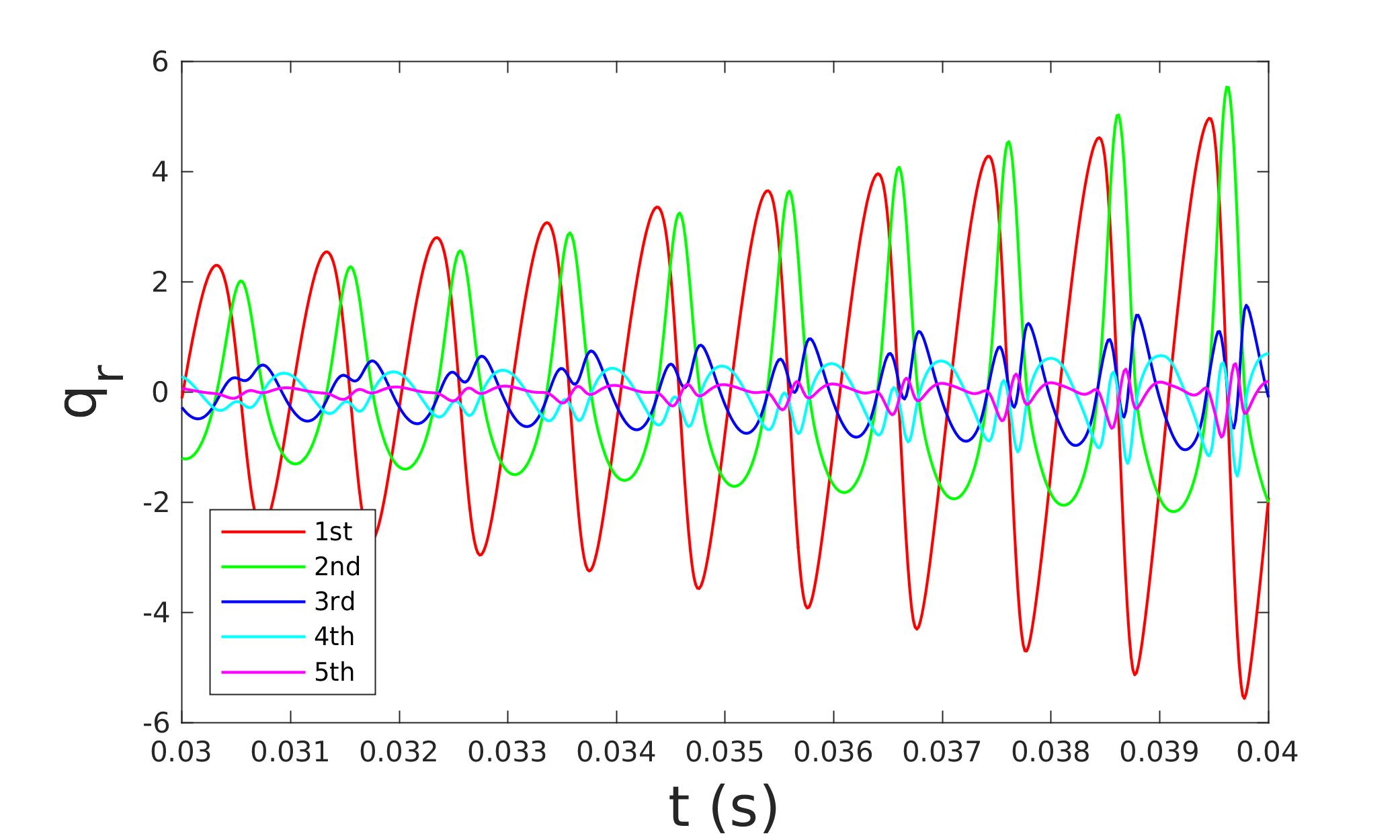}}
	\subfloat[FD-MD]{
		\includegraphics[width=0.33\textwidth]{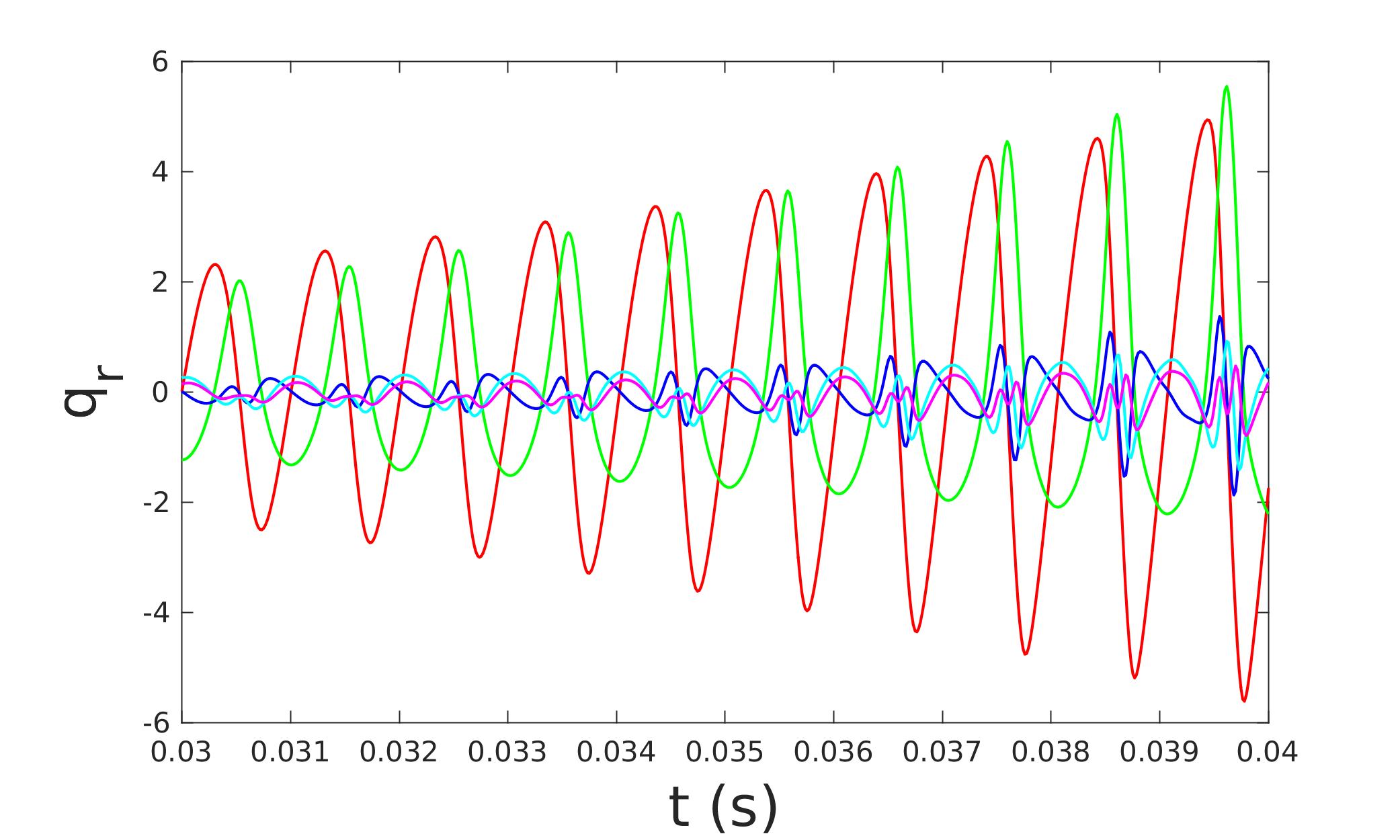}}
	\subfloat[FD-FD]{
		\includegraphics[width=0.33\textwidth]{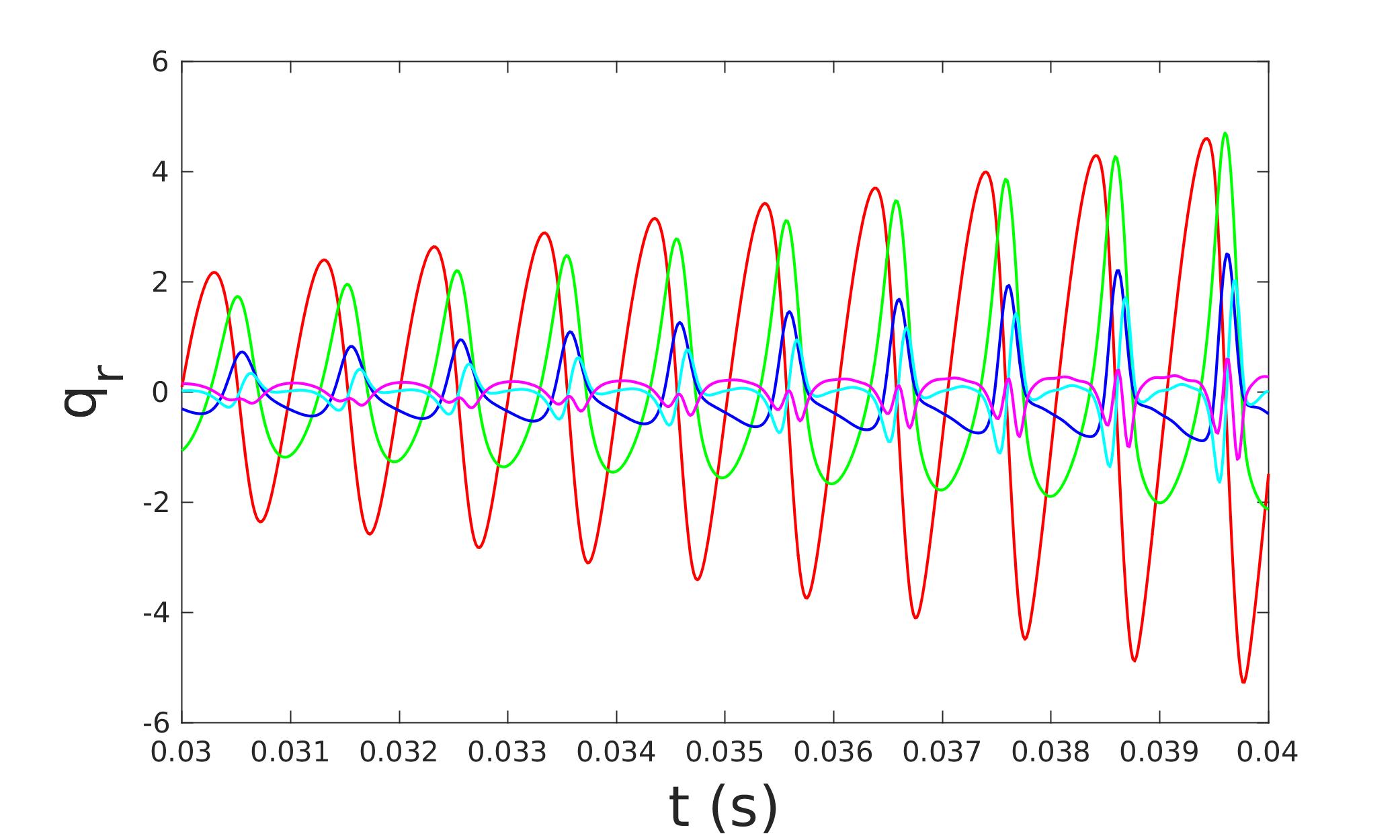}}
	\caption{Evolution of coefficients $\mb{q}_r^\rho$ at $\alpha=3.25, L_c=0.508$ m}\label{fig coefficient} 
\end{figure}

The coefficients for the five leading modes present in Fig.~\ref{fig mode} are shown in Fig. \ref{fig coefficient}. For better visualization, only a period of $t=0.3-0.4$ s is used. It can be seen that the first mode evolves at a single frequency corresponding to the large-scale instability characterized by the pressure oscillation described in the previous section. This frequency is exhibited by the latter modes, while there are also additional frequencies corresponding to the high-frequency dynamics contained in the modes.  

\begin{figure}
\centering
	\subfloat[Spatial]{
	\includegraphics[width=.45\textwidth]{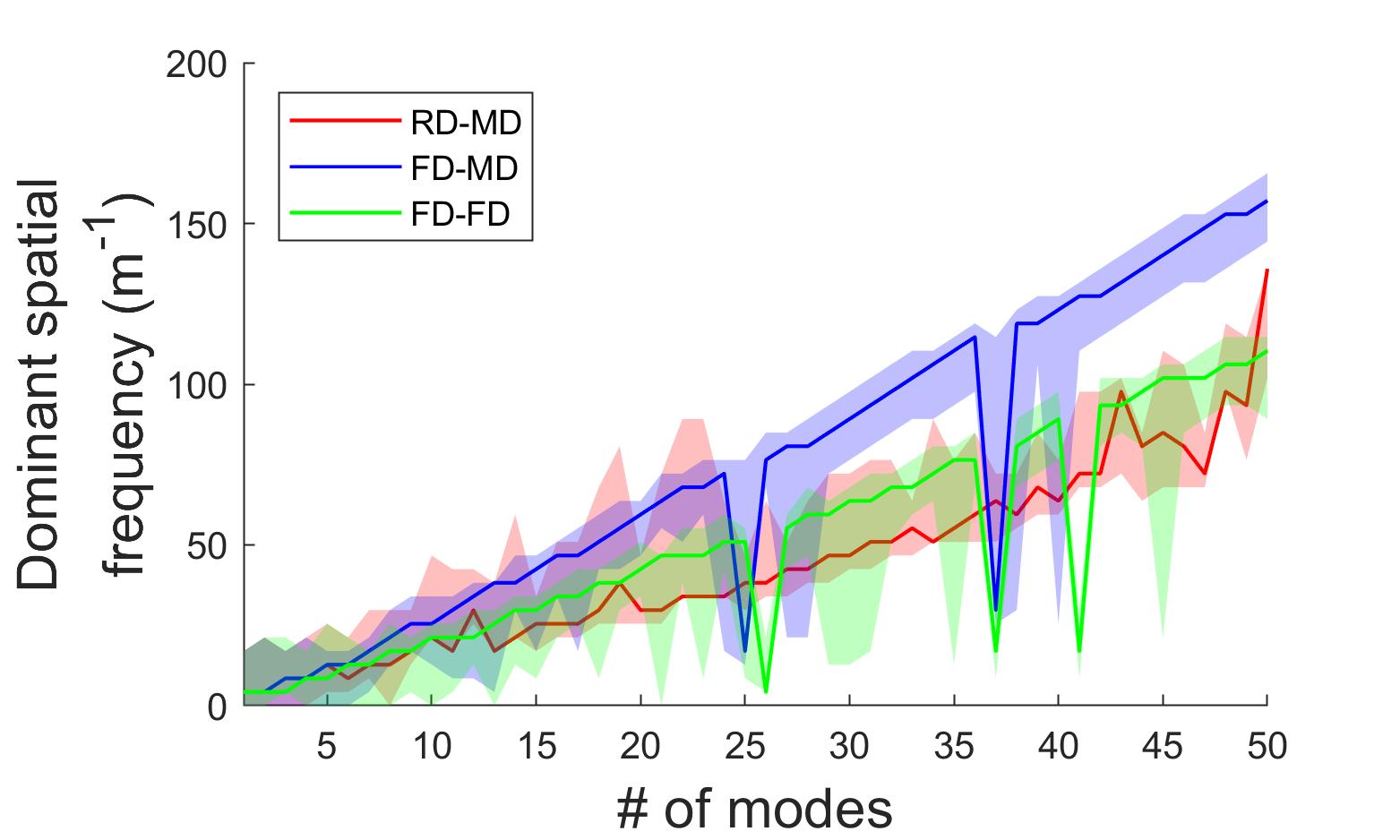}}
	\subfloat[Temporal]{
	\includegraphics[width=.45\textwidth]{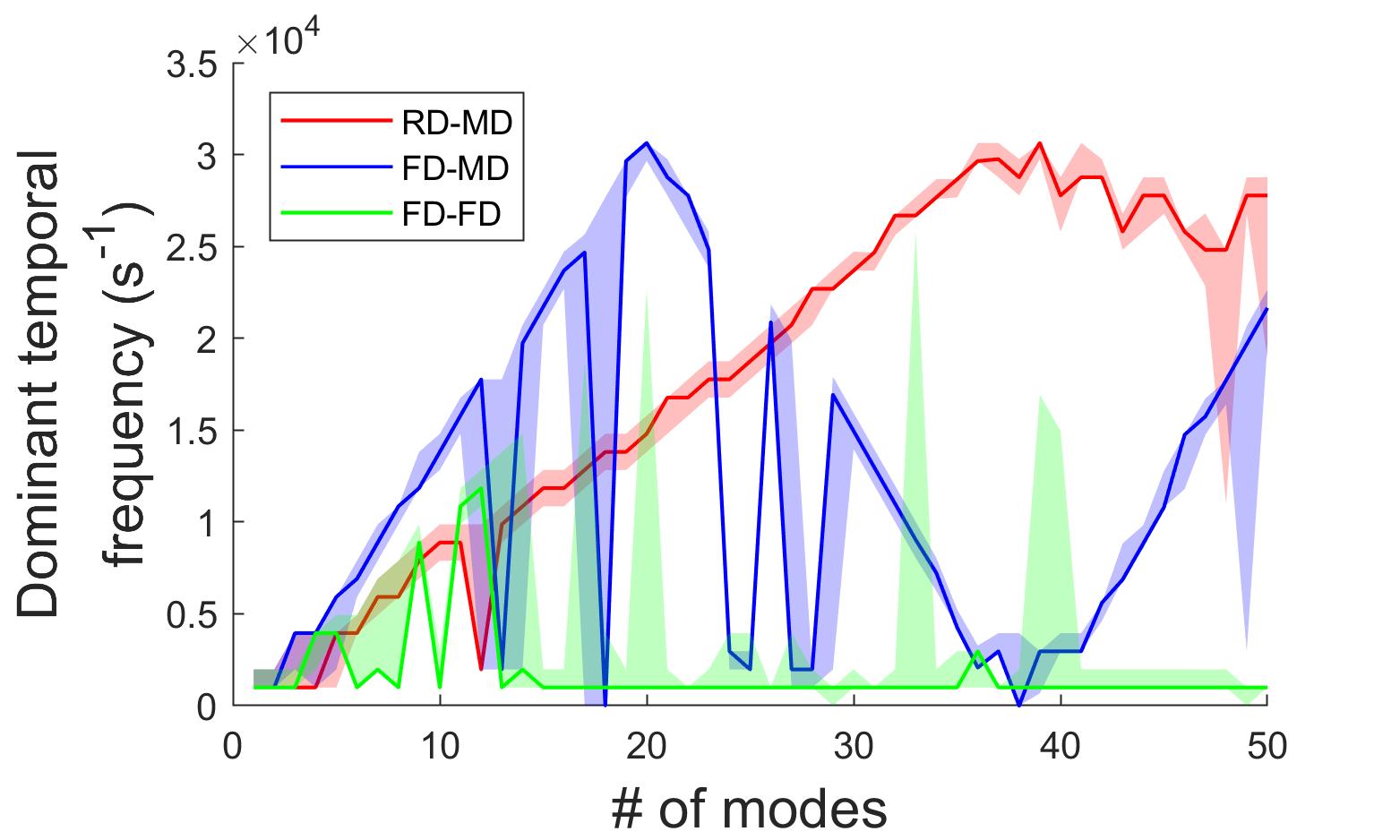}}
	\caption{Dominant spatial/temporal frequency of POD modes for $\rho$ at $\alpha=3.25, L_c=0.508$ m. 
	\\Solid line: first mode, shaded: band for the first five frequencies}\label{fig dominant_freq} 
\end{figure}

It can be noticed from Fig.~\ref{fig mode} that the  spatial modes that exhibit high-frequency temporal oscillations also have a higher spatial frequency. To make this trend clearer, the dominant spatial frequencies of the 50 leading modes of $\rho$ are plotted in Fig.~\ref{fig dominant_freq}. While showing a similar trend, the growth rates are different between the methods. The reasons are two-fold: 1) using the same single-frequency training, FD-MD shows a faster decay in the singular values, which indicates that the low-frequency high-energy dynamics is better captured in the leading modes, and thus the high-frequency modes are realized at smaller mode numbers than in FD-FD. 2) There are multiple frequencies in the reduced-domain training. These frequencies are still significantly higher than those  of the corresponding POD mode numbers in FD-FD and FD-MD. The corresponding modes representing these frequencies cause the RD-MD to have the slowest increase in the dominant spatial frequencies.  Also presented in Fig.~\ref{fig dominant_freq} are the dominant temporal frequencies exhibited in the evolution of the POD modes. For RD-MD, the frequency increases with mode number steadily before the 35-th mode. This demonstrate a good correlation between the temporal and spatial frequency captured in the modes. In contrast, many high-spatial-frequency modes from the two full-domain training methods actually evolve at a low temporal frequency. This inconsistency indicates a less efficient basis compared to RD-MD. 

\subsubsection{Projection error}
\begin{figure}
	\centering
	\subfloat[$L_c=0.254$ m]{
		\includegraphics[width=0.33\textwidth]{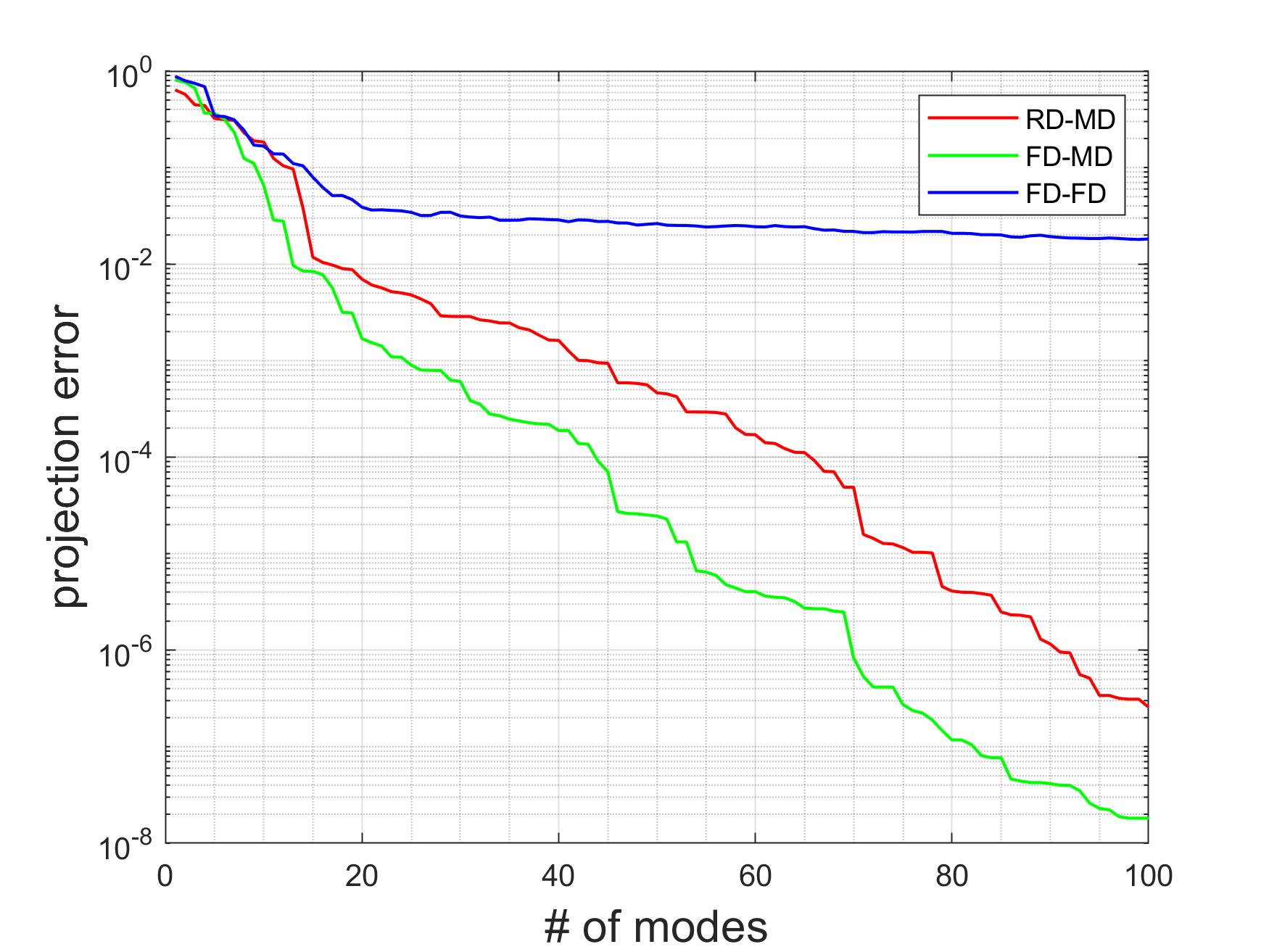}}
	\subfloat[$L_c=0.508$ m]{
		\includegraphics[width=0.33\textwidth]{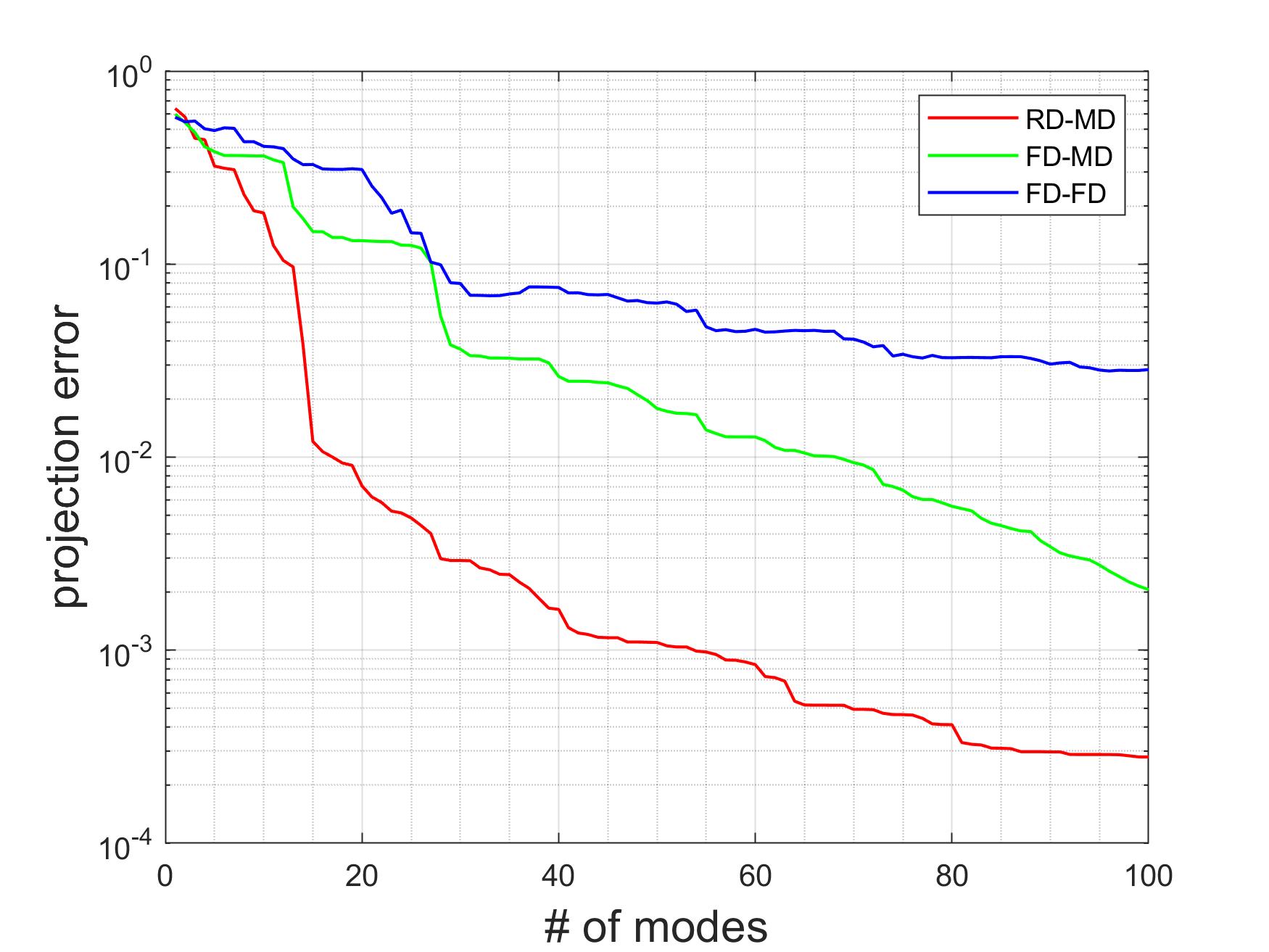}}
	\subfloat[$L_c=0.762$ m]{
		\includegraphics[width=0.33\textwidth]{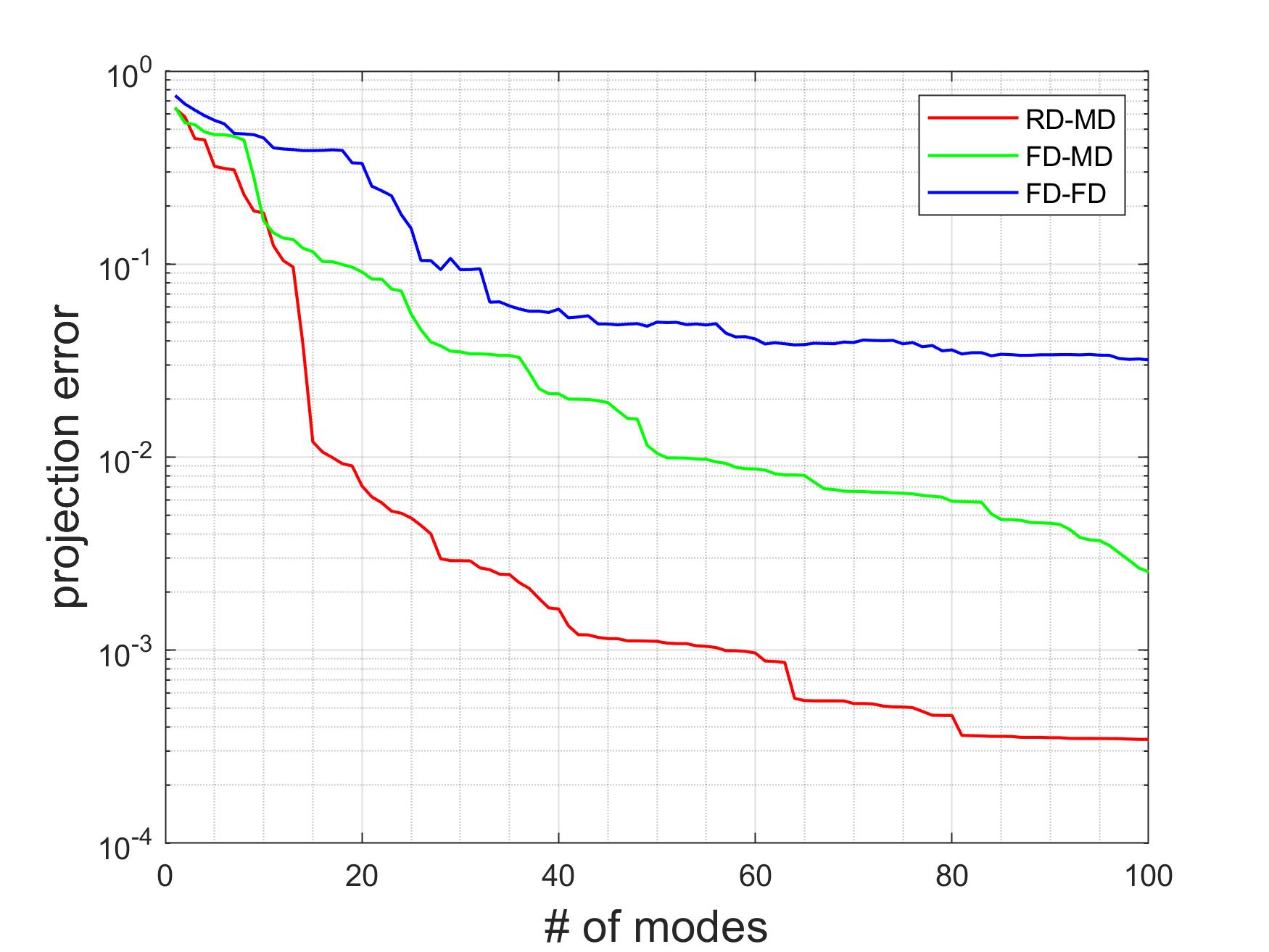}}
	\caption{Projection error in $\rho$ for $\alpha = 3.25$}\label{fig projection}
\end{figure}

For a given set of $k$ POD modes, the projection error is defined as
\beq
\epsilon  = \frac{{{{\left\| {{\bf{Q}} - {{\bf{V}}_{1:k}}{\bf{V}}_{1:k}^T{\bf{Q}}} \right\|}_2}}}{{{{\left\| {\bf{Q}} \right\|}_2}}},
\eeq
where $\mb{Q}$ is the FOM solution snapshot matrix, ${\bf{V}}_{1:k}$ is the first $k$ columns of the left singular vector from the SVD of training data. Again, at $\alpha = 3.25, L_c = 0.254,0.508,0.762$ m, the projection error of $\rho$ up to $k=100$ is shown in Fig. \ref{fig projection}. It can be observed that both RD-MD and FD-MD provide a monotonic decrease in the projection error. However for FD-FD, the error increases with $k$ at a few cases, which is not desirable for ROM development as it makes the choice of basis size more uncertain. The increase can be again attributed to the fact that the POD method solves a global minimization problem and the optimal bases for the full domain is not necessarily optimal for the reduced domain. The improved projection error property with the multi-domain method implies a higher accuracy and stability of ROM, which is confirmed in the following sections. 

The above analysis are focused on $\rho$ and results for other variables follow the same trend.

\subsection{ROM evaluations}
Before integrating the ROM into the Multi-fidelity Framework, the predictive capability of the ROM obtained from the reduced-domain simulation is evaluated. In this test, the ROM trained with broadband characteristic perturbation is tested by feeding single frequency perturbations on the characteristic boundary. This is to ensure that the ROM is able to predict the essential dynamics when the downstream section is dominated by single frequency acoustics. Aside from  the perturbation frequency, there is no difference between the conditions used in the test and the truncated domain training (Sec. \ref{sec framework}). 

\begin{figure}
\begin{center}
	\includegraphics[width=0.6\textwidth]{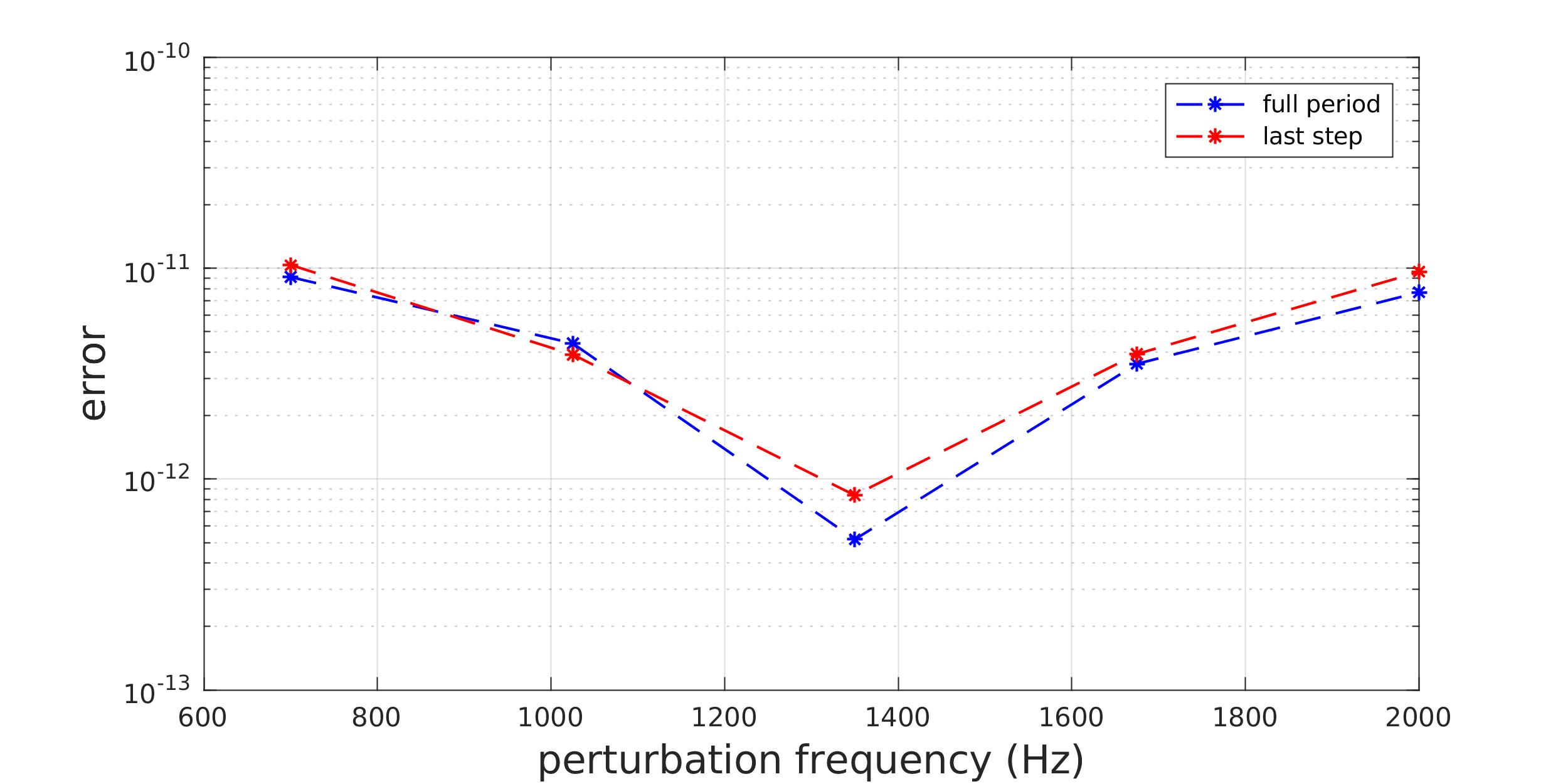}
	\caption {Sanity test for L2 error}\label{fig sanitytest} 
	\end{center}
\end{figure}

The tests are conducted over $t=0-0.05$ s at five frequencies between 700 and 2000 Hz. It should be noted that some of the frequencies are not included in the broadband training. The number of ROM modes and amplification factor are $k=100, \alpha = 3.25$, respectively. The results are compared with the reference FOMs using the same conditions, and the L2 error over the simulated period as well as at the last time step is plotted in Fig. \ref{fig sanitytest}. It can be seen that at all perturbation frequencies, the ROM is able to accurately predict the FOM behavior with an L2 error below \num{1.1e-11}. These  evaluations confirm the fact that the ROM trained with broadband perturbation is capable of predicting dynamics over a wide range of frequencies. 

\subsection{Multi-fidelity framework evaluations}
\begin{figure}
	\centering
	\subfloat[$\alpha=3.1$]{
		\includegraphics[width=0.32\textwidth]{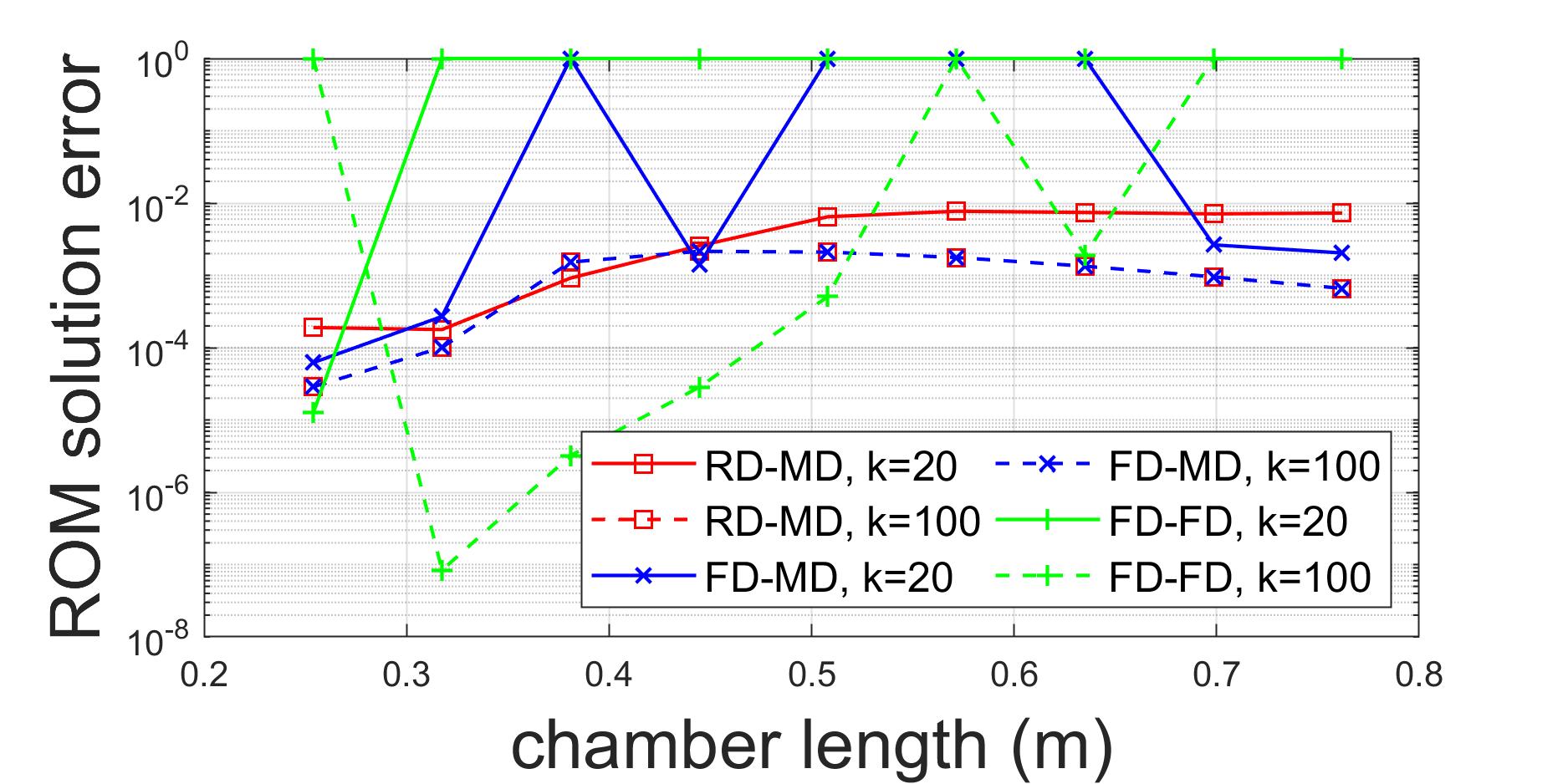}}
	\subfloat[$\alpha=3.25$]{
		\includegraphics[width=0.32\textwidth]{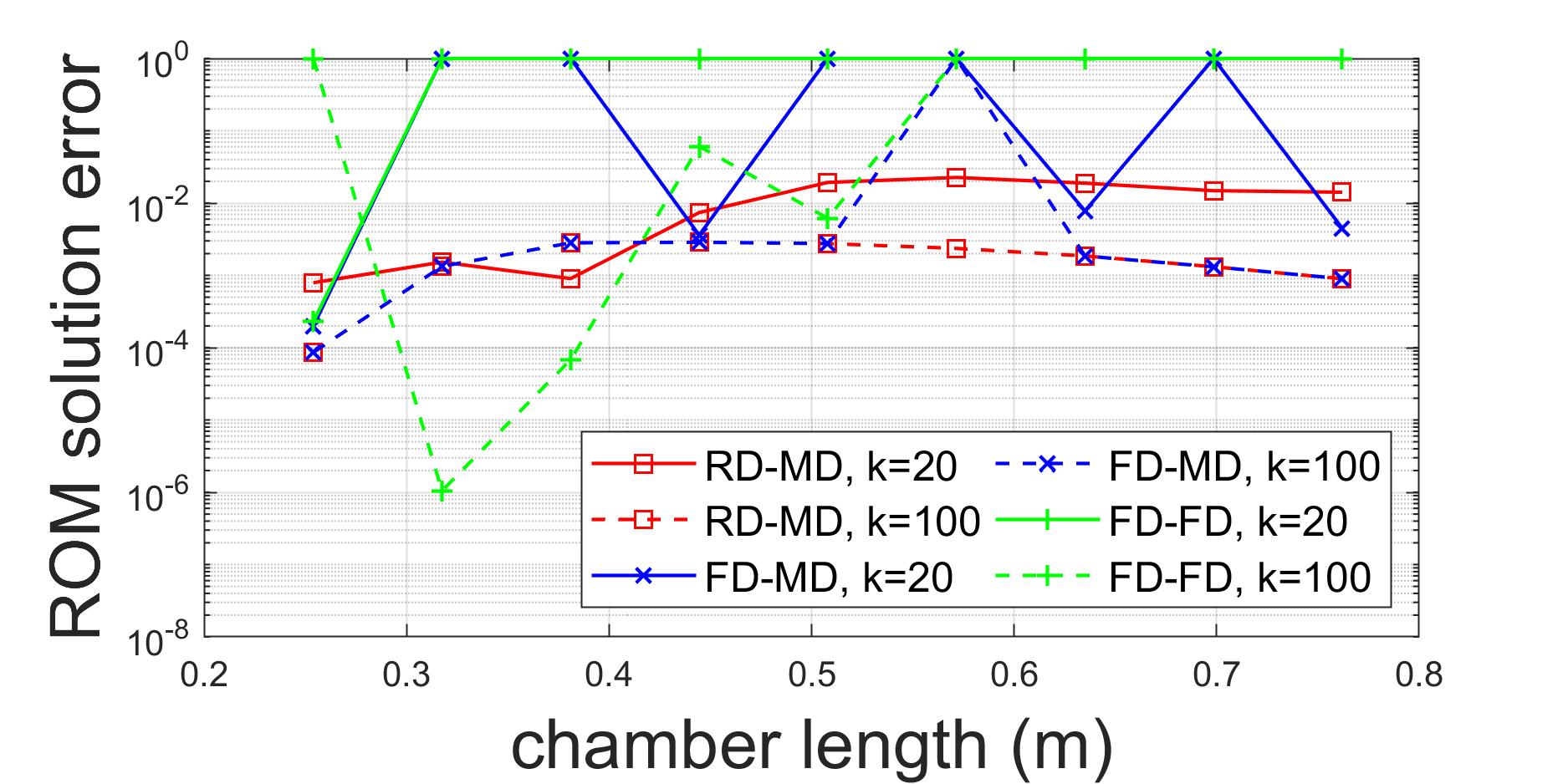}}
	\subfloat[$\alpha=3.4$]{
		\includegraphics[width=0.32\textwidth]{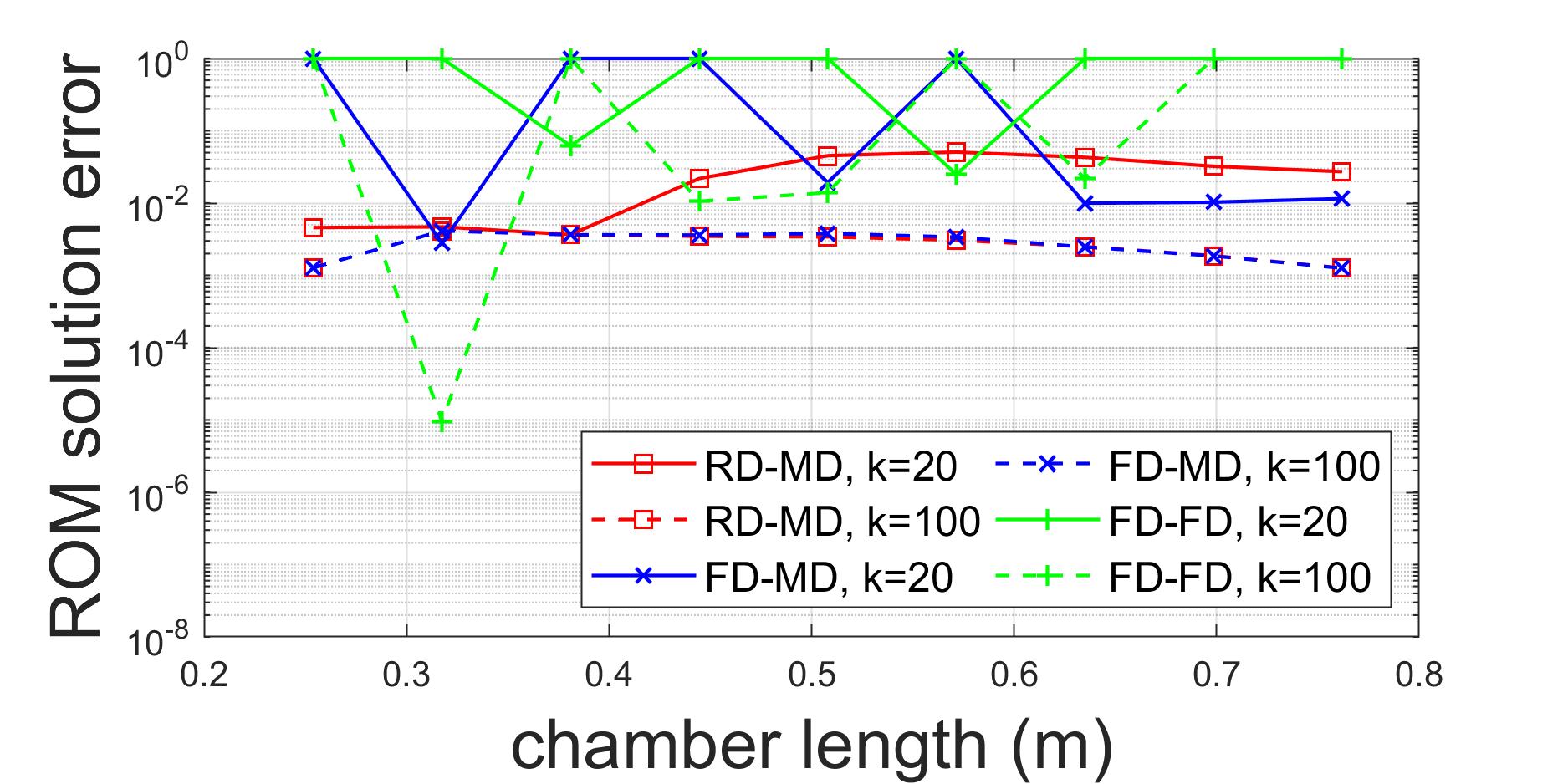}}
	\caption{Relative L2 error of ROM solutions, numerically unstable cases set to 1}\label{fig L2} 
\end{figure}

The solution from the three different techniques are compared against the FOM  over the same parameters and time period. From Fig.~\ref{fig projection} it can be seen that the projection error for RD-MD at $k=20$ is comparable to that for FD-FD at $k=100$. The three methods are compared at the two basis sizes, $k=20,100$ in this section. Comparisons of the relative L2 error of the ROM solution are shown in Fig. \ref{fig L2}. It should be mentioned that in several cases, the ROMs in the control groups are numerically unstable. The error is set to 1 under these circumstances, which is higher than the error in all stable cases. 

When the basis size is sufficiently large, i.e. at $k=100$, RD-MD performs similarly to  FD-MD, except for one case at $\alpha = 3.25, L_c = 0.5715$ m, where the latter is unstable. At high $\alpha$, where the instability is stronger, the advantage of RD-MD becomes more significant. This can be observed from the medium $L_c$ cases at $\alpha = 3.4$, where the  gap between RD-MD and FD-MD is apparent. Also, FD-FD performs worse as $\alpha$ increases, resulting in higher errors and more unstable cases. It should be noted that FD-FD sometimes outperforms the other two methods at low $L_c$, which is consistent with an improved decay in the singular values. However at $L_c = 0.254$ m, it becomes unstable. 

When the basis size is reduced to $k=20$, all the predictions deteriorate considerably. The two full-domain training methods become unstable in most conditions. In contrast, RD-MD remains stable at all combinations of parameters, although the ROM solution error is approximately one order higher than that of the $k=100$ cases.  The advantage in stability confirms the conclusion that the reduced-domain training results in a better set of basis functions. By virtue of using a smaller basis set, the computational cost, e.g. the projection computation in Eq.~\eqref{eq ROM1}, decreases linearly when an explicit time-marching scheme is used. In contrast to the traditional full-domain training methods, for which the stability is less predictable, the proposed framework provides a broader stability envelope, and consequentially a more flexible balance between ROM accuracy and efficiency.

\begin{figure}
	\centering
	\subfloat{
		\includegraphics[width=0.32\textwidth]{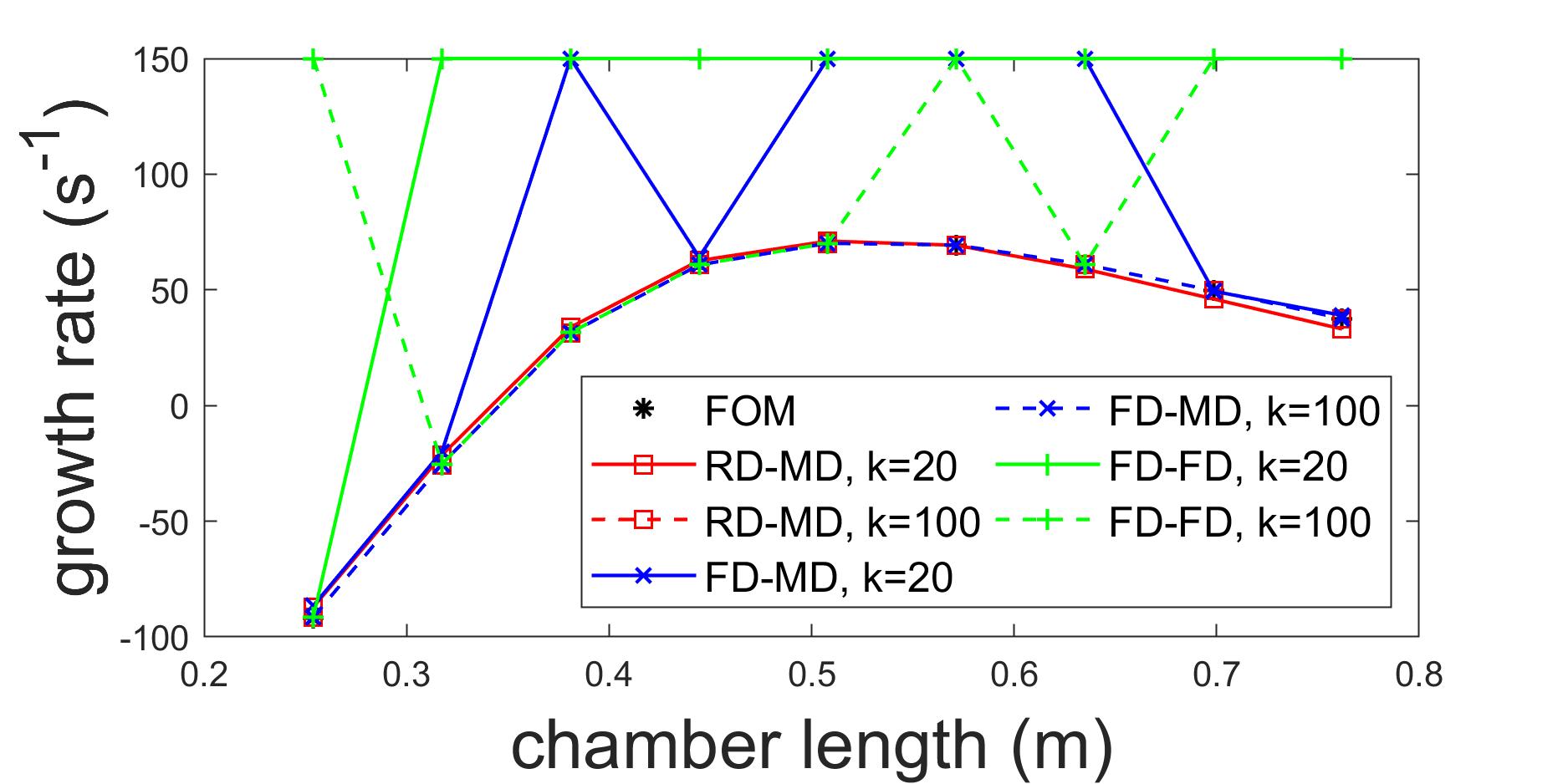}}
	\subfloat{
		\includegraphics[width=0.32\textwidth]{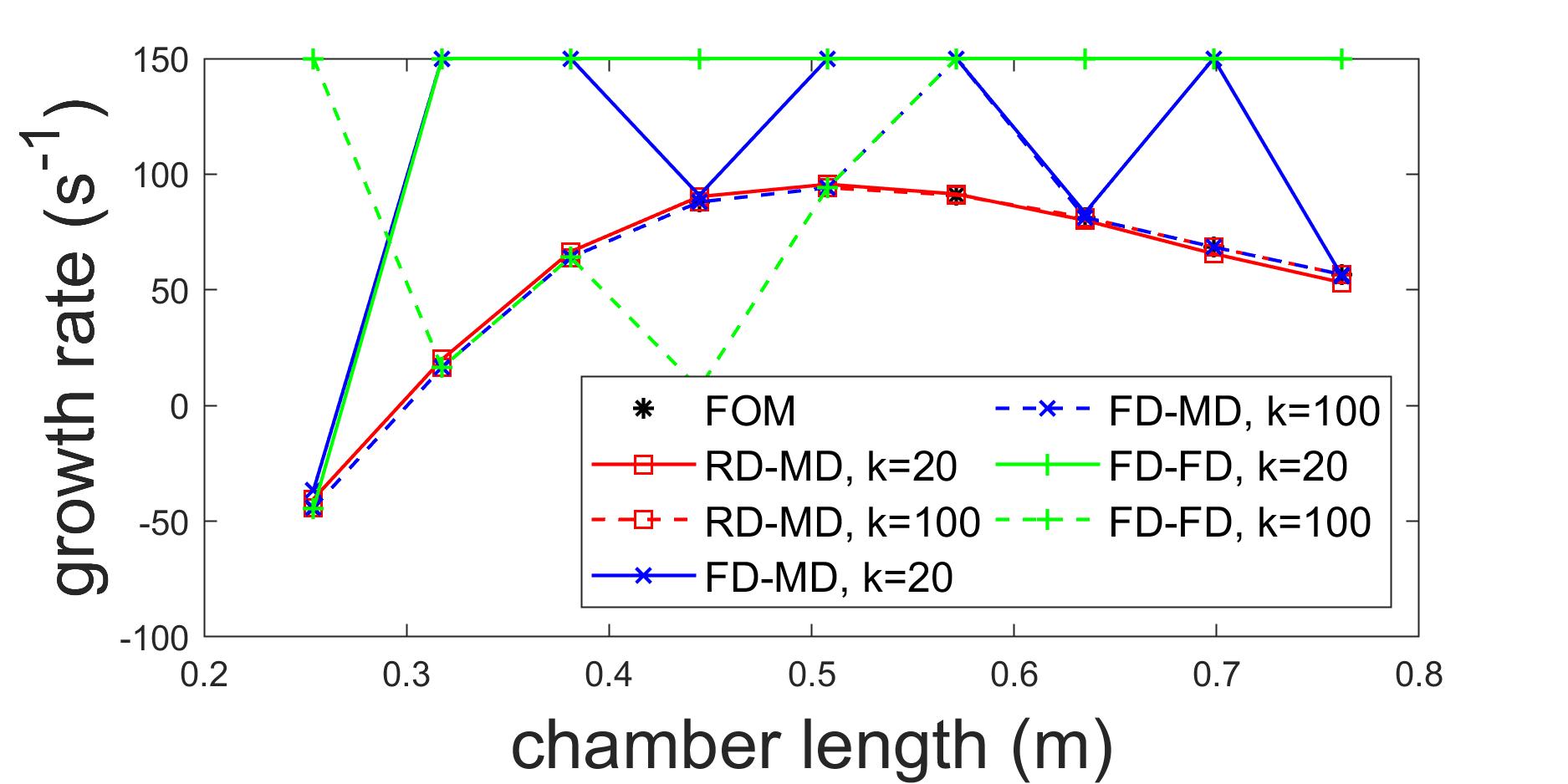}}
	\subfloat{
		\includegraphics[width=0.32\textwidth]{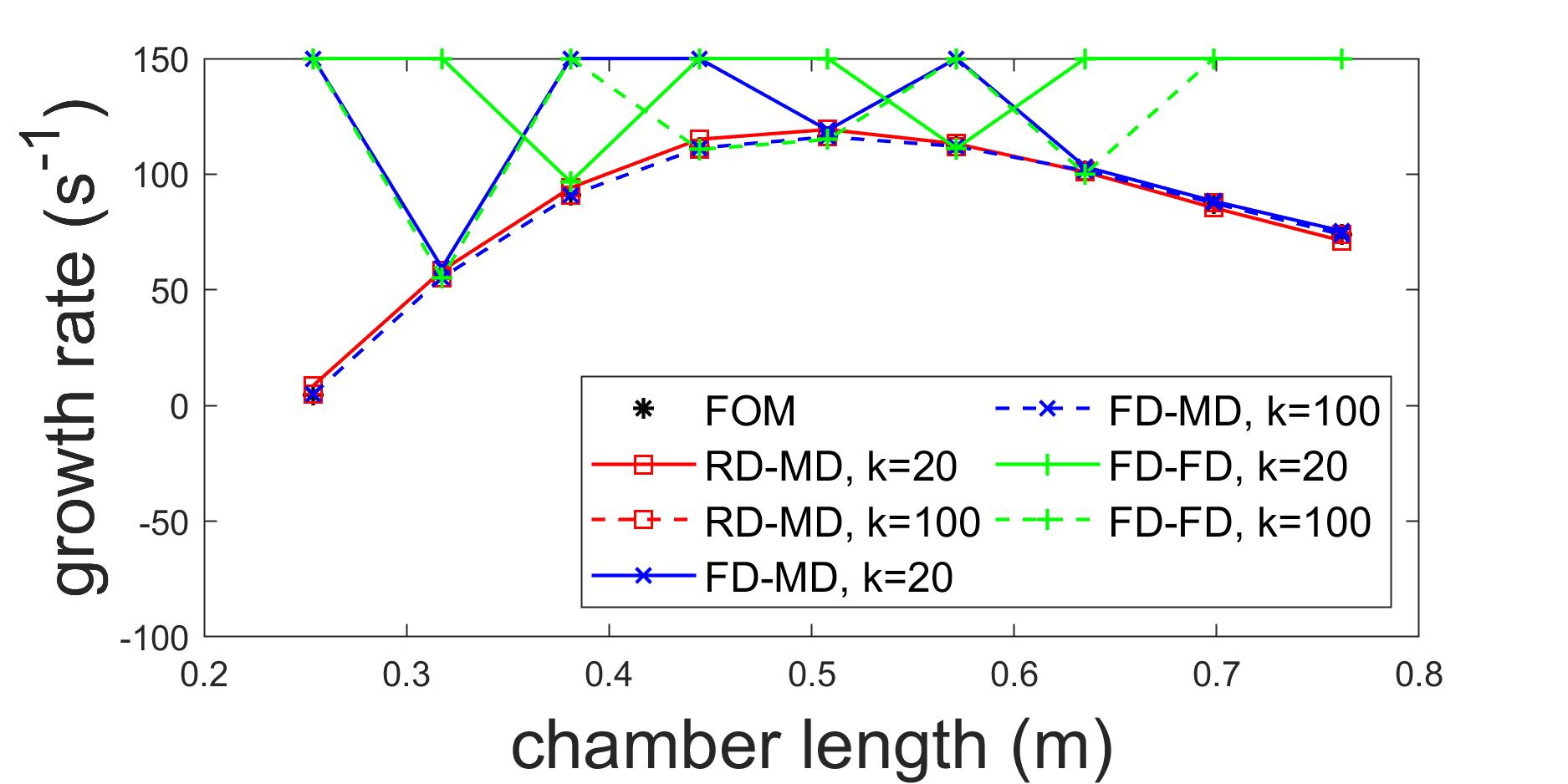}}
    \hfill
	\subfloat[$\alpha=3.1$]{
		\includegraphics[width=0.32\textwidth]{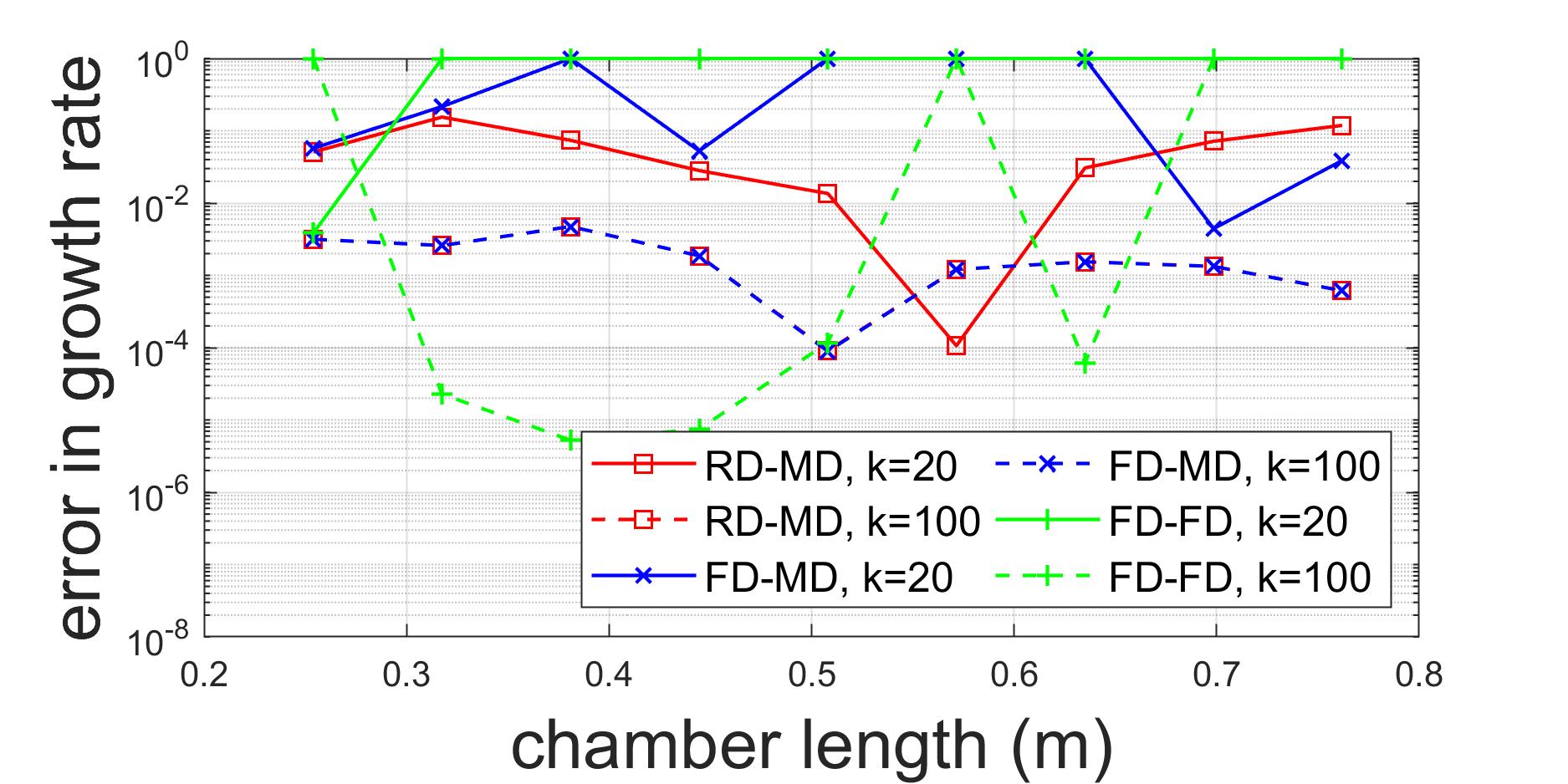}}
	\subfloat[$\alpha=3.25$]{
		\includegraphics[width=0.32\textwidth]{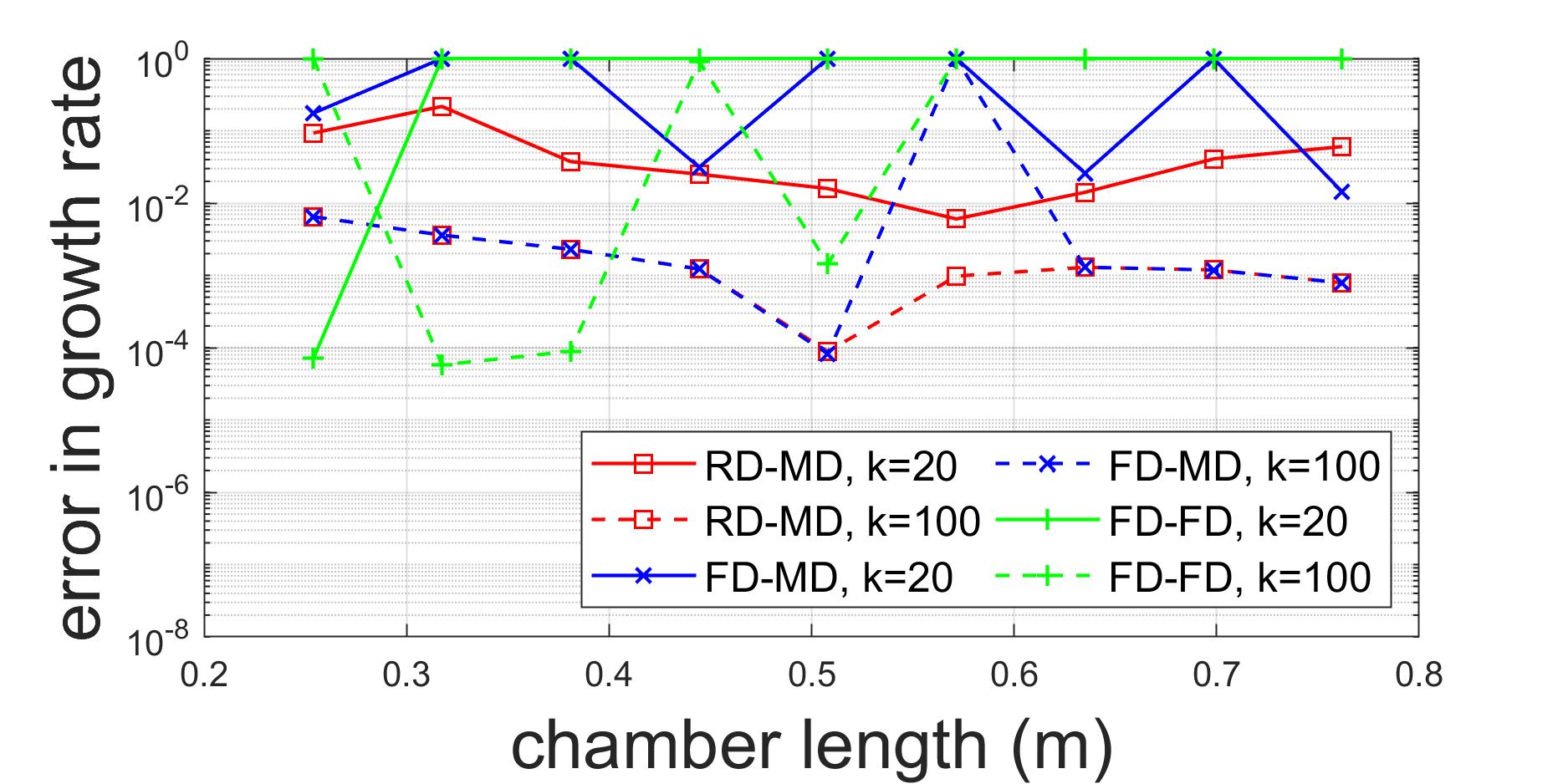}}
	\subfloat[$\alpha=3.4$]{
		\includegraphics[width=0.32\textwidth]{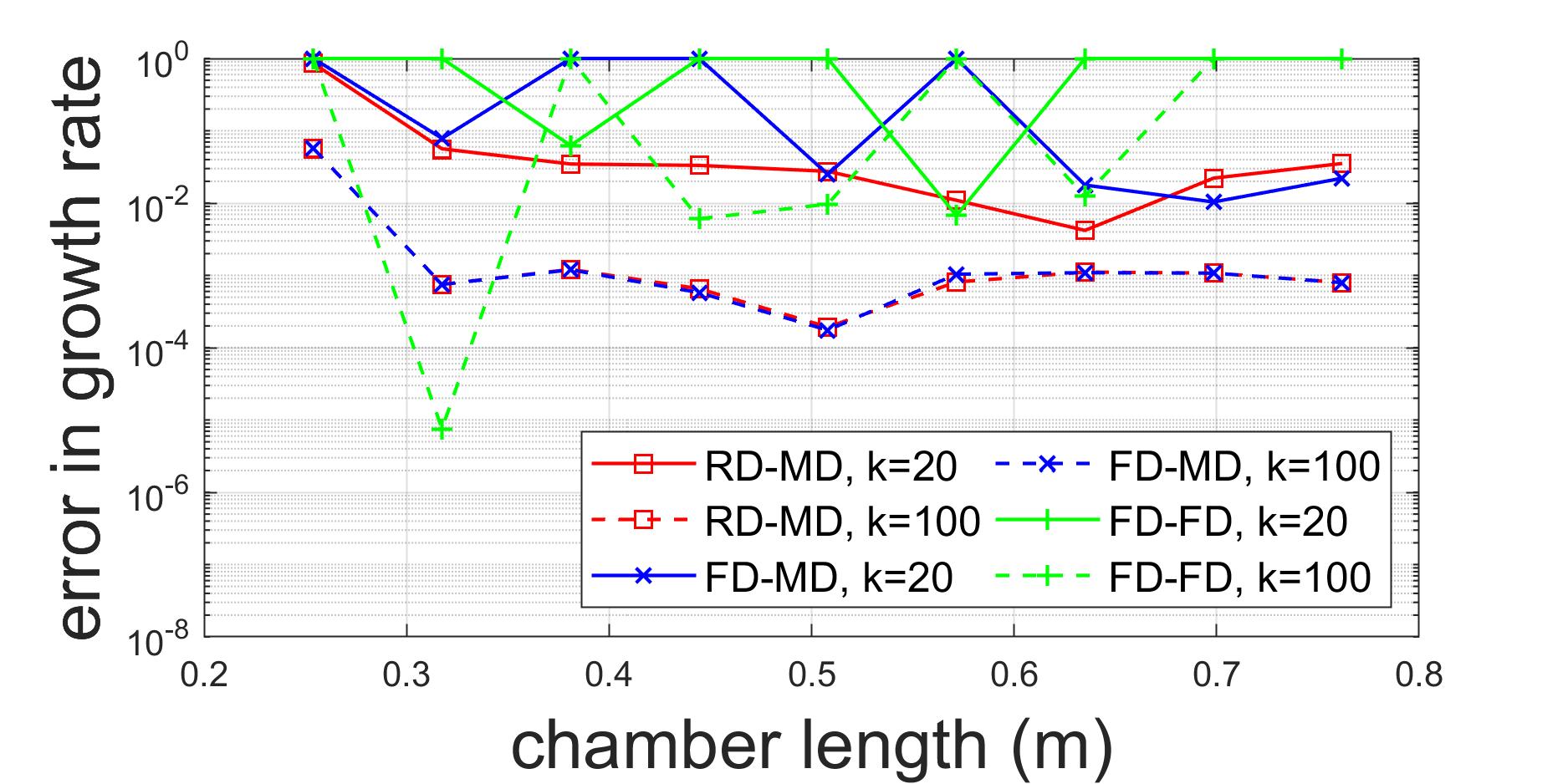}}
	\caption{Growth rate of pressure oscillation 0.0127 m upstream of  nozzle}\label{fig gr} 
\end{figure}

\begin{figure}
	\centering
	\subfloat{
		\includegraphics[width=0.32\textwidth]{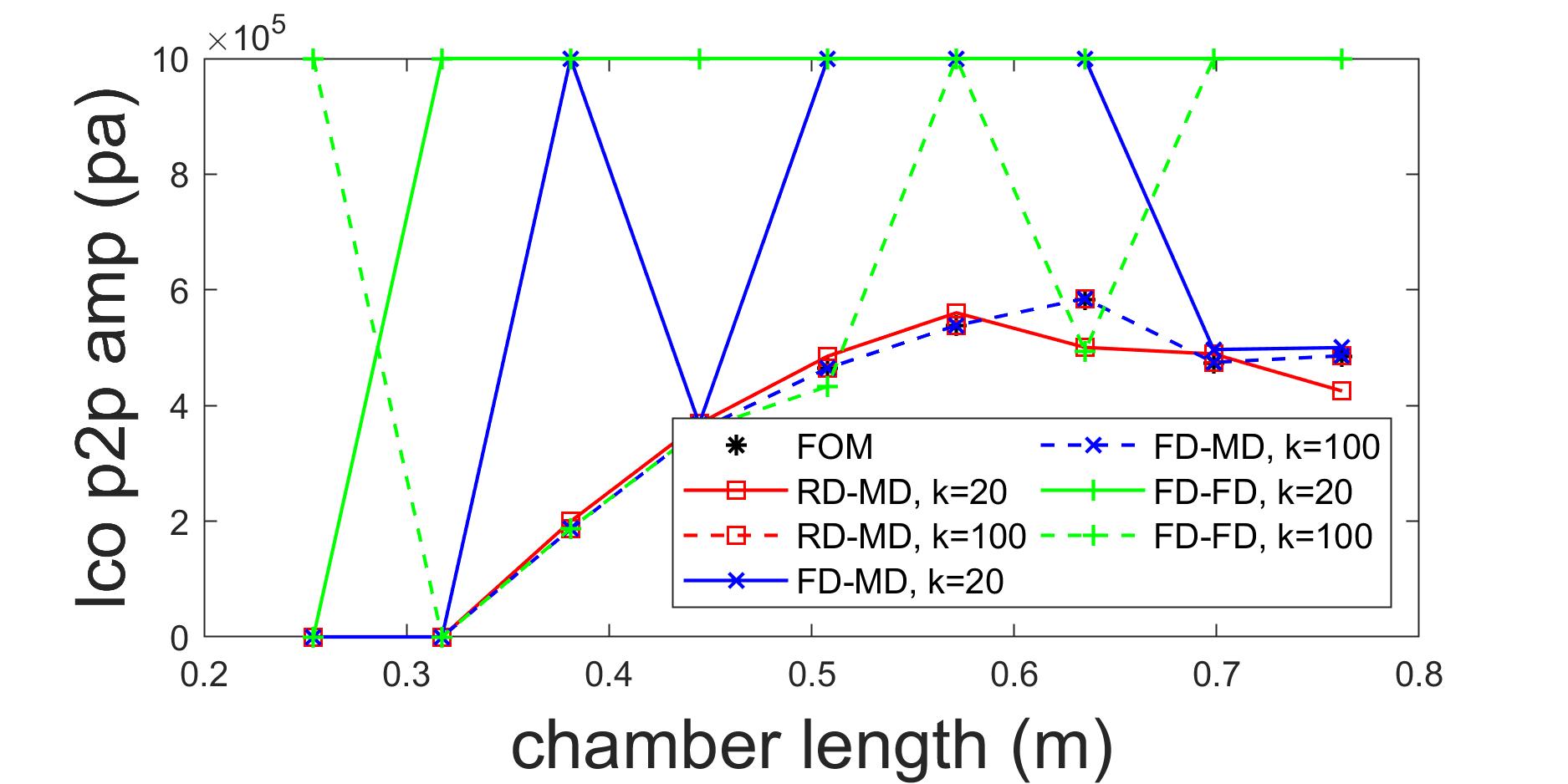}}
	\subfloat{
		\includegraphics[width=0.32\textwidth]{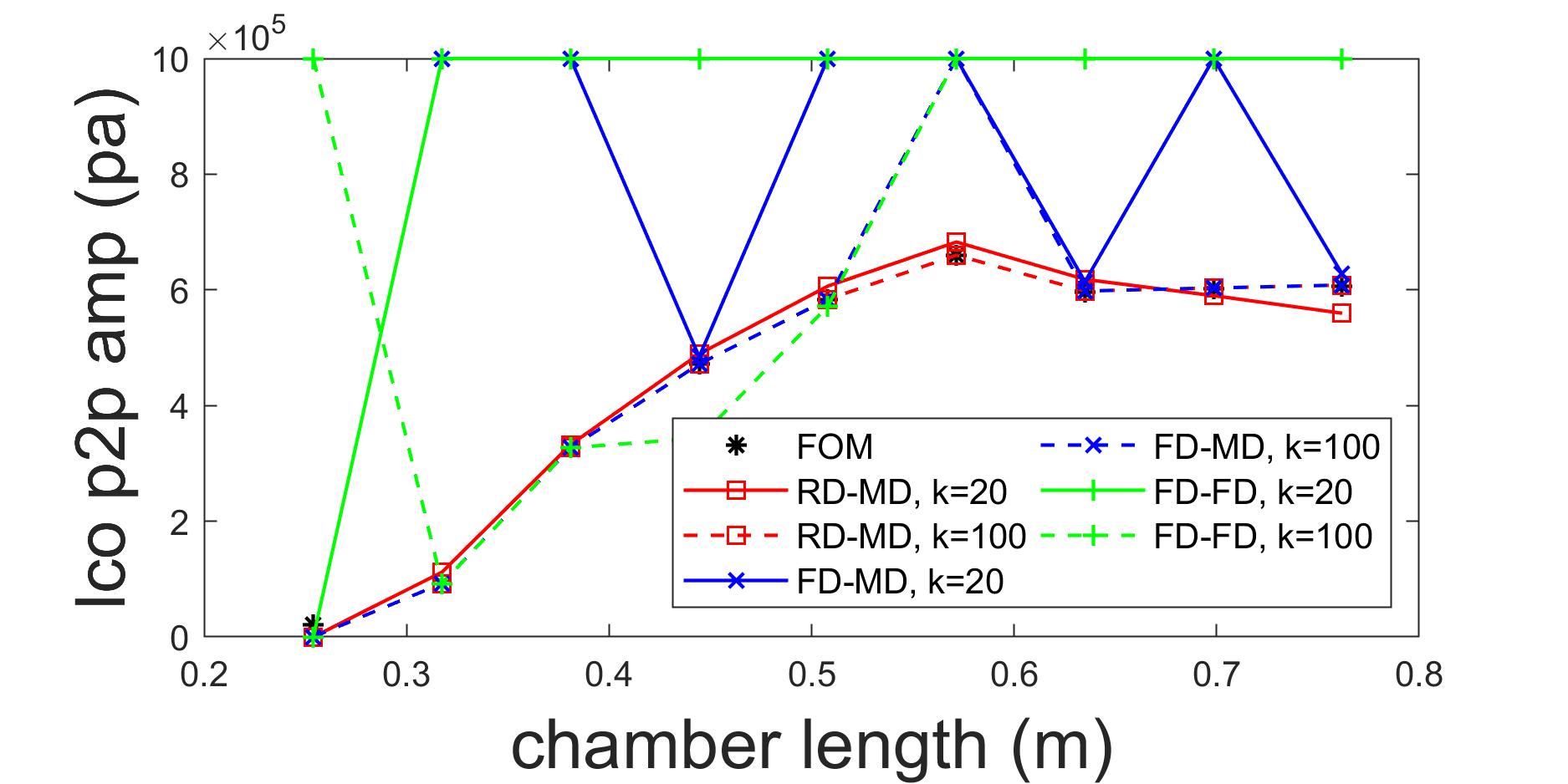}}
	\subfloat{
		\includegraphics[width=0.32\textwidth]{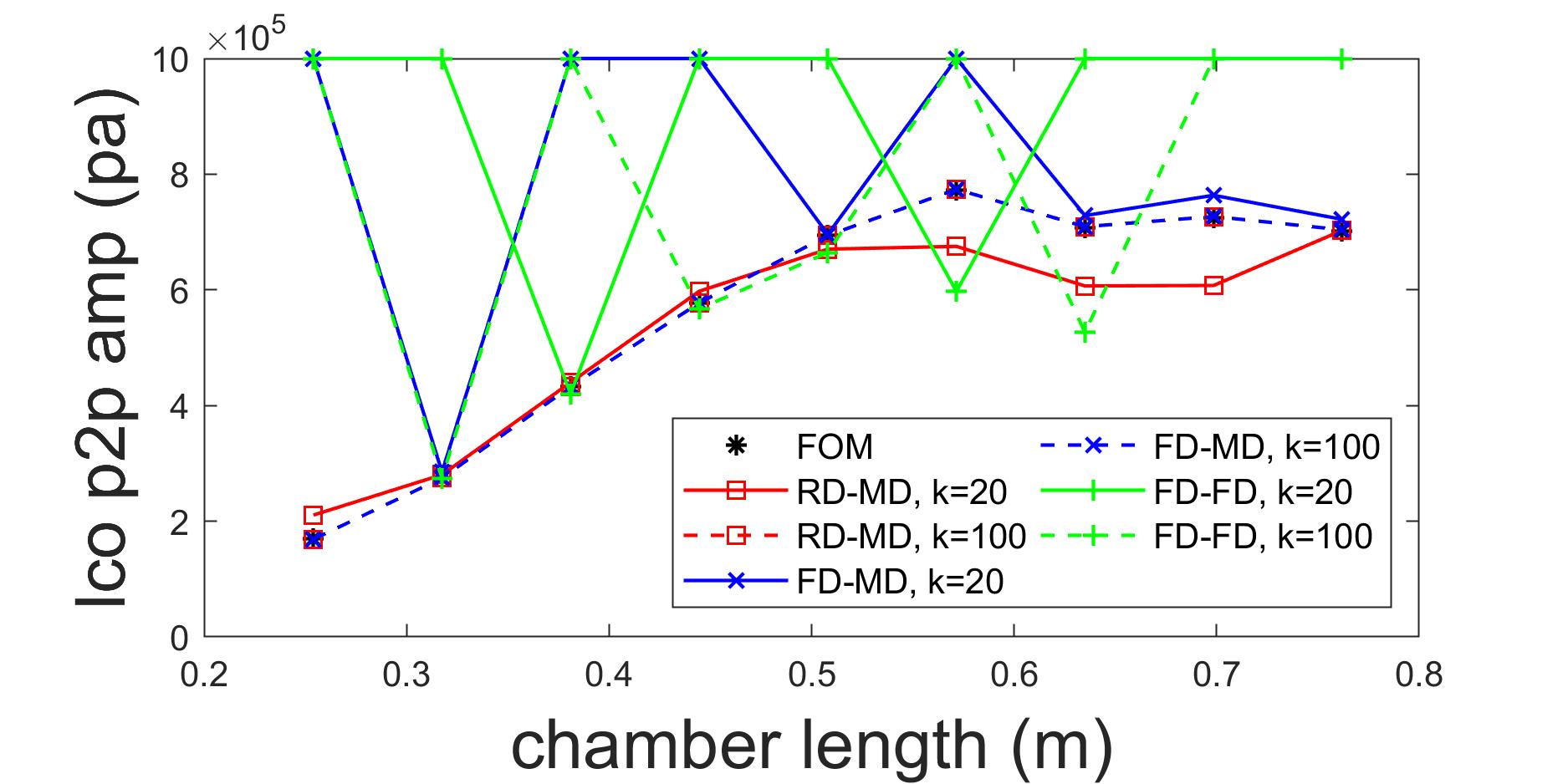}}
    \hfill
	\subfloat[$\alpha=3.1$]{
		\includegraphics[width=0.32\textwidth]{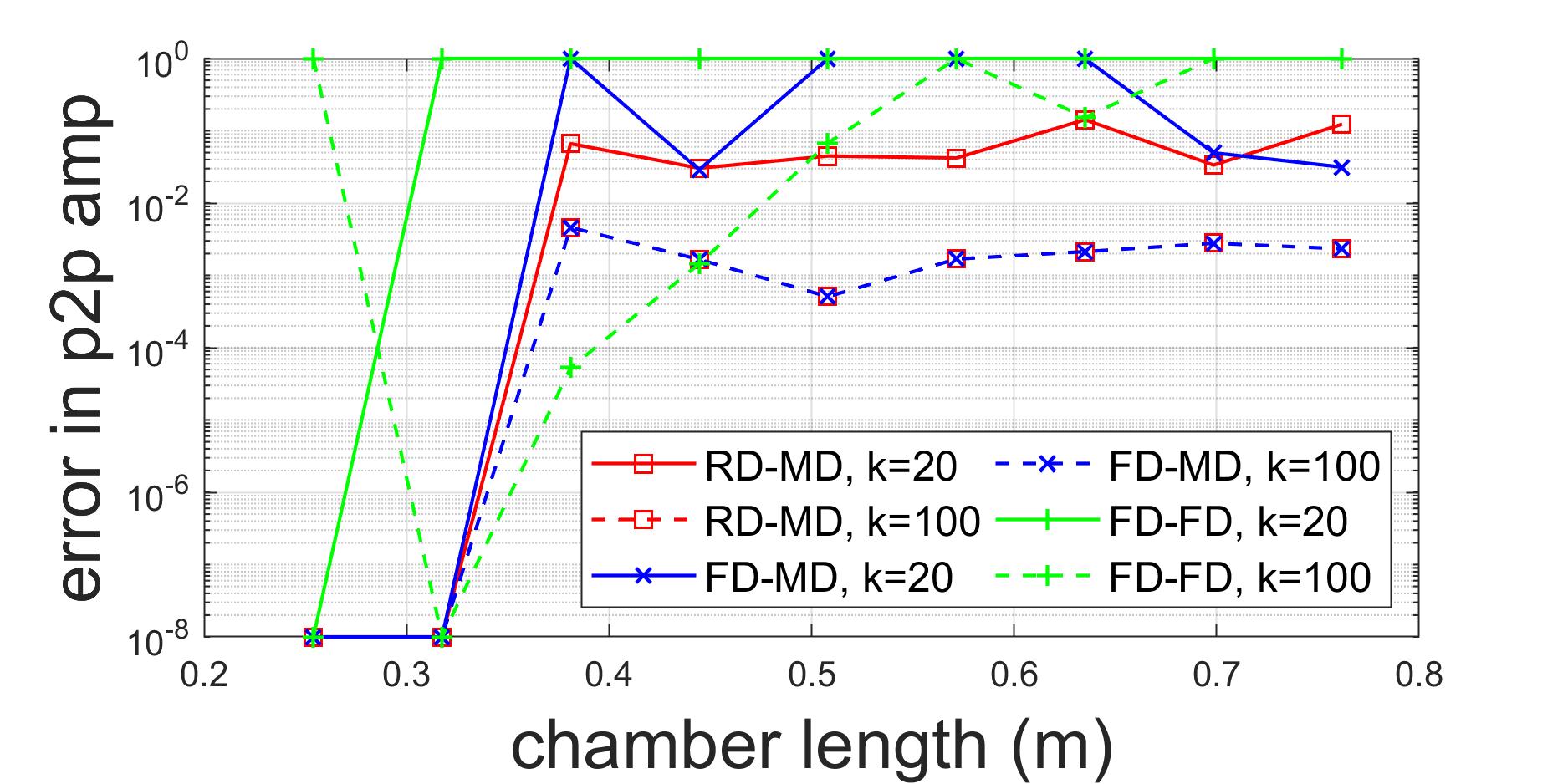}}
	\subfloat[$\alpha=3.25$]{
		\includegraphics[width=0.32\textwidth]{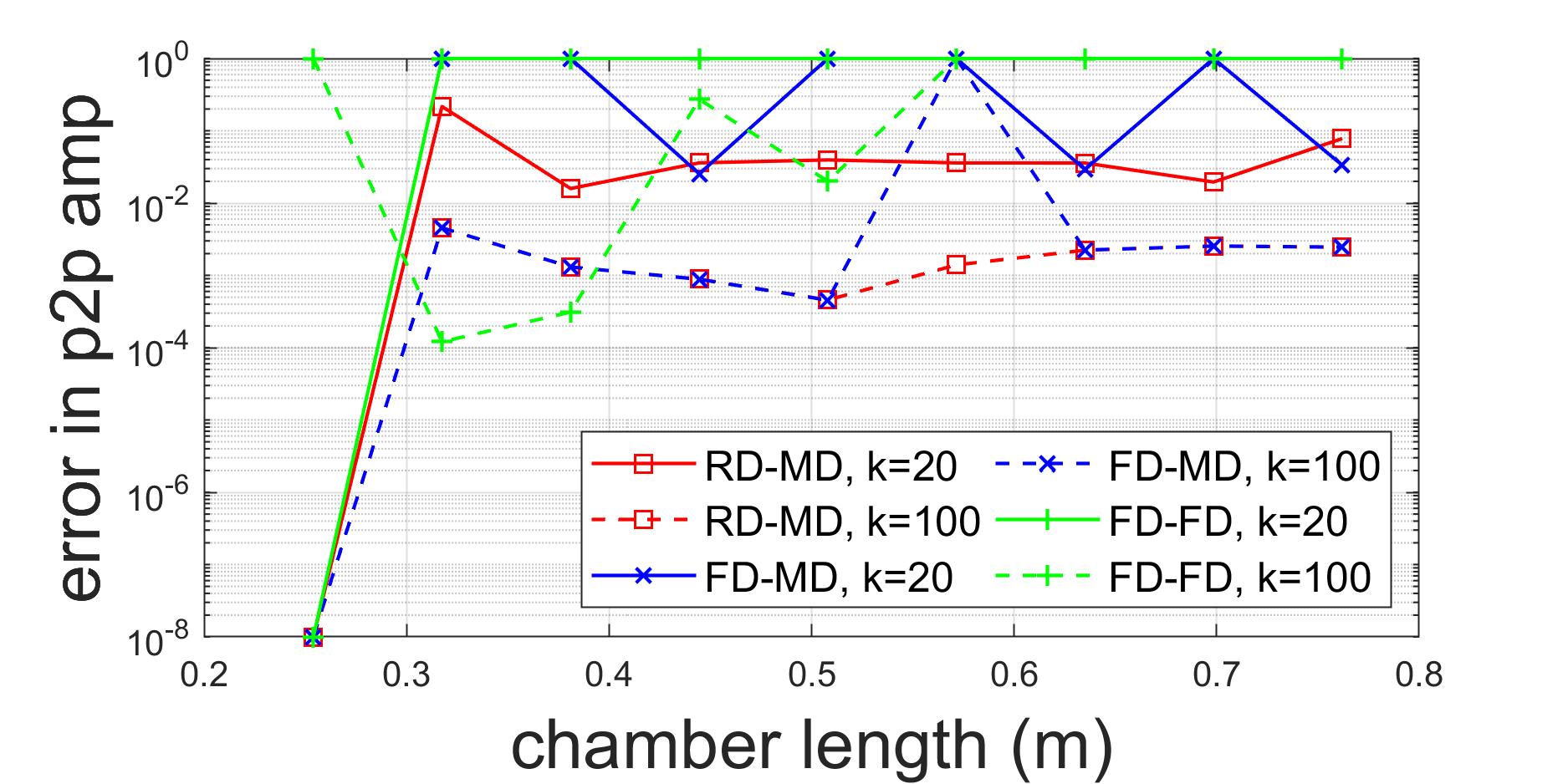}}
	\subfloat[$\alpha=3.4$]{
		\includegraphics[width=0.32\textwidth]{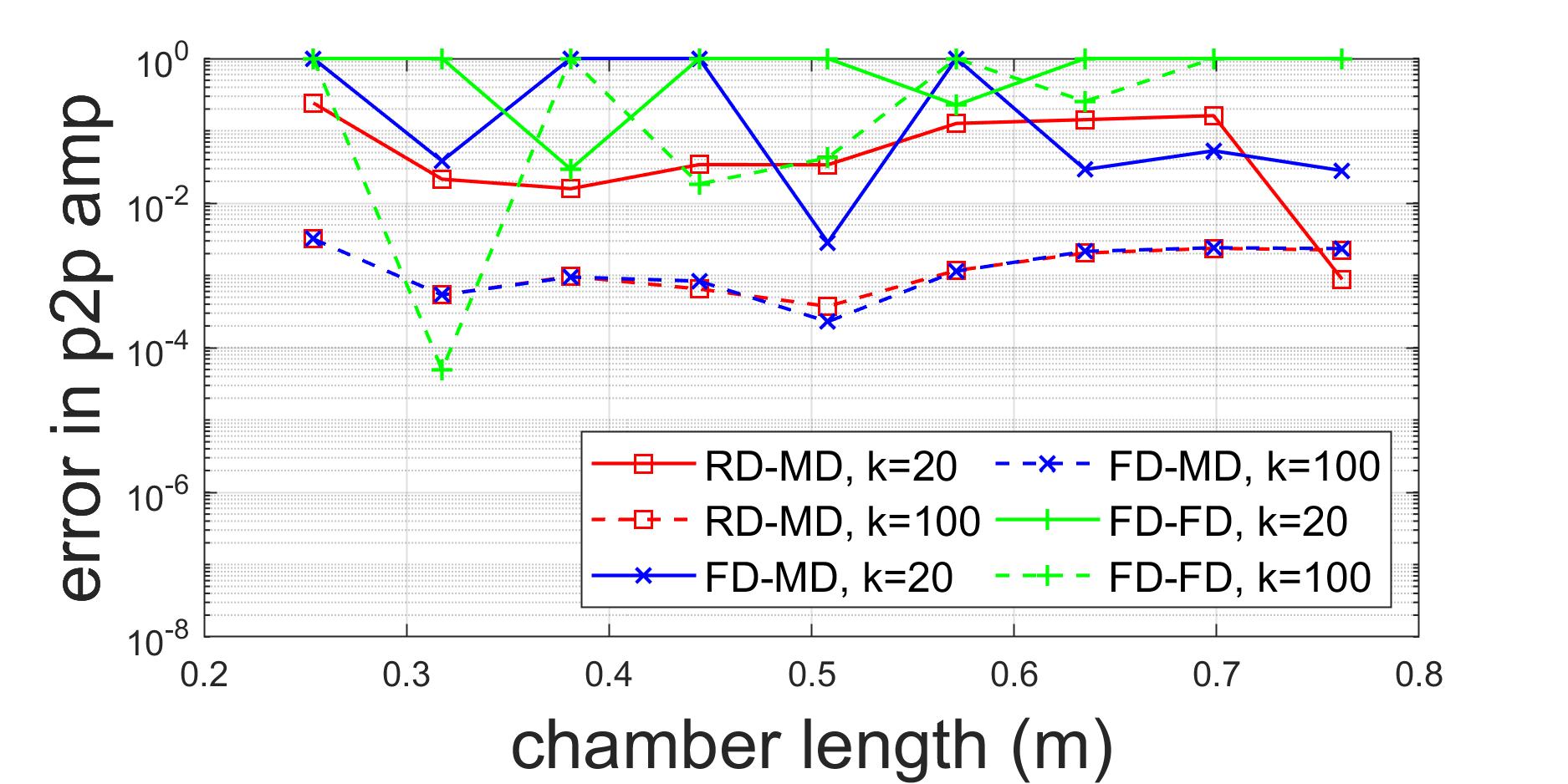}}
	\caption{LCO peak-to-peak amplitude of pressure oscillation 0.0127 m upstream of  nozzle}\label{fig lco} 
\end{figure}

Major quantities of interest in rocket combustor design include the dominant acoustic frequencies, the growth rates and LCO peak-to-peak amplitudes of the pressure oscillations. It should be noted that the dominant acoustic frequencies have been well-predicted by all three methods. Therefore only the comparisons of the other two quantities, growth rates and LCO amplitudes, are shown in the current study, as can be seen in Figs. \ref{fig gr} and \ref{fig lco} respectively for all the methods. The relative performance between different methods follows a similar relation as in the L2 error analysis. It can be seen that the proposed RD-MD framework is able to predict the relation between the growth rate and $L_c$ accurately with an error below $0.5\%$ at $k=100$ and below $5\%$ at $k=20$ in most cases, which illustrates its effectiveness.

\begin{figure}
	\centering
	\subfloat[$t=0.098$ s]{
		\includegraphics[width=0.3\textwidth]{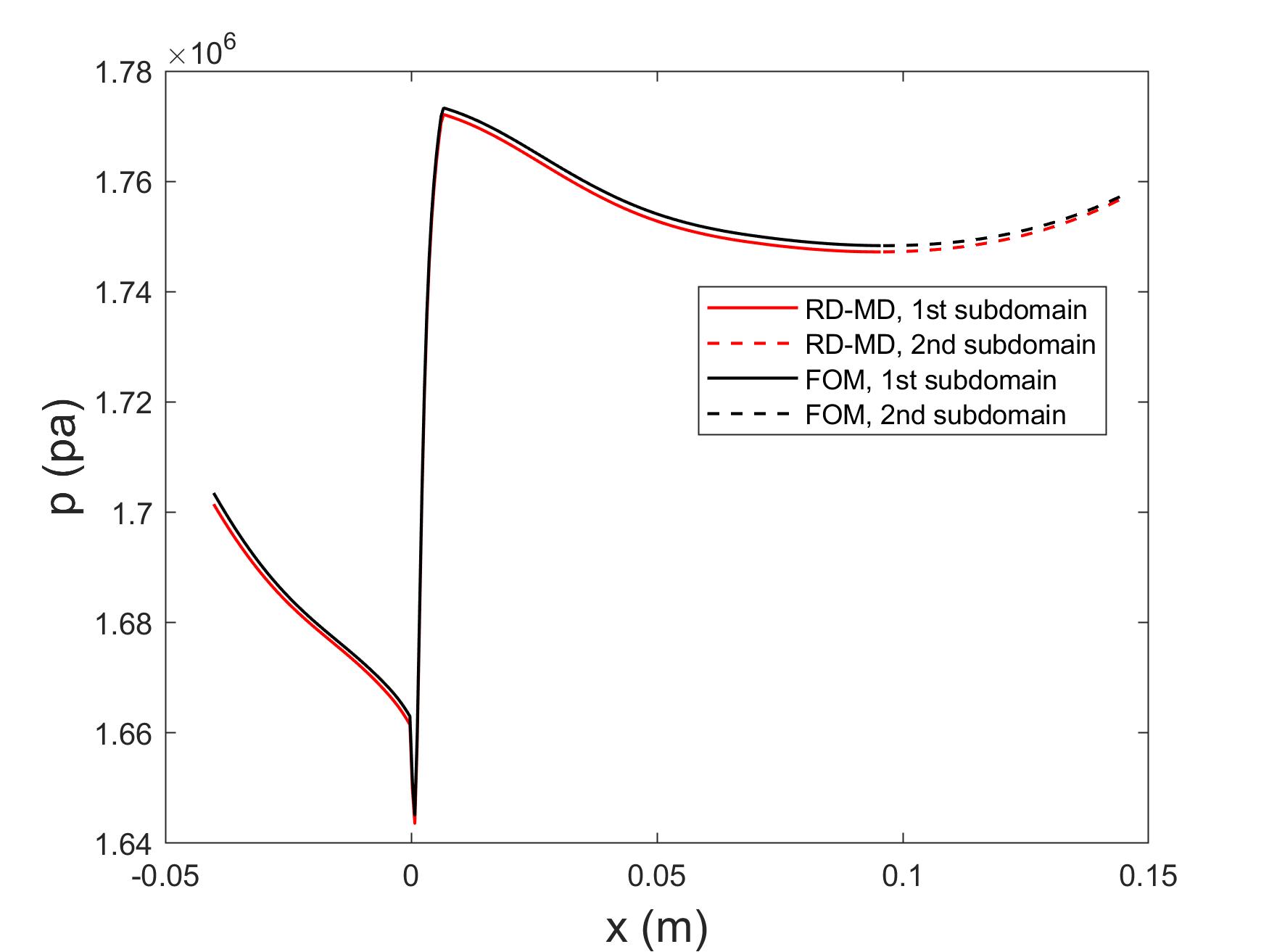}}
	\subfloat[$t=0.09825$ s]{
		\includegraphics[width=0.3\textwidth]{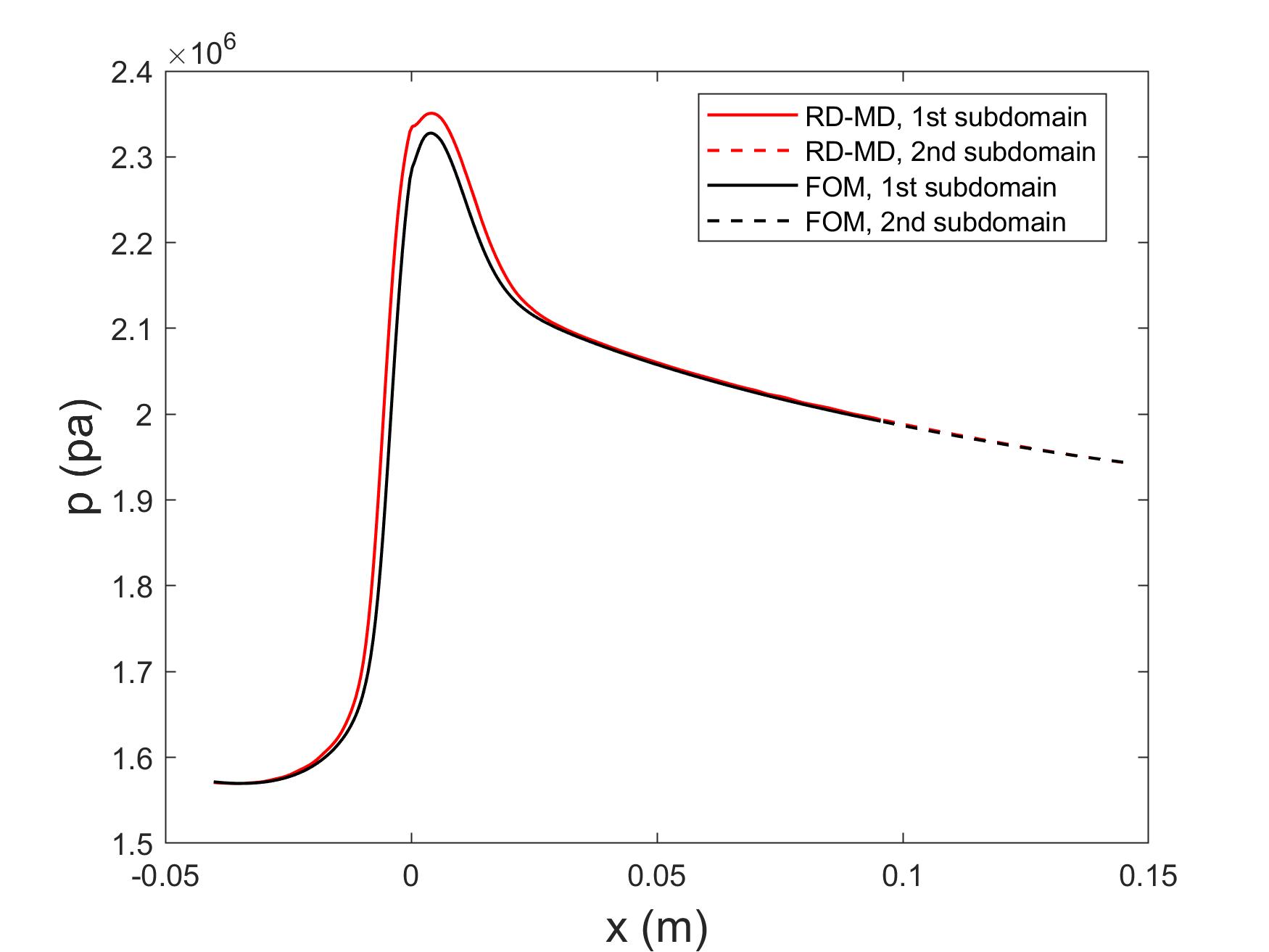}}
	\subfloat[$t=0.0985$ s]{
		\includegraphics[width=0.3\textwidth]{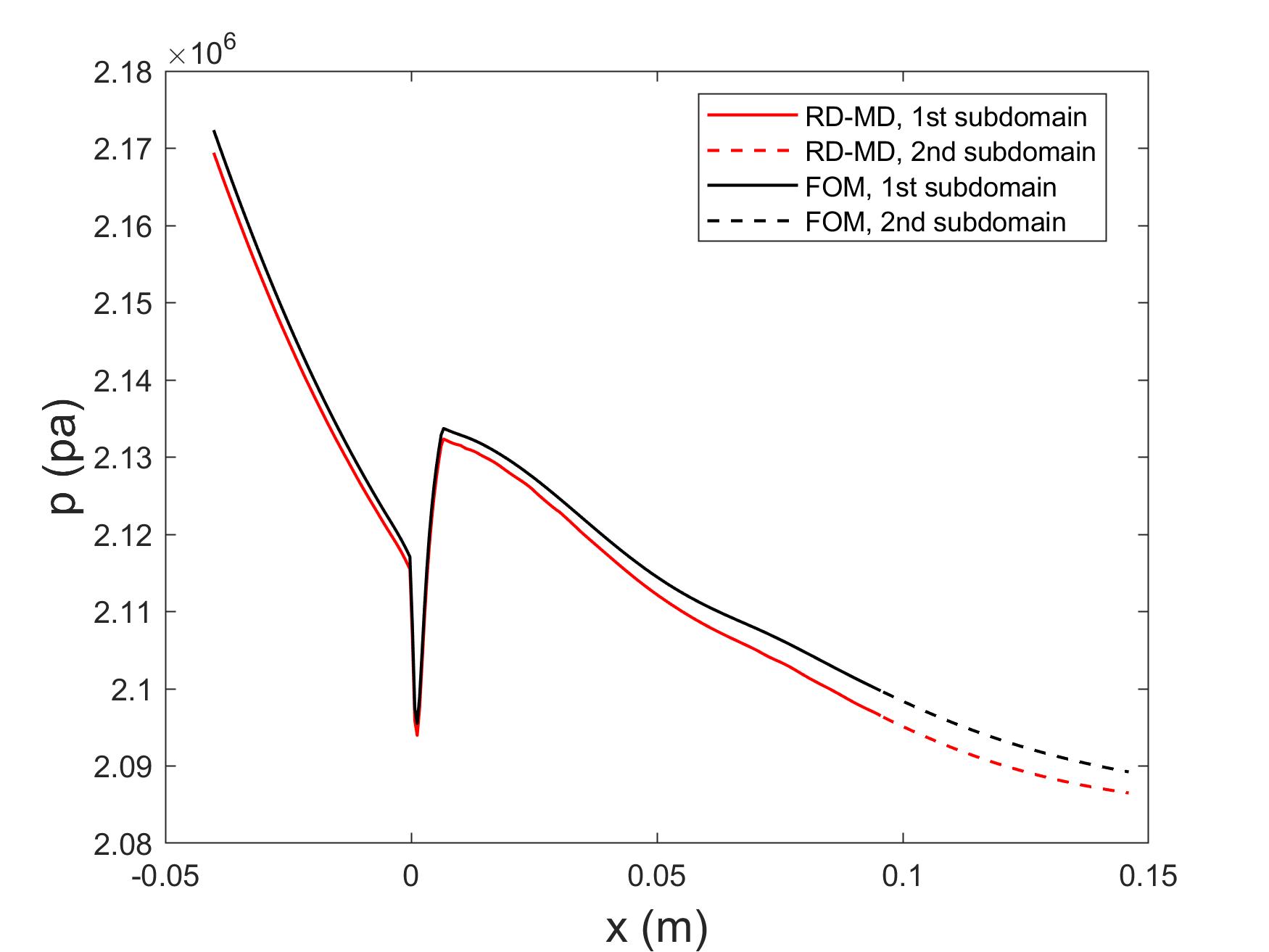}}
	\caption{Spatial profiles of pressure}\label{fig xp}
\end{figure}

Finally, to provide a more direct comparison between the ROM solution from the framework and the FOM solution, spatial pressure profiles for $L_c=0.508$ m, $\alpha = 3.25, k=100$ are provided at several time instances in Fig.~\ref{fig xp}. These plots are focused on the back-step area for better visualization, and are selected by the end of the simulation, when the error is maximum. It is observed that the RD-MD solution correlates well with the FOM, and connects smoothly across the interface.
\subsection{Off-design condition performance}
\begin{figure}
	\centering
	\subfloat{
		\includegraphics[width=0.49\textwidth]{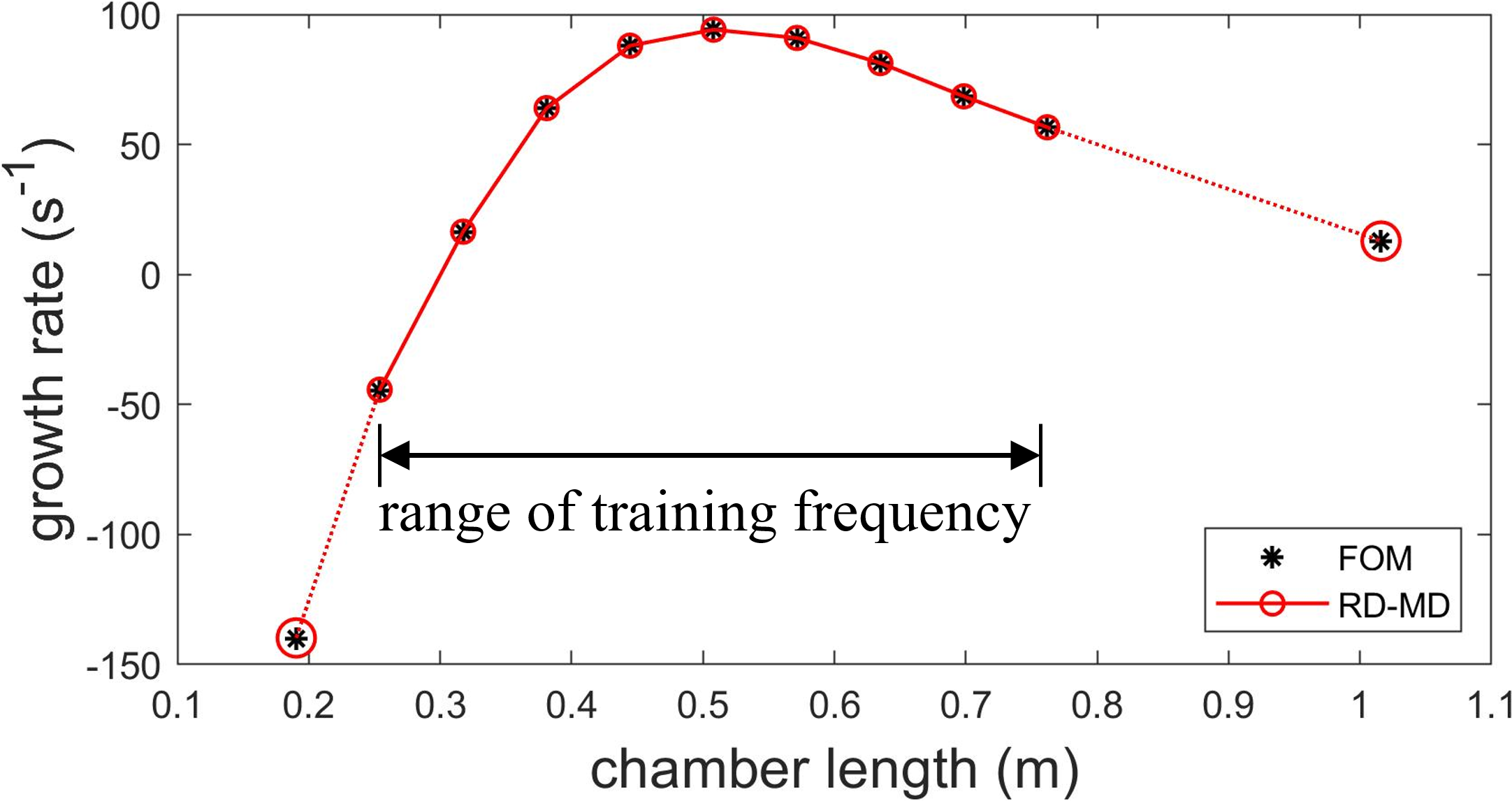}}
	\subfloat{
		\includegraphics[width=0.49\textwidth]{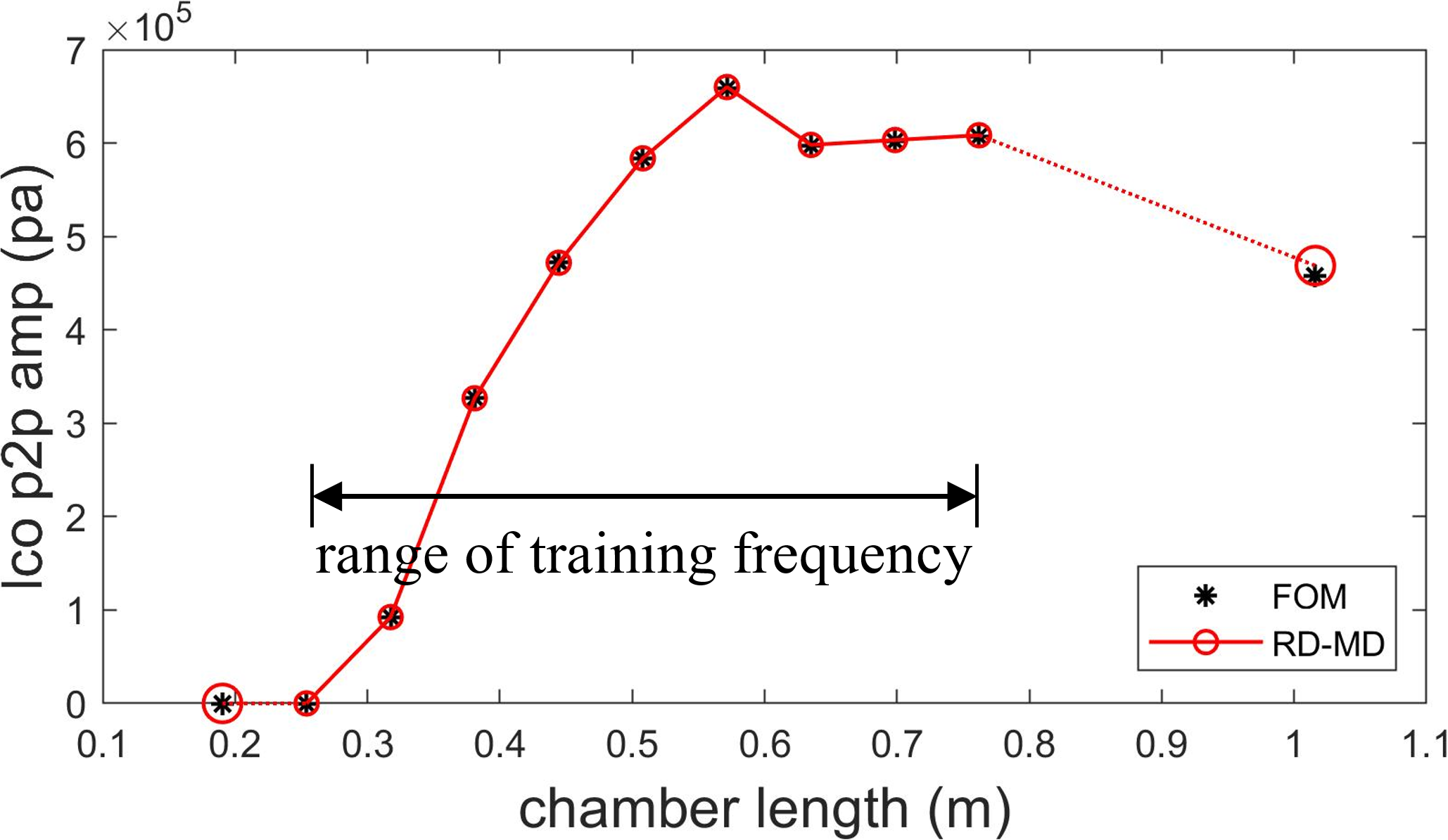}}
	\caption{Results at off-design $L_c$}\label{fig off-design lc} 
\end{figure}

\begin{figure}
	\centering
	\subfloat{
		\includegraphics[width=0.49\textwidth]{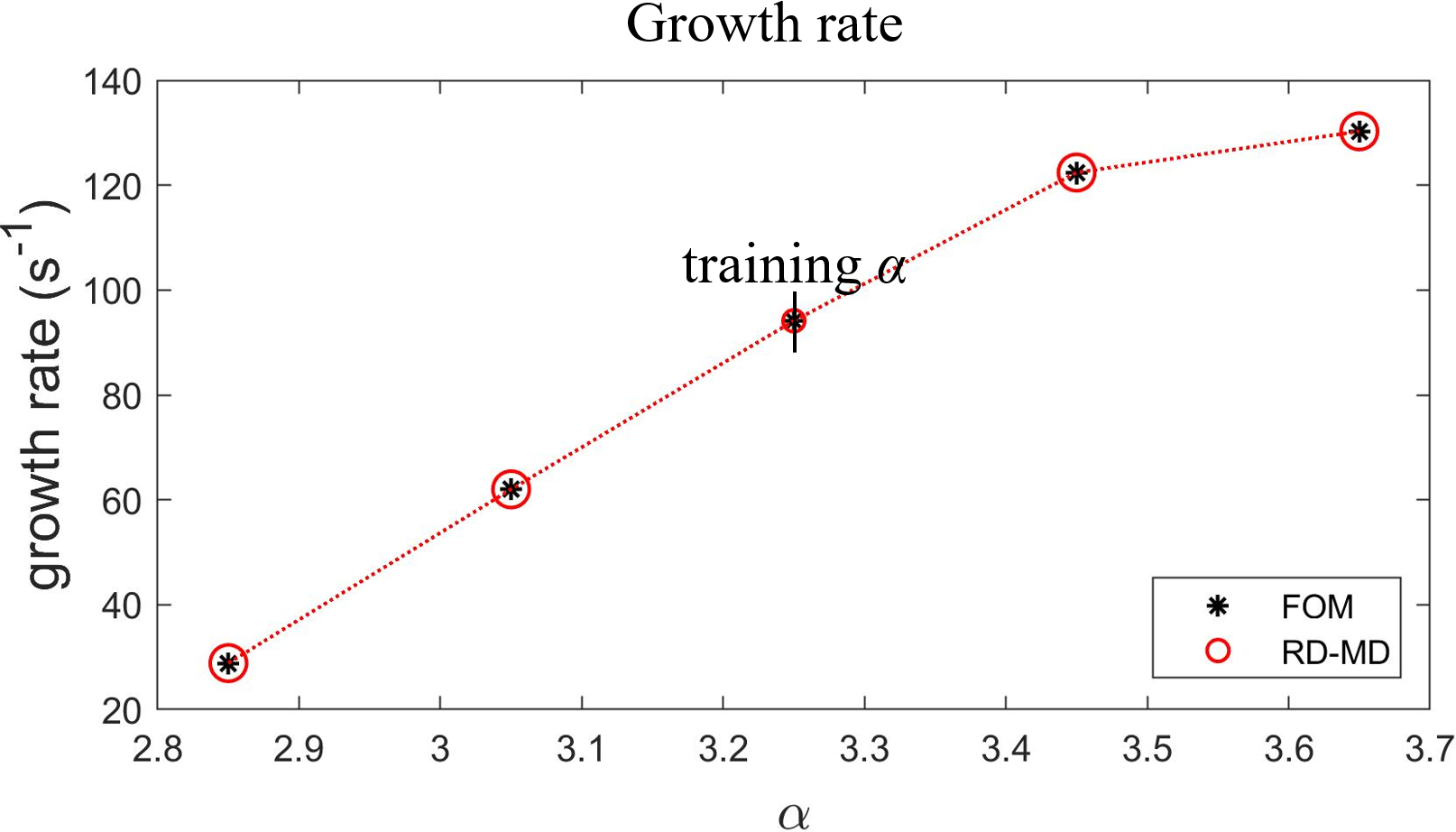}}
	\subfloat{
		\includegraphics[width=0.49\textwidth]{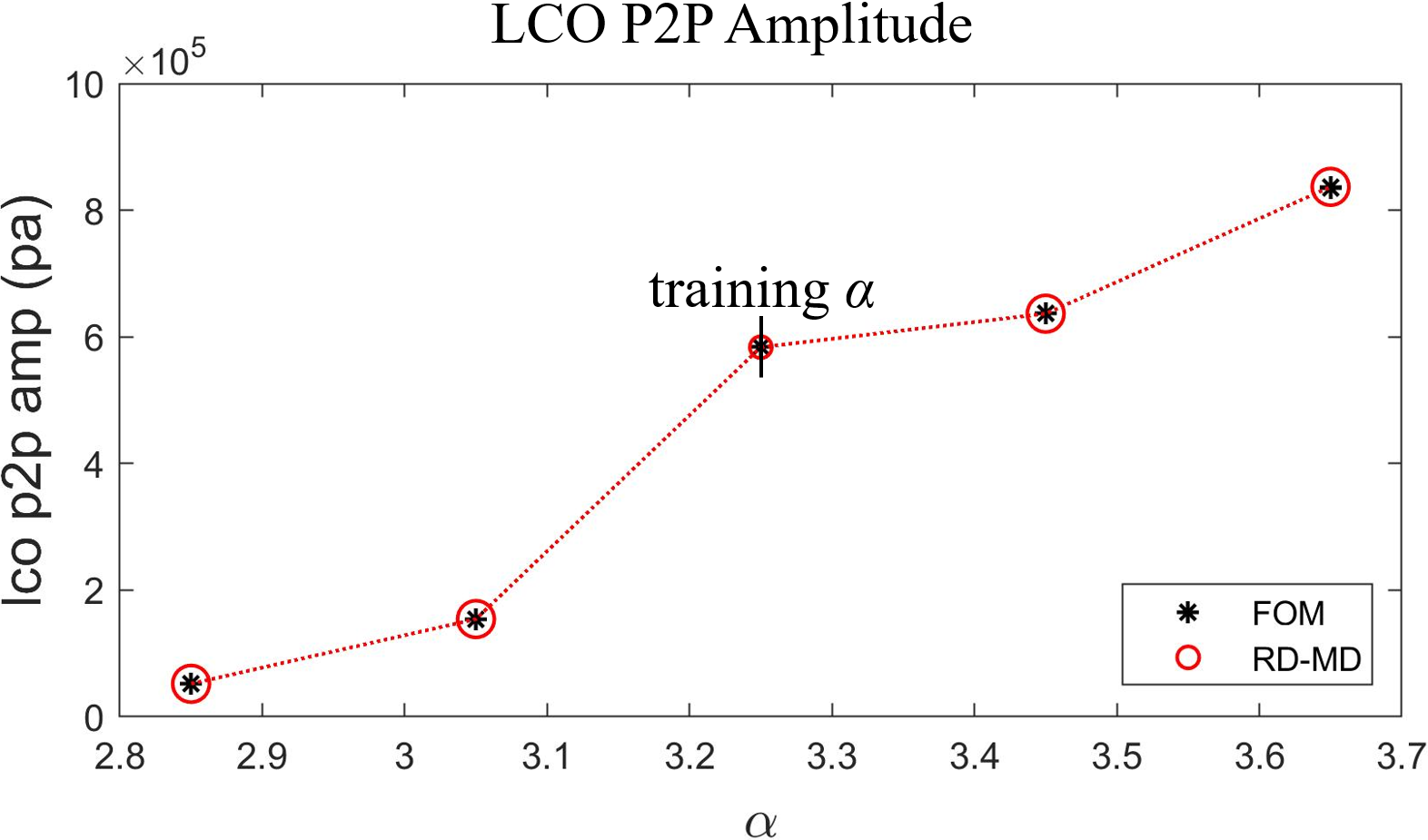}}
	\caption{Results at off-design $\alpha$}\label{fig off-design alpha} 
\end{figure}

To further assess the RD-MD framework, the ROM trained using $\alpha = 3.25, k = 100$ is evaluated at several off-design conditions. The evaluations include two cases at chamber lengths $L_c = 0.1905$ m and 1.016 m. These cases are characterized by dominant acoustic frequencies 2550 and 500 Hz, respectively, which are outside the range of the training frequencies given by Eq. \eqref{eq train signal}. The other conditions have amplification factors that deviate significantly from the training simulation, including $\alpha = 2.85,3.05,3.45,3.65$. The results for the L2 error ($\epsilon_2$), frequency ($f$), growth rate ($gr$), LCO peak-to-peak amplitude ($amp$) and their relative errors are summarized in Table \ref{table off-design}. It should be noted that due to the characteristic training method, the training case does not have a growing response as in the predictions, thus no growth rate or LCO amplitude is reported for the training set. Moreover, the results are plotted along with the designed conditions in Fig. \ref{fig off-design lc} and \ref{fig off-design alpha} for a better illustration of their relation.

It is observed that, at all the off-design conditions, the relative L2 errors are below $\num{1.2e-3}$. For the case with shorter $L_c$ and higher instability frequency (OD $L_{c1}$), and the four cases with deviated $\alpha$ (OD $\alpha_1$ to OD $\alpha_4$), the ROM performance is comparable to that in the designed conditions. However for the case with longer $L_c$ and lower instability frequency (OD $L_{c2}$), the error in LCO amplitude is $2.5\%$, which is more than 5 times higher than the other cases. The result indicates that for this problem, when operating within the frequency range of the training perturbation, the RD-MD method is not only independent of the chamber lengths, but also insensitive to changes in the unsteady heat release term. When beyond the training range, the instability frequency  influences  the predictive capabilities of the framework,  and demonstrates the importance of the multi-frequency perturbation in the characteristic training. 

\begin{table}
	\begin{center}
	\caption {Off-design condition results \\($f,gr,amp$ are listed in a ``FOM/ROM" style, all errors are relative)}\label{table off-design}
    \begin{small}
	\begin{tabular}{ |c| c c c c c c c c c|}
	\hline
		Case 	&$L_c$ (m)	&$\alpha$ &$\epsilon_2$ &$f$ (Hz) & $\epsilon_{f}$ &$gr$ ($s^{-1}$) & $\epsilon_{gr}$ & $amp $ (Mpa) & $\epsilon_{amp}$\\
		\hline
		Training &N/A & 3.25 & N/A & 700 to 2100 & N/A & N/A & N/A & N/A & N/A\\
		\hline
		OD $L_{c1}$ &0.1905 & 3.25 &\num{9.6e-5} & 2516/2516 & 0 &-139.98/-139.79 &\num{1.4e-3} &0/0 &0 \\
		OD $L_{c2}$ &1.016 & 3.25 &\num{2.7e-4} & 490/490 &0 &12.91/12.84 &\num{5.0e-3} &0.4576/0.4688 &\num{2.5e-2} \\
		OD $\alpha_1$ &0.508 & 2.85 &\num{1.2e-4} & 1000/1000 &0 &28.87/28.87 &\num{2.3e-5} &0.0522/0.0521 &\num{1.4e-3} \\
		OD $\alpha_2$ &0.508 & 3.05 &\num{3.3e-4} & 1000/1000 &0 &61.99/62.00 &\num{9.1e-5} &0.1545/0.1543 &\num{1.2e-3} \\
		OD $\alpha_3$ &0.508 & 3.45 &\num{1.2e-3} & 980/980 &0 &122.38/122.41 &\num{2.5e-4} &0.6366/0.6366 &\num{8.5e-5} \\
		OD $\alpha_4$ &0.508 & 3.65 &\num{1.6e-3} & 980/980 &0 &130.19/130.27 &\num{6.3e-4} &0.8368/0.8368 &\num{4.4e-5} \\
		\hline
	\end{tabular}
    \end{small}
	\end{center} 
\end{table}

\section{Conclusions}\label{sec conclusion}
In this work, a multi-fidelity ROM development framework   is investigated on a quasi-1D version of the CVRC model rocket combustor. An $\alpha-\tau$ model that couples the pressure oscillation and heat release is used to represent the unstable behavior of the combustor. In the proposed framework, the domain is split into two sub-domains with the first sub-domain containing the main reaction dynamics and the second covering the adjustable chamber of the CVRC. Characteristic training is conducted in the first sub-domain, resulting in a ROM that can be directly integrated with a FOM solver of the second sub-domain with any length. 

Numerical tests are conducted at different chamber lengths $L_c$ and amplification factors $\alpha$. The framework is compared against the FOM solution, and other ROM approaches where traditional training approaches are taken. A major advantage of the current framework is that it significantly reduces the number of FOM simulations required to train ROMs while the traditional method requires a separate FOM simulation for each individual target chamber lengths, which does not fit in the needs for more efficient rocket engine design. Moreover, the proposed method shows faster decay of the singular value spectrum at medium-to-high chamber lengths, which implies better basis quality and ROM reliability at a low number of modes. The advantage is especially distinct in the tests with $k=20$, where the proposed method shows significantly improved stability.

In predictive tests at conditions outside  the training set, the framework showed significant improvement in both stability and accuracy over the traditional methods,  especially when the instability is more pronounced at high $\alpha$, and when number of modes is low. Comparison of L2 error, growth rate and LCO peak-to-peak amplitude shows that the framework is able to predict the these quantities accurately at all combinations of $\alpha$ and $L_c$ at $k=100$. 

In summary, the multi-fidelity framework  proves to be promising approach for modeling rocket combustion instability. The modularity of the framework will be useful in efficient training and integration of components such as injector elements. The flexibility of the multi-domain approach also enables potential developments such as the application of different time-marching schemes, basis sizes and acceleration methods in different parts of the computational domain to achieve a better balance between efficiency and reliability in the future ROM development.

While the results are encouraging, it should be recognized that the flow and combustion models used in this study are highly simplified and further studies are required on more complex flow problems to evaluate the capabilities of the multi-fidelity framework. 

\section{Acknowledgments}
The authors acknowledge  support from the Air Force under the  Center of Excellence grant FA9550-17-1-0195, titled ``Multi-Fidelity Modeling of Rocket Combustor Dynamics.''


\begin{thebibliography}{47}
\newcommand{\enquote}[1]{``#1''}
\providecommand{\natexlab}[1]{#1}
\providecommand{\url}[1]{\texttt{#1}}
\providecommand{\urlprefix}{URL }
\expandafter\ifx\csname urlstyle\endcsname\relax
  \providecommand{\doi}[1]{doi:\discretionary{}{}{}#1}\else
  \providecommand{\doi}{doi:\discretionary{}{}{}\begingroup
  \urlstyle{rm}\Url}\fi

\bibitem[{Harvazinski et~al.(2015)Harvazinski, Huang, Sankaran, Feldman,
  Anderson, Merkle, and Talley}]{harvazinski2015coupling}
Harvazinski, M.~E., Huang, C., Sankaran, V., Feldman, T.~W., Anderson, W.~E.,
  Merkle, C.~L., and Talley, D.~G., \enquote{Coupling between hydrodynamics,
  acoustics, and heat release in a self-excited unstable combustor,}
  \emph{Physics of Fluids (1994-present)}, Vol.~27, No.~4, 2015, p. 045102.

\bibitem[{Matsuyama et~al.(2016)Matsuyama, Hori, Shimizu, Tachibana, Yoshida,
  and Mizobuchi}]{matsuyama2016large}
Matsuyama, S., Hori, D., Shimizu, T., Tachibana, S., Yoshida, S., and
  Mizobuchi, Y., \enquote{Large-Eddy Simulation of High-Frequency Combustion
  Instability in a Single-Element Atmospheric Combustor,} \emph{Journal of
  Propulsion and Power}, 2016, pp. 628--645.

\bibitem[{Domingo et~al.(2005)Domingo, Vervisch, and
  R{\'e}veillon}]{domingo2005dns}
Domingo, P., Vervisch, L., and R{\'e}veillon, J., \enquote{DNS analysis of
  partially premixed combustion in spray and gaseous turbulent flame-bases
  stabilized in hot air,} \emph{Combustion and Flame}, Vol. 140, No.~3, 2005,
  pp. 172--195.

\bibitem[{Ihme et~al.(2012)Ihme, Zhang, He, and Dally}]{ihme2012large}
Ihme, M., Zhang, J., He, G., and Dally, B., \enquote{Large-eddy simulation of a
  jet-in-hot-coflow burner operating in the oxygen-diluted combustion regime,}
  \emph{Flow, turbulence and combustion}, 2012, pp. 1--16.

\bibitem[{Huang et~al.(2014)Huang, Yoon, Gejji, Anderson, and
  Sankaran}]{huang2014computational}
Huang, C., Yoon, C., Gejji, R.~M., Anderson, W., and Sankaran, V.,
  \enquote{Computational study of combustion dynamics in a single-element lean
  direct injection gas turbine combustor,} \emph{52nd Aerospace Sciences
  Meeting}, 2014, p. 0620.

\bibitem[{Lacaze et~al.(2009)Lacaze, Cuenot, Poinsot, and
  Oschwald}]{lacaze2009large}
Lacaze, G., Cuenot, B., Poinsot, T., and Oschwald, M., \enquote{Large eddy
  simulation of laser ignition and compressible reacting flow in a rocket-like
  configuration,} \emph{Combustion and Flame}, Vol. 156, No.~6, 2009, pp.
  1166--1180.

\bibitem[{Hern{\'a}ndez-P{\'e}rez et~al.(2011)Hern{\'a}ndez-P{\'e}rez, Yuen,
  Groth, and G{\"u}lder}]{hernandez2011laboratory}
Hern{\'a}ndez-P{\'e}rez, F., Yuen, F., Groth, C., and G{\"u}lder, {\"O}.,
  \enquote{LES of a laboratory-scale turbulent premixed Bunsen flame using FSD,
  PCM-FPI and thickened flame models,} \emph{Proceedings of the Combustion
  Institute}, Vol.~33, No.~1, 2011, pp. 1365--1371.

\bibitem[{Srinivasan et~al.(2015)Srinivasan, Ranjan, and
  Menon}]{srinivasan2015flame}
Srinivasan, S., Ranjan, R., and Menon, S., \enquote{Flame dynamics during
  combustion instability in a high-pressure, shear-coaxial injector combustor,}
  \emph{Flow, Turbulence and Combustion}, Vol.~94, No.~1, 2015, pp. 237--262.

\bibitem[{Urbano et~al.(2016)Urbano, Selle, Staffelbach, Cuenot, Schmitt,
  Ducruix, and Candel}]{urbano2016exploration}
Urbano, A., Selle, L., Staffelbach, G., Cuenot, B., Schmitt, T., Ducruix, S.,
  and Candel, S., \enquote{Exploration of combustion instability triggering
  using large eddy simulation of a multiple injector liquid rocket engine,}
  \emph{Combustion and Flame}, Vol. 169, 2016, pp. 129--140.

\bibitem[{Staffelbach et~al.(2009)Staffelbach, Gicquel, Boudier, and
  Poinsot}]{staffelbach2009large}
Staffelbach, G., Gicquel, L., Boudier, G., and Poinsot, T., \enquote{Large Eddy
  Simulation of self excited azimuthal modes in annular combustors,}
  \emph{Proceedings of the Combustion Institute}, Vol.~32, No.~2, 2009, pp.
  2909--2916.

\bibitem[{Wolf et~al.(2012)Wolf, Balakrishnan, Staffelbach, Gicquel, and
  Poinsot}]{wolf2012using}
Wolf, P., Balakrishnan, R., Staffelbach, G., Gicquel, L.~Y., and Poinsot, T.,
  \enquote{Using LES to study reacting flows and instabilities in annular
  combustion chambers,} \emph{Flow, turbulence and combustion}, Vol.~88, No.
  1-2, 2012, pp. 191--206.

\bibitem[{Baukal~Jr et~al.(2000)Baukal~Jr, Gershtein, and
  Li}]{baukal2000computational}
Baukal~Jr, C.~E., Gershtein, V., and Li, X.~J., \emph{Computational fluid
  dynamics in industrial combustion}, CRC press, 2000.

\bibitem[{Frezzotti et~al.(2014)Frezzotti, Terracciano, Nasuti, Hester, and
  Anderson}]{frezzotti2014low}
Frezzotti, M.~L., Terracciano, A., Nasuti, F., Hester, S., and Anderson, W.~E.,
  \enquote{Low-order model studies of combustion instabilities in a DVRC
  combustor,} \emph{50th AIAA/ASME/SAE/ASEE Joint Propulsion Conference}, 2014,
  p. 3485.

\bibitem[{Sirignano and Popov(2013)}]{sirignano2013two}
Sirignano, W.~A., and Popov, P.~P., \enquote{Two-dimensional model for
  liquid-rocket transverse combustion instability,} \emph{AIAA Journal},
  Vol.~51, No.~12, 2013, pp. 2919--2934.

\bibitem[{Popov and Sirignano(2016)}]{popov2016transverse}
Popov, P.~P., and Sirignano, W.~A., \enquote{Transverse combustion instability
  in a rectangular rocket motor,} \emph{Journal of Propulsion and Power},
  Vol.~32, No.~1, 2016, pp. 620--627.

\bibitem[{You et~al.(2005)You, Huang, and Yang}]{you2005generalized}
You, D., Huang, Y., and Yang, V., \enquote{A generalized model of acoustic
  response of turbulent premixed flame and its application to gas-turbine
  combustion instability analysis,} \emph{Combustion Science and Technology},
  Vol. 177, No. 5-6, 2005, pp. 1109--1150.

\bibitem[{Lieu et~al.(2006)Lieu, Farhat, and Lesoinne}]{lieu2006reduced}
Lieu, T., Farhat, C., and Lesoinne, M., \enquote{Reduced-order fluid/structure
  modeling of a complete aircraft configuration,} \emph{Computer methods in
  applied mechanics and engineering}, Vol. 195, No. 41-43, 2006, pp.
  5730--5742.

\bibitem[{Lucia et~al.(2004)Lucia, Beran, and Silva}]{lucia2004reduced}
Lucia, D.~J., Beran, P.~S., and Silva, W.~A., \enquote{Reduced-order modeling:
  new approaches for computational physics,} \emph{Progress in Aerospace
  Sciences}, Vol.~40, No. 1-2, 2004, pp. 51--117.

\bibitem[{Durox et~al.(2009)Durox, Schuller, Noiray, and
  Candel}]{durox2009experimental}
Durox, D., Schuller, T., Noiray, N., and Candel, S., \enquote{Experimental
  analysis of nonlinear flame transfer functions for different flame
  geometries,} \emph{Proceedings of the Combustion Institute}, Vol.~32, No.~1,
  2009, pp. 1391--1398.

\bibitem[{Noiray et~al.(2008)Noiray, Durox, Schuller, and
  Candel}]{noiray2008unified}
Noiray, N., Durox, D., Schuller, T., and Candel, S., \enquote{A unified
  framework for nonlinear combustion instability analysis based on the flame
  describing function,} \emph{Journal of Fluid Mechanics}, Vol. 615, 2008, pp.
  139--167.

\bibitem[{Chatterjee(2000)}]{chatterjee2000introduction}
Chatterjee, A., \enquote{An introduction to the proper orthogonal
  decomposition,} \emph{Current science}, 2000, pp. 808--817.

\bibitem[{Lucia and Beran(2003)}]{lucia2003projection}
Lucia, D.~J., and Beran, P.~S., \enquote{Projection methods for reduced order
  models of compressible flows,} \emph{Journal of Computational Physics}, Vol.
  188, No.~1, 2003, pp. 252--280.

\bibitem[{Xu and Duraisamy(2017)}]{xu2017reduced}
Xu, J., and Duraisamy, K., \enquote{Reduced-Order Modeling of Model Rocket
  Combustors,} \emph{53rd AIAA/SAE/ASEE Joint Propulsion Conference}, 2017, p.
  4918.

\bibitem[{Xu et~al.(2018)Xu, Huang, and Duraisamy}]{xu2018multi}
Xu, J., Huang, C., and Duraisamy, K., \enquote{Multi-Domain Reduced-Order
  Modeling with Sparse Acceleration of Combustion Instability,} \emph{2018
  Joint Propulsion Conference}, 2018, p. 4680.

\bibitem[{Huang et~al.(2018{\natexlab{a}})Huang, Xu, Duraisamy, and
  Merkle}]{huang2018exploration}
Huang, C., Xu, J., Duraisamy, K., and Merkle, C., \enquote{Exploration of
  Reduced-Order Models for Rocket Combustion Applications,} \emph{2018 AIAA
  Aerospace Sciences Meeting}, 2018{\natexlab{a}}, p. 1183.

\bibitem[{Huang et~al.(2016{\natexlab{a}})Huang, Anderson, Merkle, and
  Sankaran}]{huang2016multi}
Huang, C., Anderson, W.~E., Merkle, C.~L., and Sankaran, V.,
  \enquote{Multi-Fidelity Framework for Modeling Combustion Instability,} Tech.
  rep., AFRL/RQR Edwards AFB United States, 2016{\natexlab{a}}.

\bibitem[{Huang et~al.(2017)Huang, Anderson, and Merkle}]{huang2017multi}
Huang, C., Anderson, W.~E., and Merkle, C., \enquote{Multi-fidelity Framework
  Explorations for Nonlinear Euler Equations,} \emph{53rd AIAA/SAE/ASEE Joint
  Propulsion Conference}, 2017.

\bibitem[{Rowley et~al.(2004)Rowley, Colonius, and Murray}]{rowley2004model}
Rowley, C.~W., Colonius, T., and Murray, R.~M., \enquote{Model reduction for
  compressible flows using POD and Galerkin projection,} \emph{Physica D:
  Nonlinear Phenomena}, Vol. 189, No. 1-2, 2004, pp. 115--129.

\bibitem[{Ravindran(2000)}]{ravindran2000reduced}
Ravindran, S.~S., \enquote{A reduced-order approach for optimal control of
  fluids using proper orthogonal decomposition,} \emph{International journal
  for numerical methods in fluids}, Vol.~34, No.~5, 2000, pp. 425--448.

\bibitem[{Amsallem and Farhat(2008)}]{amsallem2008interpolation}
Amsallem, D., and Farhat, C., \enquote{Interpolation method for adapting
  reduced-order models and application to aeroelasticity,} \emph{AIAA Journal},
  Vol.~46, No.~7, 2008, pp. 1803--1813.

\bibitem[{Lieu and Farhat(2007)}]{lieu2007adaptation}
Lieu, T., and Farhat, C., \enquote{Adaptation of aeroelastic reduced-order
  models and application to an F-16 configuration,} \emph{AIAA Journal},
  Vol.~45, No.~6, 2007, pp. 1244--1257.

\bibitem[{Barbagallo et~al.(2012)Barbagallo, Dergham, Sipp, Schmid, and
  Robinet}]{barbagallo2012closed}
Barbagallo, A., Dergham, G., Sipp, D., Schmid, P.~J., and Robinet, J.-C.,
  \enquote{Closed-loop control of unsteadiness over a rounded backward-facing
  step,} \emph{Journal of Fluid Mechanics}, Vol. 703, 2012, pp. 326--362.

\bibitem[{Barbagallo et~al.(2011)Barbagallo, Sipp, and
  Schmid}]{barbagallo2011input}
Barbagallo, A., Sipp, D., and Schmid, P.~J., \enquote{Input--output measures
  for model reduction and closed-loop control: application to global modes,}
  \emph{Journal of Fluid Mechanics}, Vol. 685, 2011, pp. 23--53.

\bibitem[{Huang et~al.(2019{\natexlab{a}})Huang, Anderson, Merkle, and
  Sankaran}]{huang2019multifidelity}
Huang, C., Anderson, W.~E., Merkle, C., and Sankaran, V.,
  \enquote{Multifidelity Framework for Modeling Combustion Dynamics,}
  \emph{AIAA Journal}, 2019{\natexlab{a}}.

\bibitem[{Smith et~al.(2008)Smith, Ellis, Xia, Sankaran, Anderson, and
  Merkle}]{smith2008computational}
Smith, R., Ellis, M., Xia, G., Sankaran, V., Anderson, W., and Merkle, C.,
  \enquote{Computational investigation of acoustics and instabilities in a
  longitudinal-mode rocket combustor,} \emph{AIAA Journal}, Vol.~46, No.~11,
  2008, pp. 2659--2673.

\bibitem[{Yu et~al.(2012)Yu, Sisco, Rosen, Madhav, and
  Anderson}]{yu2012spontaneous}
Yu, Y., Sisco, J.~C., Rosen, S., Madhav, A., and Anderson, W.~E.,
  \enquote{Spontaneous longitudinal combustion instability in a
  continuously-variable resonance combustor,} \emph{Journal of Propulsion and
  Power}, Vol.~28, No.~5, 2012, pp. 876--887.

\bibitem[{Yu(2009)}]{yu2009experimental}
Yu, Y.~C., \enquote{Experimental and analytical investigations of longitudinal
  combustion instability in a continuously variable resonance combustor
  (CVRC),} Ph.D. thesis, Purdue University, 2009.

\bibitem[{Huang et~al.(2019{\natexlab{b}})Huang, Duraisamy, and
  Merkle}]{huang2019investigations}
Huang, C., Duraisamy, K., and Merkle, C., \enquote{Investigations and
  Improvement of Robustness of Reduced-Order Models of Reacting Flow,}
  \emph{AIAA Scitech 2019 Forum}, 2019{\natexlab{b}}, p. 2012.

\bibitem[{Huang et~al.(2018{\natexlab{b}})Huang, Duraisamy, and
  Merkle}]{huang2018challenges}
Huang, C., Duraisamy, K., and Merkle, C., \enquote{Challenges in Reduced Order
  Modeling of Reacting Flows,} \emph{2018 Joint Propulsion Conference},
  2018{\natexlab{b}}, p. 4675.

\bibitem[{Frezzotti et~al.(2015{\natexlab{a}})Frezzotti, Nasuti, Huang, Merkle,
  and Anderson}]{frezzotti2015response}
Frezzotti, M.~L., Nasuti, F., Huang, C., Merkle, C., and Anderson, W.~E.,
  \enquote{Response Function Modeling in the Study of Longitudinal Combustion
  Instability by a Quasi-1D Eulerian Solver,} \emph{51st AIAA/SAE/ASEE Joint
  Propulsion Conference}, 2015{\natexlab{a}}, p. 3840.

\bibitem[{Frezzotti et~al.(2015{\natexlab{b}})Frezzotti, Nasuti, Huang, Merkle,
  and Anderson}]{frezzotti2015parametric}
Frezzotti, M.~L., Nasuti, F., Huang, C., Merkle, C., and Anderson, W.~E.,
  \enquote{Parametric Analysis of Response Function in Modeling Combustion
  Instability by a Quasi-1D Solver,} \emph{6th EUROPEAN CONFERENCE FOR
  AEROSPACE SCIENCES}, 2015{\natexlab{b}}.

\bibitem[{Crocco et~al.(1958)Crocco, Grey, and Harrje}]{crocco1958importance}
Crocco, L., Grey, J., and Harrje, D., \enquote{On the importance of the
  sensitive time lag in longitudinal high-frequency rocket combustion
  instability,} \emph{Jet Propulsion}, Vol.~28, No.~12, 1958, pp. 841--843.

\bibitem[{Parish et~al.(2018)Parish, Wentland, and Duraisamy}]{ericpaper}
Parish, E., Wentland, C., and Duraisamy, K., \enquote{A Residual-Based
  Petrov-Galerkin Reduced Order Model with Memory Effects,} \emph{Submitted,
  Journal of Computational Physics}, 2018.

\bibitem[{Roe(1986)}]{roe1986characteristic}
Roe, P.~L., \enquote{Characteristic-based schemes for the Euler equations,}
  \emph{Annual review of fluid mechanics}, Vol.~18, No.~1, 1986, pp. 337--365.

\bibitem[{Grenda et~al.(1995)Grenda, Venkateswaran, and
  Merkle}]{grenda1995application}
Grenda, J., Venkateswaran, S., and Merkle, C., \enquote{Application of
  computational fluid dynamics techniques to engine instability studies,}
  \emph{PROGRESS IN ASTRONAUTICS AND AERONAUTICS}, Vol. 169, 1995, pp.
  503--528.

\bibitem[{Wang et~al.(2018)Wang, Hesthaven, and Ray}]{wang2018non}
Wang, Q., Hesthaven, J.~S., and Ray, D., \enquote{Non-intrusive reduced order
  modeling of unsteady flows using artificial neural networks with application
  to a combustion problem,} Tech. rep., 2018.

\bibitem[{Huang et~al.(2016{\natexlab{b}})Huang, Anderson, Harvazinski, and
  Sankaran}]{huang2016analysis}
Huang, C., Anderson, W.~E., Harvazinski, M.~E., and Sankaran, V.,
  \enquote{Analysis of self-excited combustion instabilities using
  decomposition techniques,} \emph{AIAA Journal}, 2016{\natexlab{b}}, pp.
  2791--2807.

\end{thebibliography}
\end{document}